\documentclass[acmtog]{acmart}

\usepackage{algorithm}
\usepackage{subcaption}
\usepackage{graphicx}
\usepackage{amsmath}
\usepackage[noend]{algpseudocode} 
\usepackage{booktabs} 
\usepackage{prettyref}
\usepackage{wrapfig}
\usepackage{multirow}
\usepackage{changepage}
\newrefformat{fig}{Fig.~\ref{#1}}
\newrefformat{par}{Sec.~\ref{#1}}
\newrefformat{appen}{App.~\ref{#1}}
\newrefformat{sec}{Sec.~\ref{#1}}
\newrefformat{sub}{Sec.~\ref{#1}}
\newrefformat{table}{Tab.~\ref{#1}}
\newrefformat{ass}{Ass.~\ref{#1}}
\newrefformat{alg}{Alg.~\ref{#1}}
\newrefformat{def}{Def.~\ref{#1}}
\newrefformat{thm}{Thm.~\ref{#1}}
\newrefformat{cor}{Cor.~\ref{#1}}
\newrefformat{lem}{Lem.~\ref{#1}}
\newrefformat{step}{Step~\ref{#1}}
\newrefformat{ln}{Line~\ref{#1}}
\newrefformat{eq}{Eq.~\ref{#1}}
\newrefformat{pb}{Prob.~\ref{#1}}
\newrefformat{it}{Item~\ref{#1}}
\newrefformat{te}{Term~\ref{#1}}
\def\Eqref Eq:#1:{\eqref{eq:#1}}

\algnewcommand{\LineComment}[1]{\State \(\triangleright\) #1}
\newcommand{\TWORCell}[2]{\begin{tabular}{@{}c@{}}#1 \\ #2\end{tabular}}

\acmPrice{15.00}

\setcopyright{acmlicensed}
\acmJournal{TOG}
\acmYear{2023} \acmVolume{42} \acmNumber{6} \acmArticle{} \acmMonth{12} \acmPrice{15.00}\acmDOI{10.1145/3618348}

\setcitestyle{square}

\begin{document}

\title{Learning based 2D Irregular Shape Packing}

\author{Zeshi Yang}
\orcid{0000-0001-9680-8132}
\email{zs243@mail.ustc.edu.cn}
\affiliation{%
 \institution{LightSpeed Studios}
 \city{Seattle}
 \state{WA}
 \country{USA}
}

\author{Zherong Pan}
\orcid{0000-0001-9348-526X}
\email{zrpan@global.tencent.com}
\affiliation{%
  \institution{LightSpeed Studios}
  \city{Seattle}
  \state{WA}
  \country{USA}
}

\author{Manyi Li}
\orcid{0000-0002-5251-0462}
\email{manyili@sdu.edu.cn}
\affiliation{%
 \institution{Shandong University}
 \city{Jinan}
 \state{Shandong}
 \country{China}
}
\author{Kui Wu}
\orcid{0000-0003-3326-7943}
\email{kwwu@global.tencent.com}
\affiliation{%
  \institution{LightSpeed Studios}
  \city{Los Angeles}
  \state{CA}
  \country{USA}
}
\author{Xifeng Gao}
\orcid{0000-0003-0829-7075}
\email{xifgao@global.tencent.com}
\affiliation{%
  \institution{LightSpeed Studios}
  \city{Seattle}
  \state{WA}
  \country{USA}
}

\definecolor{myBlack}{rgb}{0,0.,0}
\definecolor{myRed}{rgb}{1,0.,0.}
\definecolor{myLightRed}{rgb}{0.7,0.,0.}
\definecolor{myOrange}{rgb}{0.8,0.3,0.}
\definecolor{myGreen}{rgb}{0,0.5,0.27}
\definecolor{myBlue}{rgb}{0,0.0,0.9}
\definecolor{myPurple}{rgb}{0.6,0.,0.6}
\definecolor{myBluegreen}{rgb}{0.2,0.6,0.8}
\newcommand{\highlightA}[1]{\textcolor{myOrange}{#1}}
\newcommand{\highlightB}[1]{\textcolor{myBlue}{#1}}
\newcommand{\tmp}[1]{\textcolor{myRed}{#1}}
\newcommand{\zeshi}[1]{\textcolor{myBlack}{#1}}
\newcommand{\zherong}[1]{\textcolor{myOrange}{ZP: #1}}
\newcommand{\manyi}[1]{\textcolor{myBlue}{Manyi: #1}}
\newcommand{\xifeng}[1]{\textcolor{myLightRed}{XF: #1}}
\newcommand{\kui}[1]{\textcolor{myBluegreen}{Kui: #1}}
\newcommand{\TODO}[1]{\textcolor{myRed}{#1}}

\begin{abstract}
2D irregular shape packing is a necessary step to arrange UV patches of a 3D model within a texture atlas for memory-efficient appearance rendering in computer graphics. Being a joint, combinatorial decision-making problem involving all patch positions and orientations, this problem has well-known NP-hard complexity. Prior solutions either assume a heuristic packing order or modify the upstream mesh cut and UV mapping to simplify the problem, which either limits the packing ratio or incurs robustness or generality issues. Instead, we introduce a learning-assisted 2D irregular shape packing method that achieves a high packing quality with minimal requirements from the input. Our method iteratively selects and groups subsets of UV patches into near-rectangular super patches, essentially reducing the problem to bin-packing, based on which a joint optimization is employed to further improve the packing ratio. In order to efficiently deal with large problem instances with hundreds of patches, we train deep neural policies to predict nearly rectangular patch subsets and determine their relative poses, leading to linear time scaling with the number of patches. We demonstrate the effectiveness of our method on three datasets for UV packing, where our method achieves a higher packing ratio over several widely used baselines with competitive computational speed.
\end{abstract}



\begin{teaserfigure}
\centering
\resizebox{\linewidth}{!}{
\begin{tabular}{ccccccccc}
\includegraphics[scale=.1]{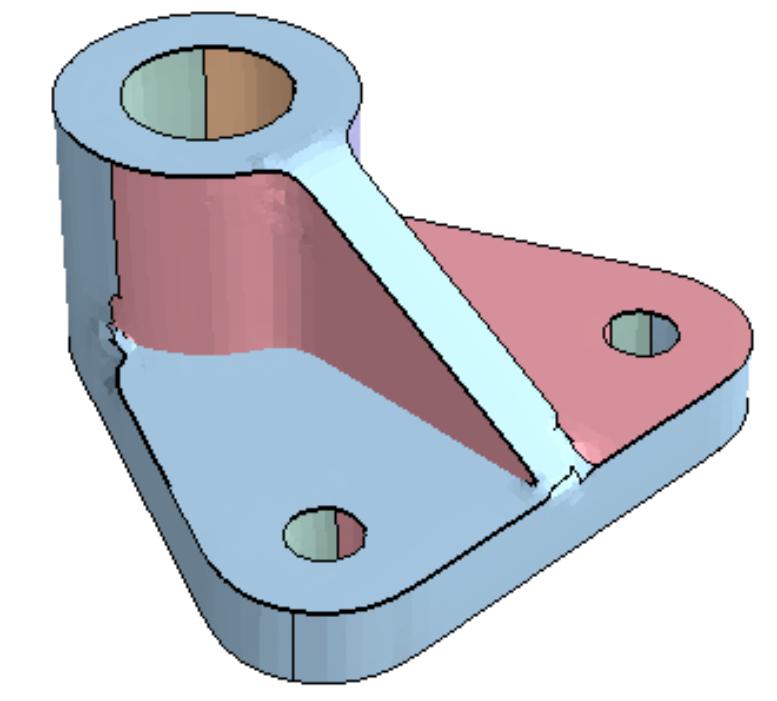}&
\includegraphics[scale=.1]{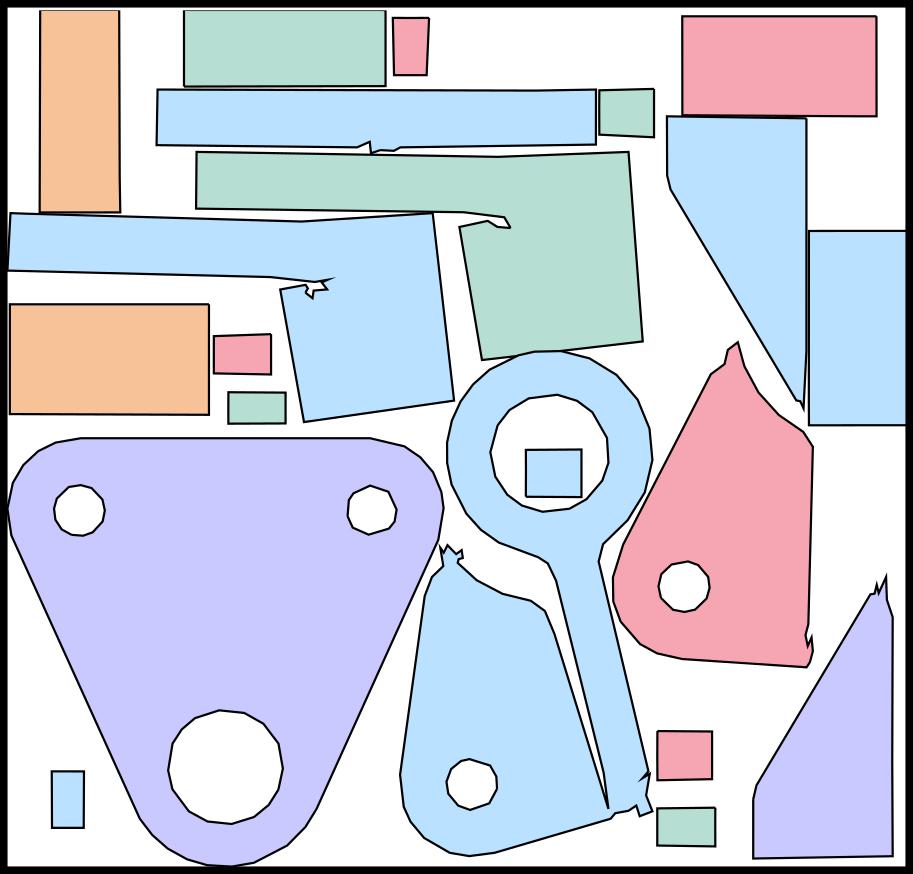}&
\includegraphics[scale=.1]{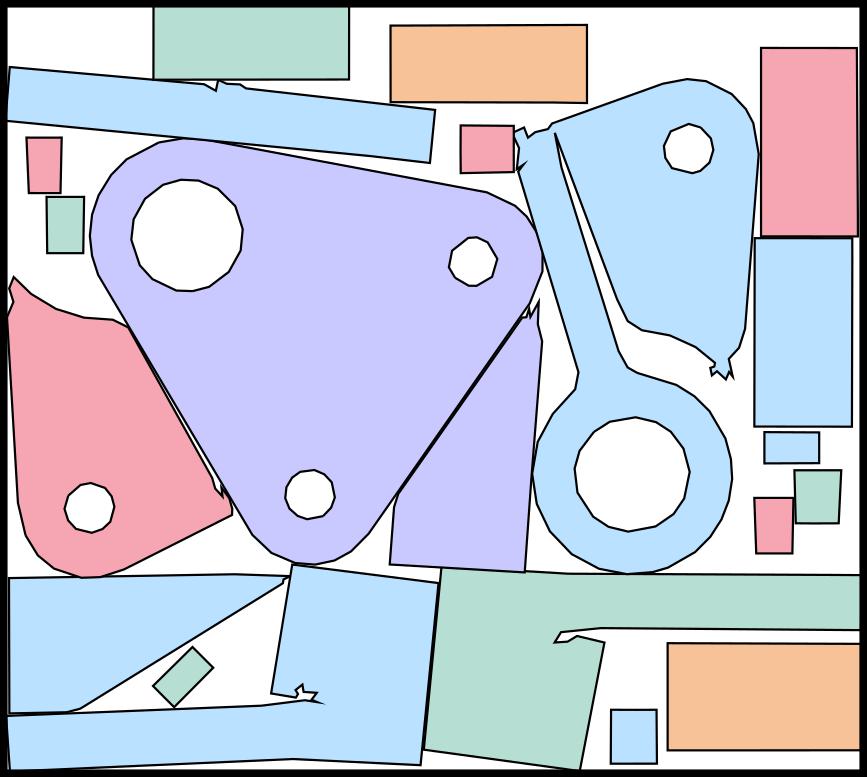}&
\includegraphics[scale=.15]{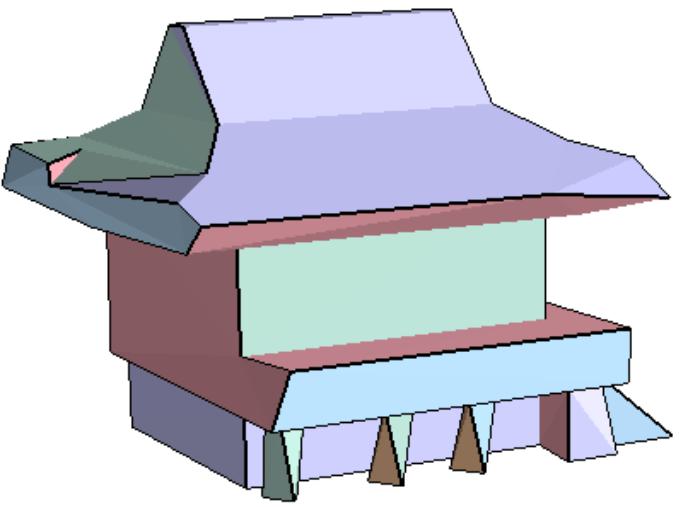}&
\includegraphics[scale=.1]{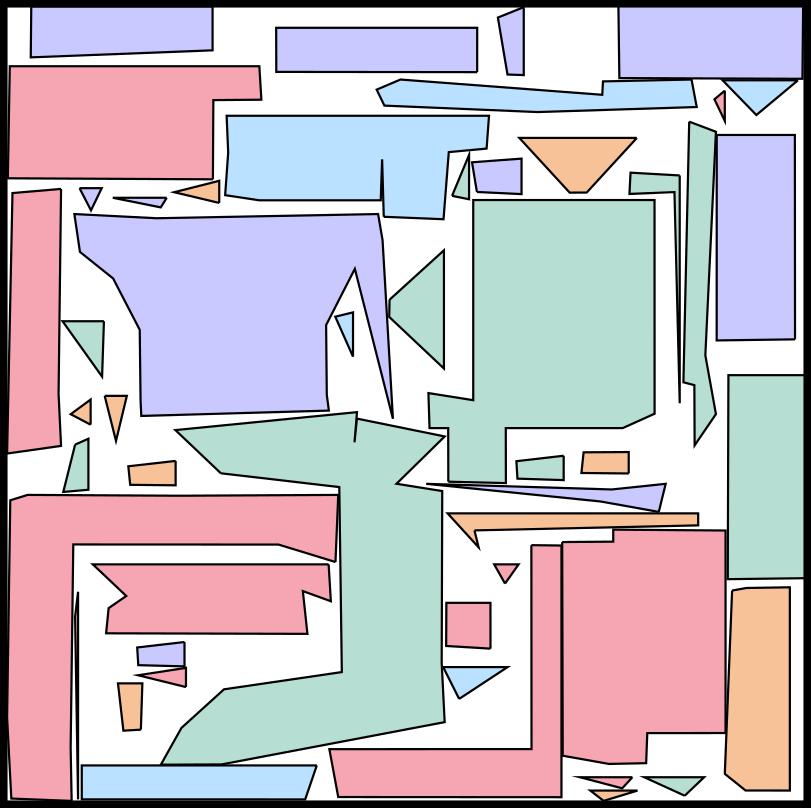}&
\includegraphics[scale=.1]{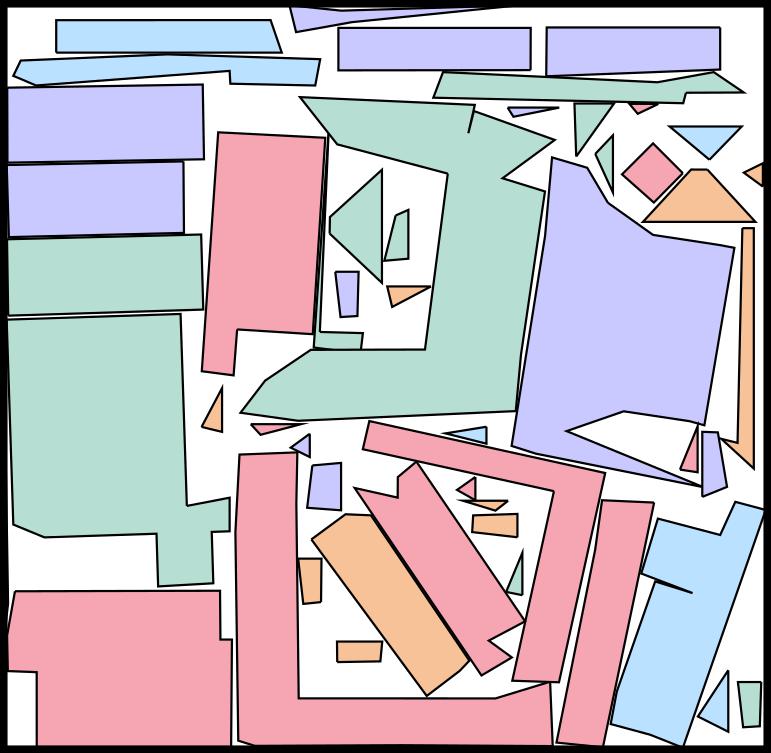}&
\includegraphics[scale=.1]{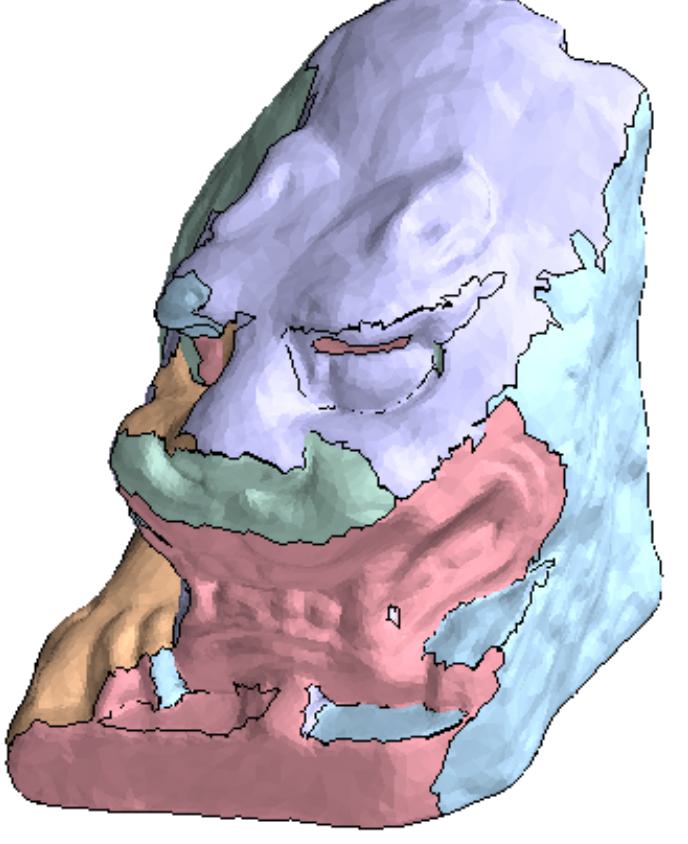}&
\includegraphics[scale=.1]{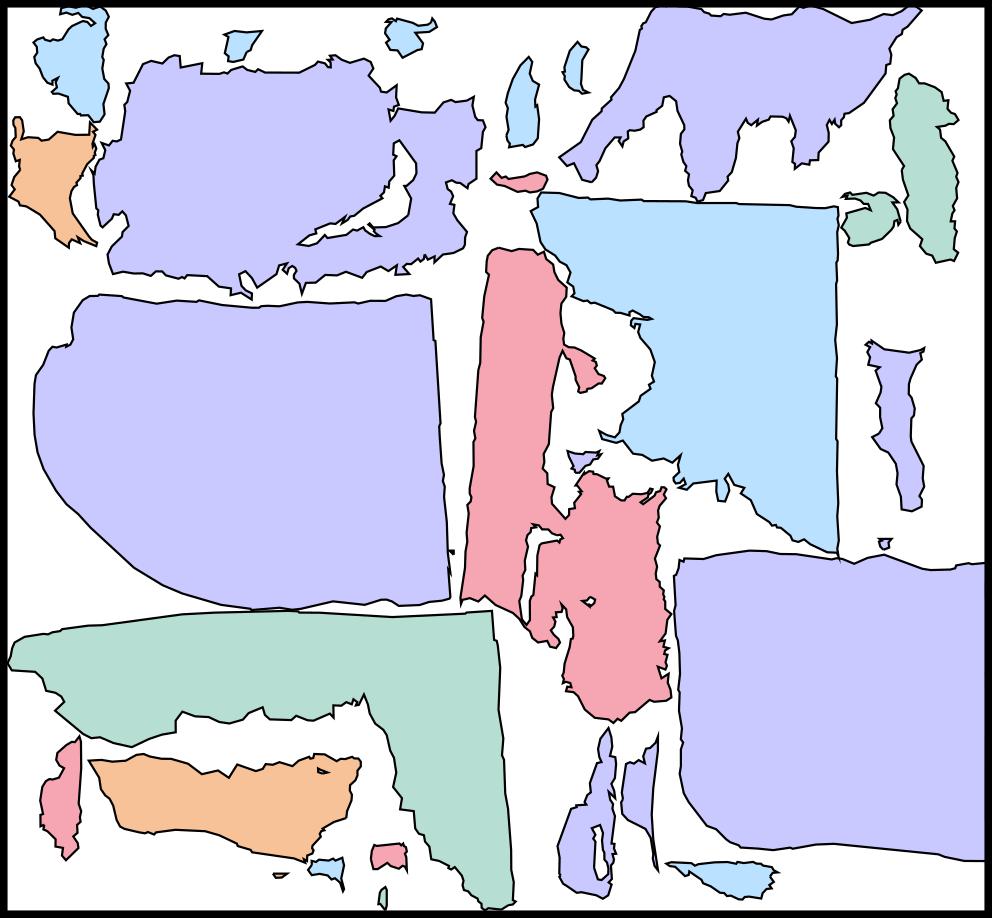}&
\includegraphics[scale=.1]{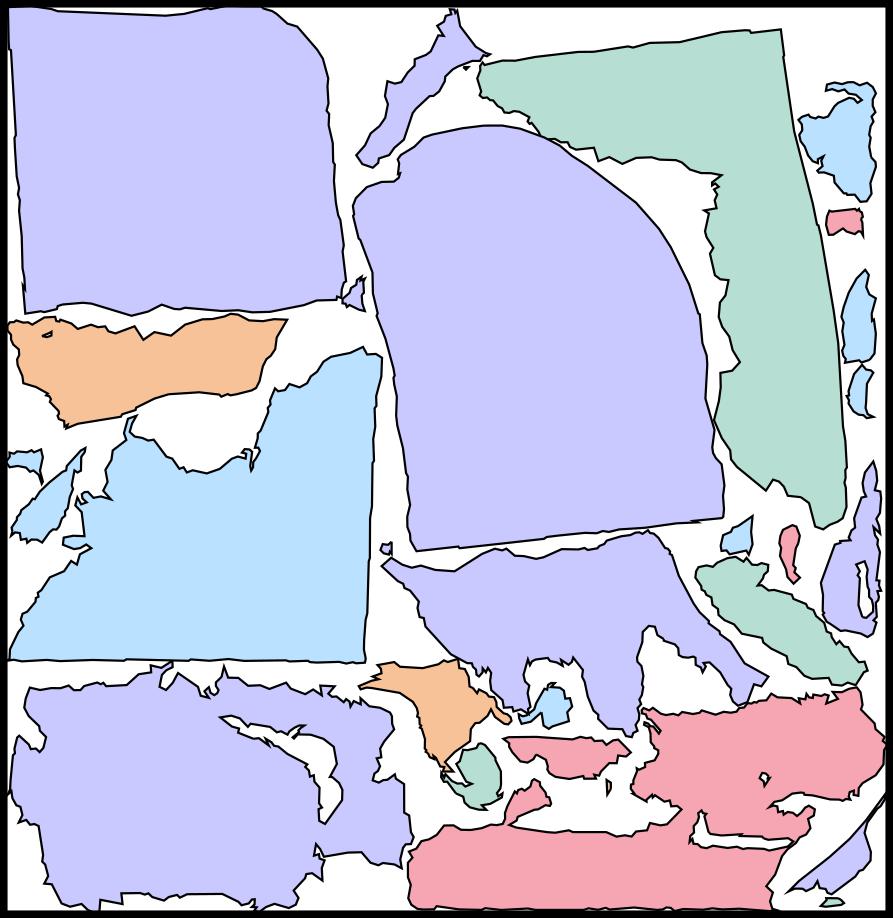}\\
 & XAtlas pr: 69.8\% & Ours pr: 77.4\% &
 & XAtlas pr: 66.2\% & Ours pr: 76.4\% &
 & XAtlas pr: 61.0\% & Ours pr: 72.9\%
\end{tabular}}
\vspace{-10px}
\caption{\label{fig:teaser}Three UV-packing results generated using XAtlas and our learning-assisted method.}
\end{teaserfigure}
\maketitle
\section{Introduction\label{sec:intro}}
2D irregular shape packing lies the theoretical foundation for a wide spectrum of applications across various industrial areas, e.g. material parts assembly~\cite{ke2020electric}, VLSI module placements~\cite{cheng2005floorplanning}, and 2D shelf arrangements in automatic warehouses~\cite{onut2008particle}. The most prominent packing application in computer graphics is UV chart packing, where a set of 2D shapes corresponding to patches of 3D models are packed together into a single texture atlas for mapping various appearance details onto the 3D surface for downstream rendering. In the realm of digital games where many 3D models are rendered in the virtual world, improving the UV packing ratio could significantly save the graphical memory, improve the loading speed, and reduce rendering overhead. This problem was recognized decades ago, e.g., by~\citet{soucy1996texture} and~\citet{sander2001texture}, which has recently revived as an active area of graphic research~\cite{limper2018box,Liu_AAAtlas_2019,zhang2020robust}.

\zeshi{An packing algorithm needs to search for both the position and orientation of each patch. Due to its combinatorial nature, finding optimal 2D shape packing has been well-understood as being NP-complete and APX-hard. Prior research efforts are focused on heuristic or genetic algorithms to determine the order of objects to be packed, and then search for the pose of each shape, e.g., using no-fit polygon (NFP) and the ``Tetris'' algorithm. However, the sub-optimality gap of these heuristics can be substantial. On the other hand, when the problem size is small, globally optimal stochastic optimization algorithms have been proposed to search for the near-optimal packing orders. Although these techniques can scale to tens of 2D shapes, UV unwrapping approaches and commercial 3D modeling software can oftentimes generate hundreds of patches for one 3D model, which is far beyond the capabilities of the global search algorithms.}

\zeshi{We propose a learning-assisted 2D packing pipeline for general irregular-shaped patches. The design of our approach is based on two observations: 1) Prior works generate sub-optimal solutions by assuming objects are packed sequentially. 2) Prior works try all possible poses and orientations for each to-be-packed object to find the best solution, leading to slow computation. To tackle these issues, we propose to hierarchically group patches into nearly rectangular super-patches, allowing a larger search space for patch combination and a smaller optimality gap. Instead of exhaustively trying all possible combinations, we train a high-level group selector network (HSN) to efficiently predict how likely a candidate patch subset can be grouped into a rectangular super-patch. Given an identified patch subset, we use the sequential ordering technique, similar to prior works. Specifically, we use a low-level sorter network (LSN) to determine the suitable order of packing within the subset. Next, a low-level pose network (LPN) infers the rough initial pose of each patch. The ultimate pose of these patches is determined by local numerical optimization. These two networks are trained using reinforcement learning to efficiently infer near optimal sequential packing policies.}

We have evaluated our method on 3 datasets, containing highly irregular UV patches generated from both organic and man-made 3D models. We show the effectiveness of our approach by comparing against widely used algorithm baselines, including piecewise linear NFP, XAtlas~\cite{Xatlas}, and~\cite{sander2003multi}, where we achieve a $5\%-10\%$ packing ratio improvement across all the tested datasets. We further highlight the generality of our learned policies, by training them on one dataset and evaluating them on all three datasets. We find our method still outperforming baselines on unseen datasets, proving that our method does not need to be retrained for each problem domain. We include the full datasets and results with statistics in the supplementary material.
\begin{figure*}[ht]
\centering
\includegraphics[trim=1cm 3cm 2cm 4cm,clip,width=\linewidth]{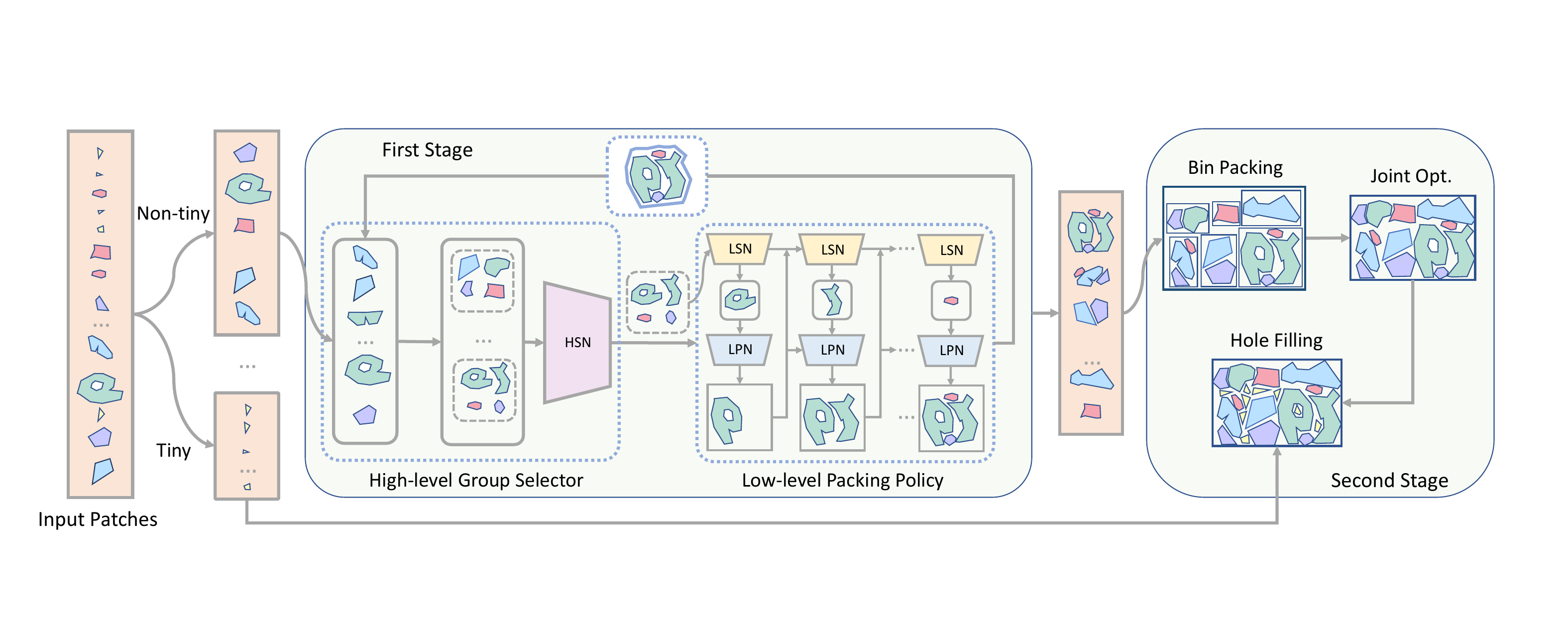}
\caption{\label{fig:pipeline} The overall pipeline of our learning-assisted packing approach.} 
\end{figure*}
\section{Related Work}
We discuss representative related works on UV-atlas generation, irregular shape packing, and learning-based packing algorithms.

\subsection{UV-atlas Generation}
The standard industrial pipeline for generating UV-atlas involves three stages: cutting, parameterization, and packing. Mesh parameterization minimizes various distortion energies as summarized in~\cite{rabinovich2017scalable}. 
To realize strict distortion bounds, \citet{sorkine2002bounded} proposed greedy, simultaneous cutting and parameterization that compromises between the distortion and cut length. Similarly, \citet{levy2002least} proposed greedy mesh cutting to avoid overlapping. More recent approaches~\cite{poranne2017autocuts} formulate the cut and parameterization under a joint optimization framework, leading to more optimal solutions. Under a strict distortion bound, however, these joint optimization techniques can lead to a large number of patches being handled by the packer. 

To handle the resultant large packing problems, \citet{sander2001texture} used the bounding box approximation for each patch, simplifying the problem to a bin packing. \citet{levy2002least} proposed to solve the irregular shape packing by always putting new patches on top of existing ones, where the exact horizontal position is chosen to minimize the wasted space between vertical boundaries. \citet{noll2011efficient} revised the irregular packing technique~\cite{levy2002least} and considered both horizontal and vertical boundaries, including holes. Although these techniques are less accurate than exact NFP-based methods~\cite{bennell2008comprehensive}, their computational costs are much lower for large problem instances. On the downside, all these packing algorithms are myopic and assume a fixed packing order, potentially limiting their solution quality.

Instead of packing UV patches at a separate stage, several works propose to optimize mesh cuts and patch shapes for higher packing ratio. \citet{limper2018box} proposed to progressively cut along void box boundaries to improve packing ratio. \citet{schertler2018generalized} generated motorcycle graph for quad-dominant meshes consisting of rectangular patches, leading to perfect packing ratios. \citet{Liu_AAAtlas_2019} generalized the idea of motorcycle graph to arbitrary meshes by deforming arbitrarily shaped patches to be nearly axis-aligned. Although these techniques are appealing for a high packing ratio, they change either the topology or geometry of the input UV patches. We choose to solve packing by maintaining the original shapes of UV patches for better conformity with existing pipelines.

\subsection{General Irregular Shape Packing}
Due to the theoretical hardness~\cite{hartmanis1982computers,10.1287/moor.1050.0168} and practical importance, the irregular packing problem has garnered research attention over the past decades. In the community of manufacturing design~\cite{cheng2005floorplanning,onut2008particle,attene2015shapes,wang20223d}, researchers have developed algorithms to compute highly efficient packing solutions via a bilevel pipeline, where the high-level algorithm computes an order of the shapes for packing and the low-level algorithm determines the affine transformation for each shape regarding the already packed shapes. Early research works focus on low-level geometric algorithms that determine all possible collision-free poses for placing a new shape, resulting in the useful tool of NFP~\cite{bennell2008comprehensive} and fast collision checking~\cite{riff2009revision}. The most efficient method for computing the NFP is through the Minkowski sum~\cite{bennell2008comprehensive}, which incurs a complexity of at least $\mathcal{O}(N^2)$ with $N$ being the number of edges in each polygon. The ultimate pose of each shape is then chosen from the potential pose set using simple heuristic rules as summarized in~\cite{10.3389/fmech.2022.966691}, where prominent rules include bottom-left-first, maximal-packing-ratio, and minimal-boundary-length. Unfortunately, even computing the exact NFP for hundreds and thousands of UV patches is intractable, and practical algorithms~\cite{levy2002least,noll2011efficient} only consider horizontal or vertical boundaries.

Built off of the low-level algorithm, a high-level shape selector further optimizes the packing ratio by manipulating the order of packing. This is a critical step to high-quality packing and researchers propose a series of global optimization algorithms, including brute force search~\cite{crainic2009ts2pack}, genetic algorithm~\cite{ke2020electric}, and simulated annealing~\cite{gomes2006solving}. However, each search step of these algorithms involves calling the low-level algorithms, which are intractable for UV packing, so existing UV packing algorithms such as~\cite{noll2011efficient} simply sort the patches in area-descending order.

\subsection{Learned Packing Policy}
Machine learning bears significant potential in solving combinatorial optimization~\cite{bengio2021machine}. Several recent works have applied learning approaches to various packing problems, most of which~\cite{hu2020tap,jiang2021learning,zhao2021learning,zhao2022learning,10018146} deal with 3D bin packing problems. In particular, \citet{zhao2021learning} and \citet{10018146} only deal with online packing problems where the high-level packing order is fixed. In contrast, \citet{hu2020tap} focuses on learning the physically realizable packing order. They trained an attention network to locate the next object to be packed, given their precedence constraints. Similar to our technique, \citet{jiang2021learning} and \citet{zhao2022learning} presented a complete pipeline involving learned high-level object selectors and low-level pose inference.

Learning irregular packing skills is a much more challenging problem, where the learning model has to recognize the complex object shapes and handle the continuous decision space of object poses. Two recent works~\cite{goyal2020packit,fang2023hybrid} considered packing of irregular shapes. \citet{goyal2020packit} proposed a problem set for 3D packing problems and showed that a neural shape selector trained via reinforcement learning outperforms heuristic baselines. \citet{fang2023hybrid} also trained a neural shape selector via reinforcement learning. Both algorithms use heuristic rules for the low-level pose computation and \citet{fang2023hybrid} even used NFP-based pose computation. However, only learning the high-level selector is not enough for UV packing, since low-level pose selection can still be a major computational bottleneck. 
\section{Overview}
Our method takes as input a set of UV patches, each represented as an irregular, planar, manifold triangular mesh in their local frame of reference. We allow an arbitrary number of holes within each UV patch. The output of our algorithm is a set of rigid transformations, one for each patch, such that, after being transformed, the UV patches are tightly packed into a rectangular texture domain in the global frame of reference and in a collision-free manner. The pipeline of our method is illustrated in~\prettyref{fig:pipeline}. 

Given the set of UV patches, we filter out the tiny patches and firstly process the non-tiny ones in two stages. Our first stage aims to turn the non-tiny patches into nearly rectangular super-patches (\prettyref{sec:rectagular}). We maintain the set $\mathcal{S}$ of super-patches that is initialized to be the non-tiny patches. We first employ a sampling-based group selector network (HSN~\prettyref{sec:highLevelSelect}) to identify a potential subset $\mathcal{S}'$ of at most $H$ patches that can form a nearly rectangular super-patch. To compute the super-patch, we utilize a low-level sorter network (LSN~\prettyref{sec:lowLevelOrder}) and a low-level pose network (LPN~\prettyref{sec:lowLevelPack}). LSN re-orders the patches within the subset and sequentially outputs the next patch to be packed. LPN selects the rough initial pose for each incoming patch, whose ultimate pose is adjusted using a local numerical optimization. This procedure is denoted as the low-level function $\text{LL}(\mathcal{S}')$. Once the super-patch is generated, it will be inserted back into $\mathcal{S}$ to replace the subset of patches, essentially updating $\mathcal{S}$ to $\mathcal{S}-\mathcal{S}'\cup\{\text{LL}(\mathcal{S}')\}$. The above process is repeated until each (super-)patch in $\mathcal{S}$ is nearly rectangular. Our second stage (\prettyref{sec:global}) assembles all the patches using a heuristic bin-packing algorithm. A joint local optimization is then performed to squeeze all the patches as tightly as possible. Finally, the tiny patches filtered out in the beginning will be put into gaps between non-tiny patches. In the following, we describe the two stages in detail. We summarize our network architectures and pseudo-code of our algorithm in our supplementary material. 

\section{\label{sec:rectagular}Near Rectangular Patch Generation}
Although we use high-level selector networks before low-level policies, we need to train low-level policies first due to data dependency and we introduce our method in the order of training.

\subsection{\label{sec:lowLevelPack}Low-Level Pose Network (LPN)}
Given a subset of $H$ patches that are selected by HSN, let us assume the first $i-1$ patches have been packed into a super-patch $P_{i-1}$ in the global frame, and $p_i$ is the geometric domain of the $i$-th patch in the local frame. Given $p_i$ and $P_{i-1}$, the low-level packing algorithm needs to select the translation $t_i$ and rotation $\theta_i$ for $p_i$ such that the packed shape $P_i\triangleq\left[R(\theta_i)p_i+t_i\right]\cup P_{i-1}$ is collision-free with a high packing ratio. Conventional packing algorithms~\cite{10.3389/fmech.2022.966691} would consider each patch independently and evenly sample $K$ rotations and consider all the possible translations under each rotation using NFP algorithm, leading to at least $\mathcal{O}(KN^2)$ complexity with $N$ being the total number of edges in $p_i$ and $P_{i-1}$, which is a major bottleneck of packing algorithms. And due to its myopic nature, the packing ratio is sub-optimal. 

To address the above two shortcomings, we propose to model the packing procedure as a Markov Decision Process (MDP) and train LPN to maximize the packing ratio via reinforcement learning. Being aware of not only the current patch but also the future incoming patches, our LPN policy exhibits a small optimality gap. Briefly, the MDP is identified with a tuple $<S,A,\tau,r>$, which models the procedure of a decision-maker iteratively observing the current system state in the state space $S$ and taking an action in the action space $A$ to change the environment. The environment would then update its state via a state transition function $\tau$ and the decision-maker receives a reward $r$. We refer readers to~\cite{sutton2018reinforcement} for more details on this model. For the packing problem, however, the action space could involve all possible patch translations and rotations, which is notoriously difficult to handle for reinforcement learning. Instead, we propose to restrict the action space to a small discrete subset, and then use local optimization to fine-tune the final pose of each patch.

\setlength{\columnsep}{10pt}
\begin{wrapfigure}{r}{0.5\linewidth}
\centering
\includegraphics[trim=24.5cm 7.2cm 24.7cm 11.6cm,clip,width=\linewidth]{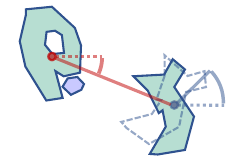}
\put(-9,38){\small $\theta_i$}
\put(-68,48){\small $\phi_i$}
\put(-100,23){\small{$P_{i-1}$}}
\put(-23,10){\small{$p_i$}}
\caption{\label{fig:action}An illustration of our action space. Given the already packed patch $P_{i-1}$ and the current patch $p_i$, our action needs to determine the rough rotation $\theta_i$ of $p_i$ from local (dashed) to global frame, as well as the relative rotation $\phi_i$ with respect to $P_{i-1}$.}
\end{wrapfigure}
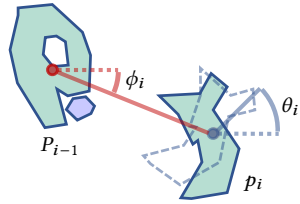
\subsubsection{State Space $S$}
During the $i$-th iteration, the LPN observes the current system state $s_i$ in the state space $S$. In our problem, we assume LPN can observe the current packing patch $P_{i-1}$ and a set of at most $H$ future patches to be packed, i.e., $s_i\triangleq(P_{i-1},p_i,\cdots,p_{i+H-1})$. This is because future patches can affect the pose of the current patch in order to achieve joint optimality. Unlike the myopic algorithms that consider a single future patch $p_i$, we feed the entire ordered sequence of $H$ future patches to the network. Empirically, we find this strategy can effectively guide the network to avoid myopic local minima. 

Practically, each patch can be of arbitrary geometric shapes, so we rasterize each patch in $s_i$ (including $P_{i-1}$) to a $50 \times 50$ 2D image. For each patch, we move their center-of-mass (COM) to the image center and encode each patch using a shared Fully Convolutional Network (FCN) into a $432$-dimensional latent code $\bar{p}_i\triangleq\text{FCN}(p_i)$, which is concatenated as $\bar{s}_i=\text{FCN}(s_i)$ for short. Since patches are of drastically different sizes, we scale all the patches before using the FCN, such that the area of all the $H$ patches is equal to $60\%$ of the area of the 2D image. Note that such global scaling will not change the packing ratio, so it should not change the optimal packing policy. 
When there are not enough patches to fill up the $H$ channels of patches, we feed the FCN with blank images. 

\subsubsection{Action Space $A$}
Having observed $s_i$, our LPN can be represented as a policy function $a_i=\pi^\text{LPN}(s_i)$ that maps the state to an action $a_i$ in the action space $A$. Unlike prior works for learning-based regular shape packing~\cite{zhao2021learning} or shape ordering~\cite{fang2023hybrid}, the design of action space $A$ for irregular packing is much more challenging. On the one hand, a valid action space should only consist of collision-free patch poses, while identifying these actions can involve extensive collision checks. On the other hand, training a decision-maker in high-dimensional action spaces can be rather data-demanding, and a promising subset of actions should be pre-selected. We tackle these two problems by having the policy $\pi^\text{LPN}$ to select a rough initial guess and then use local optimization to revise the pose. Specifically, we re-parameterize the action space $A$ under polar coordinates (\prettyref{fig:action}). We first compute the COMs for $P_{i-1}$ and $p_i$, denoted as $\text{COM}(P_{i-1})$ and $\text{COM}(p_i)$. The relative position $\text{COM}(p_i)$ with respect to $\text{COM}(P_{i-1})$ is expressed under polar coordinates with relative angle $\phi_i$ and we ignore their relative distance. Similarly, the local-to-global rotation of $p_i$ is encoded as another angle $\theta_i$. In summary, we define our action space as: $a_i\triangleq(\theta_i,\phi_i)$. \zeshi{At an early stage of this research, we parameterize our policy to use this continuous action space. However, in experiments we find the optimal state-action distribution can be multi-modal, and Gaussian distributions widely adopted to model the continuous action space cannot capture the multi-modal nature, leading to inferior performances. Therefore, we sample the range of $\theta_i$ and $\phi_i$ at $16$ angles and consider the $16\times16=256$-dimensional discrete action space.}

\subsubsection{State Transition\label{sec:transition}}
Our state transition function $s_{i+1}=\tau(s_i,a_i)$ computes the next state $s_{i+1}$ from $s_i$ and $a_i$ by converting the action $\theta_i$ and $\phi_i$ into a collision-free, tightly packed pose. Since we use a coarse discretization of the action space, we also use the state transition function to locally revise the action and improve the packing ratio. To this end, we devise the following collision-constrained local optimization:
\begin{equation}
\begin{aligned}
\label{eq:localOpt}
\theta_i^\star,t_i^\star\triangleq\underset{\theta,t}{\text{argmin}}\;&\|R(\theta)\text{COM}(p_i)+t-\text{COM}(P_{i-1})\|^2\\
\text{s.t.}\quad& \left[R(\theta)p_i+t\right]\cap P_{i-1}=\emptyset,
\end{aligned}
\end{equation}

where we initialize $\theta=\theta_i$ and $t_i$ is initialized from $\phi_i$ by elongating the relative direction (red line in~\prettyref{fig:action}) between $P_i$ and $p_i$ until they are collision free. Finally, we update the next state as $s_{i+1}\triangleq([R(\theta_i^\star)p_i+t_i^\star]\cup P_{i-1},p_{i+1},\cdots,p_{i+H})$. Note that we use the distance between center-of-mass as a surrogate measure for the packing ratio. We choose not to use the packing ratio as our objective function, because the new patch $p_i$ can oftentimes be entirely contained in the bounding box of $P_{i-1}$ and all the poses of $p_i$ inside the bounding box has the same packing ratio. The collision constraint can be realized in several ways, including the scaffold method~\cite{jiang2017simplicial} and the boundary barrier energy~\cite{smith2015bijective}, and we adopt the latter approach to avoid the costly 2D re-meshing. Although the scaffold method~\cite{jiang2017simplicial} has better solutions in large-scale problems, the barrier energy technique performs better under our small problem sizes with only $3$ decision variables. We solve the optimization using Newton's method with line-search to guarantee constraint satisfaction. During the line-search, we implement a bounding volume hierarchy to accelerate the collision check and assembly of barrier energy terms.

\subsubsection{LPN Training}
We parameterize $\pi^\text{LPN}$ as an MLP mapping $\bar{s}_i$ to the Q-value of all $256$ actions. After each state transition, the policy receives a sparse reward signal defined as:
\begin{align*}
r(s_i,a_i,s_{i+1})=\text{pr}(P_i)\mathbb{I}[i=H],
\end{align*}
where $\text{pr}(P)$ is the packing ratio of super-patch $P$ and $\mathbb{I}[i=H]$ is an indicator function of the last iteration. Note that using sparse reward signals can significantly slow down policy learning, but such reward does not pose a major problem in our application as we use a short horizon $H$, i.e. $|H| < 5$. We train our LPN policy via Q-learning algorithm by maximizing the expected cumulative reward:
\begin{align*}
\underset{\pi^\text{LPN}}{\text{argmax}}\mathbb{E}_{a_i\sim\pi^\text{LPN}}\left[\sum_{i=1}^Hr(s_i,a_i,s_{i+1})\right].
\end{align*}
To solve the stochastic optimization, we adopt the robust double deep Q-learning (DDQN) algorithm~\cite{van2016deep} and train $\pi^\text{LPN}$ to pack randomly sampled batches of at most $H$ patches of arbitrary order from our patch dataset, and we ensure each sampled batch comes from the same 3D model. 

\subsection{\label{sec:lowLevelOrder}Low-Level Sorter Network (LSN)}
Our LSN provides the optimal patch ordering for the LPN to achieve the best packing ratio, as our LPN can only pack the patches in a given order. We model the patch sorting procedure as another MDP denoted as $<S,A',\tau',r>$ with the same state space and reward signal as that of LPN. We represent LSN as another policy function $a_i'=\pi^\text{LSN}(s_i)$ that selects the next patch to be packed, i.e., $a_i'$ consists of the Q-value of the $k$ future patches. Given the selected next patch, the state transition function $\tau'(s_i,a_i')$ would invoke LSN and the state transition function $\tau$ from~\prettyref{sec:transition} to yield $s_{i+1}$. 

Neural networks need to understand the relative relationship between the future patches in order to accomplish the sorting task. Therefore, we apply a Graph Attention Network (GAT) module~\cite{velickovic2017graph}, which is known to be effective in solving sorting tasks~\cite{hu2020tap,zhao2022learning}. We organize all the patches into a fully connected graph, where the nodal input of GAT is the patch's feature $\bar{p}_i$, along with the feature of the already packed super-patch $\bar{P}_{i-1}$. GAT outputs the high-level graph feature for each of the existing patches. \zeshi{Then the patch features are converted to their corresponding Q-values using an MLP network}. This architecture parameterizes our sorting policy, denoted as $\pi^\text{LSN}$. Similar to $\pi^\text{LPN}$, $\pi^\text{LSN}$ is trained using DDQN via randomly sampled batches of at most $H$ patches coming from the same 3D model. The LSN and LPN combined define our low-level function $\text{LL}(\mathcal{S}')$.

\begin{figure}[ht]
\makebox[\linewidth]{
\begin{minipage}{\dimexpr\linewidth}
\begin{algorithm}[H]
\caption{\label{alg:HSN} \small{Iterative Selection of Super-Patches}}
\begin{small}
\begin{algorithmic}[1] 
\State $\mathcal{S}\gets$non-tiny set
\State Sample 400 subsets $\mathcal{S}'_{1,\cdots,400}$\label{ln:sample}
\State Sort $\mathcal{S}'_{1,\cdots,400}$ in $\text{HSN}(\mathcal{S}'_i)$-descending order
\State Sort $\mathcal{S}'_{1,\cdots,10}$ in $\text{pr}(\mathcal{S}-\mathcal{S}'_i\cup\text{LL}(\mathcal{S}'_i))$-descending order
\If{$\text{pr}(\mathcal{S}-\mathcal{S}'_1\cup\text{LL}(\mathcal{S}'_1))>\text{pr}(\mathcal{S})$}
\State $\mathcal{S}\gets\mathcal{S}-\mathcal{S}'_1\cup\text{LL}(\mathcal{S}'_1)$, goto~\prettyref{ln:sample}
\EndIf
\State {\algorithmicelse} Return $\mathcal{S}$
\end{algorithmic} 
\end{small}
\end{algorithm}
\end{minipage}}
\end{figure}
\subsection{\label{sec:highLevelSelect}High-Level Group Selector Network (HSN)}
Our low-level policies can only sort and pack a small subset $\mathcal{S}'$ of $H$ patches. In order to solve practical UV packing problems with hundreds of patches, we need to iteratively select $\mathcal{S}'$ from $\mathcal{S}$. To this end, we design a weighted average packing ratio $\text{pr}(\bullet)$ to evaluate the quality of the updated configuration $\text{pr}(\mathcal{S}-\mathcal{S}'\cup\text{LL}(\mathcal{S}'))$ and pick $\mathcal{S}'$ corresponding to the highest ratio. We then train our HSN to rank the quality of $\mathcal{S}'$ without actually calling the costly low-level function $\text{LL}(\mathcal{S}')$. Finally, we propose a sampling strategy to further reduce the calls to HSN.

\subsubsection{Weighted Average Packing Ratio} 
\begin{wrapfigure}{r}{0.5\linewidth}
\centering
\includegraphics[width=\linewidth]{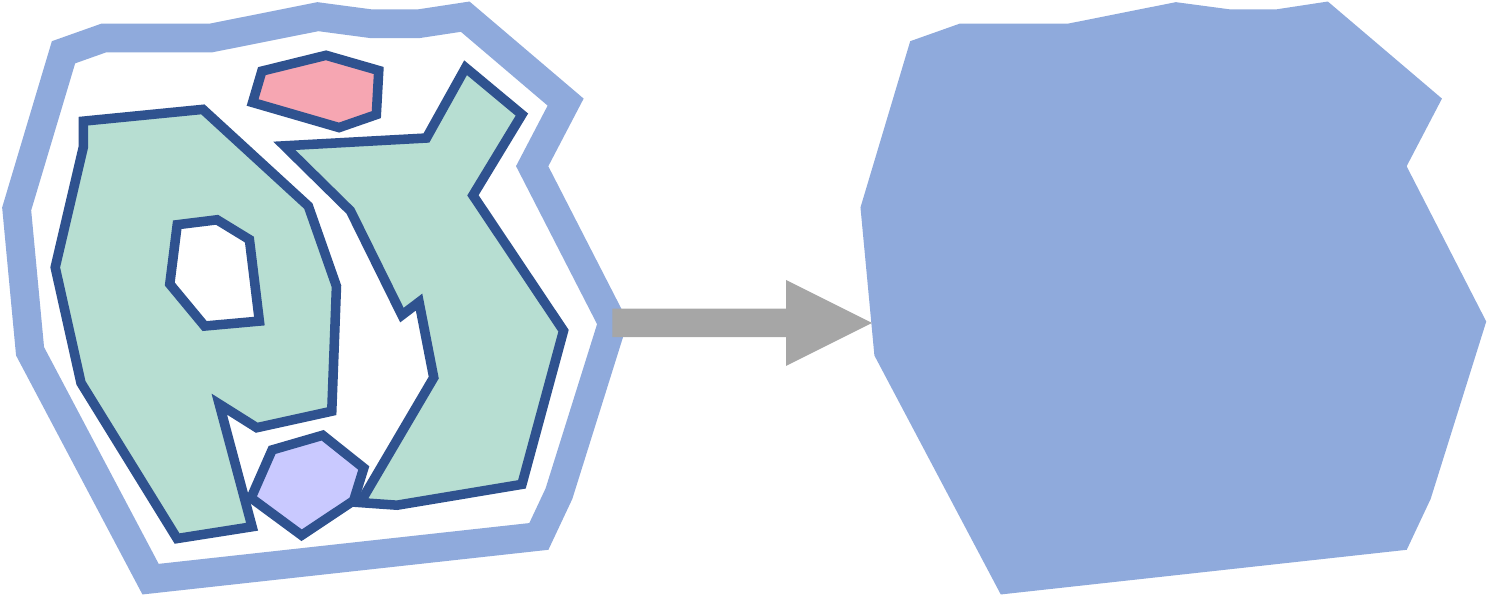}
\caption{\label{fig:alpha}After forming a super-patch (left), we will use alpha shape (right) to fill the gaps and inform the neural networks that these gaps are useless.}
\end{wrapfigure}

In order to compare different choices of $\mathcal{S}'$, we need a metric that measures their similarity to rectangles. To this end, we define the area-averaged packing ratio over all the super-patches in $\mathcal{S}$ as follows:
\begin{align*}
\text{pr}(\mathcal{S})\triangleq\frac{\sum_{p\in\mathcal{S}}\text{area}(p)\text{pr}(p)}{\sum_{p\in\mathcal{S}}|\text{bound}(p)|}\quad\text{pr}(p)\triangleq\frac{\text{area}(p)}{|\text{bound}(p)|},
\end{align*}
where $\text{area}(p)$ and $\text{bound}(p)$ are the actual area and the bounding box of the super-patch $p$, respectively. We further observe that the high geometric complexity of the super-patches is due to the interior gaps between them. However, these interior gaps are useless for low-level packing policies because, by the design of our low-level action space of LPN as shown in~\prettyref{fig:action}, we always pack a new patch $p_i$ from the outside of $P_{i-1}$. Therefore, before inserting $\text{LL}(\mathcal{S}')$ to $\mathcal{S}$, we compute the alpha shape~\cite{akkiraju1995alpha} (\prettyref{fig:alpha}) for each new super-patch to fill up the interior gaps and inform the neural networks that interior gaps are useless by design.

\setlength{\columnsep}{10pt}
\subsubsection{Packing Ratio Prediction}
Exhaustively evaluating $\text{pr}(\bullet)$ for all $C_{|\mathcal{S}|}^{H}$ choices of $\mathcal{S}'$ would require an intensive amount of calls to the low-level packing policies as well as the costly optimizations (\prettyref{eq:localOpt}). Instead, we propose a learning-based technique to predict the packing ratio using HSN. Given $\mathcal{S}'$, our HSN uses the same FCN from the low-level policies to encode each patch. The latent codes are then brought through GAT to yield the high-level graph features as in LSN. All the graph features are then brought through a max-pooling layer and a $\textrm{Sigmoid}$ layer to yield the predicted packing ratio. Note the absolution values of packing ratio is less important, since we are only interested in the relative ordering of the potential super-patches. In view of this, we train our HSN via supervised metric learning~\cite{chopra2005learning}. During each learning iteration, we randomly sample two (at most) $H$-patch groups denoted as $\mathcal{S}'$ and $\mathcal{S}''$ with groundtruth packing ratios denoted as $\text{pr}'$ and $\text{pr}''$ and HSN predicted packing ratio denoted as $\text{HSN}(\mathcal{S}')$ and $\text{HSN}(\mathcal{S}'')$. We then update HSN via the following margin ranking loss:
\begin{align}
\label{eq:metricLoss}
\mathcal{L}=\text{max}(0,-\text{sgn}(\text{pr}'-\text{pr}'')(\text{HSN}(\mathcal{S}')-\text{HSN}(\mathcal{S}''))+\epsilon),
\end{align}
where $\epsilon$ is the minimal positive margin. We have found that HSN can empirically reach a high prediction accuracy for simple patches, but its accuracy gradually deteriorates as more and more patches are packed into complex-shaped super-patches.

\subsubsection{Patch Group Sampling}
Even using the HSN to efficiently rank the packing ratio, batch evaluation of $\text{pr}(\bullet)$ for all $\mathcal{S}'$ is still time-consuming. To further alleviate the runtime cost, we randomly sample $400$ subsets of patches and predict their packing ratio via a batched HSN evaluation. The top $10$ out of $400$ best groups are then forwarded to the low-level algorithm to evaluate the groundtruth packing ratio $\text{pr}(\mathcal{S}-\mathcal{S}'\cup\text{LL}(\mathcal{S}'))$, and finally the best of the $10$ groups is adopted to form the next super-patch. As outlined in~\prettyref{alg:HSN} of the supplementary, this procedure is repeated until the updated packing ratio $\text{pr}(\mathcal{S})$ is not higher than the current value.

\subsection{\label{sec:global}Super-Patch Assembly}
After our first stage, the non-tiny patches are grouped into nearly rectangular super-patches. During our second stage, we confidently take the rectangular shape assumption and use the divide-and-conquer algorithm implemented in the Trimesh library~\cite{trimesh} to assemble all the super-patches together. 

With the initially packed result from bin-packing, we then propose to adjust the joint poses of all the patches, locally squeezing them together via numerical optimization. We denote the bounding box enclosing all the $M$ patches as $\text{bound}(p_1,\cdots,p_M)$, and our optimization problem is formulated as:
\small
\begin{equation}
\begin{aligned}
\label{eq:jointOpt}
\underset{\theta_{1,\cdots,M},t_{1,\cdots,M}}{\text{argmin}}\;&|\text{bound}(p_1,\cdots,p_M)|\\
\text{s.t.}\quad&\left[R(\theta_i)p_i+t_i\right]\cap\left[R(\theta_j)p_j+t_j\right]=\emptyset\;\forall1\leq i<j\leq M.
\end{aligned}
\end{equation}

\normalsize
We again use the barrier function technique~\cite{smith2015bijective} to handle all the collision-free guarantees and ensure that the bounding box is surrounding all the patches. Although this is a joint optimization, it is still efficient to solve, since we only allow rigid motions for all the patches.

Finally, there is another set of tiny patches that are set aside by our filter at the beginning of our pipeline. For these small patches, we adopt a conventional approach to sort them in an area-descending order and then use the scanline algorithm~\cite{hu2018efficient} to fit them into the gaps and holes of the super-patches. During this stage, we also replace alpha shapes with the original patches to expose the scanline algorithm to potentially useful gaps and holes. These final adjustments are illustrated in~\prettyref{fig:stageTwo}.
\section{Experiments}
\paragraph{Network Training} We perform all experiments on a computer with an Intel E5-1650 $12$-core CPU at 3.60GHz and 32GB RAM. We implement all learning algorithms via Pytorch~\cite{paszke2017automatic} with the GAT implemented based on~\cite{wang2019dgl}. For training the LPN, we use DDQN with an experience buffer size of $10^6$ transition tuples. We sample roughly $5\times10^6$ random packing problems with $H=4$ to populate the experience buffer and we update $\pi^\text{LPN}$ using $2\times10^4$ epochs of stochastic gradient descend (SGD). The same procedure is used for training $\pi^\text{LSN}$. Both learning rates of the LPN and LSN are set to $10^{-4}$. We train HSN using a collected dataset of $6\times10^4$ $H$-patch subsets with pre-computed groundtruth packing ratios. We update $\text{HSN}$ using $500$ epochs of SGD with a learning rate of $10^{-3}$ and a batch size of $256$. For each dataset, we use $70\%$ of data for training and the rest for testing. 

\paragraph{Runtime Setting} Given the set of input patches, we first sort them in the area-descending order, and then consider the subset of largest patches, whose sum of areas takes up $80\%$ of the area of all patches, which is denoted as the salient subset $\mathcal{S}_s$. The average area of patches in the salient subset is denoted as $\bar{a}=\sum_{p\in\mathcal{S}_s}\text{area}(p)/|\mathcal{S}_s|$. Next, we define the tiny patch set as all the patches with an area smaller than $\bar{a}/5$, with other patches classified as non-tiny. \zeshi{Note that the patches are splitted based on their relative sizes instead of absolute sizes, so the final results are insensitive to the threshold values within a reasonable range.} The numerical optimizations in~\prettyref{eq:localOpt} and~\prettyref{eq:jointOpt} are implemented in C++, where we set the maximal allowed iterations and initial step size to $10^3$ and $10^{-4}$, respectively. In the super-patch assembly, the aspect ratio between the width and height of the texture image domain ranges from $1$ to $2$. To choose the appropriate aspect ratio, we run the bin-packing algorithms $10$ times using different aspect ratios and choose the one with the highest packing ratio. \zeshi{Practical applications would create gutter space around the boundaries of patches to eliminate interpolation artifacts. In our implementations we set the minimal distance between patches to one pixel.}

\begin{figure*}[ht]
\centering
\bgroup
\def\arraystretch{.5}
\resizebox{\linewidth}{!}{
\begin{tabular}{ccccc}
\toprule
 & \scalebox{0.7}{\cite{sander2003multi}} & \scalebox{0.7}{XAtlas} & \scalebox{0.7}{NFP} & \scalebox{0.7}{Ours}\\
 \midrule
\includegraphics[scale=.1]{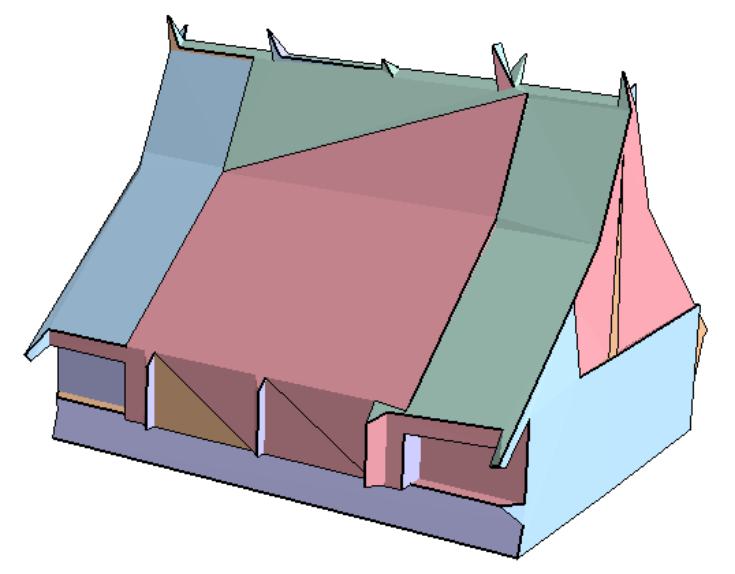}&
\includegraphics[scale=.1]{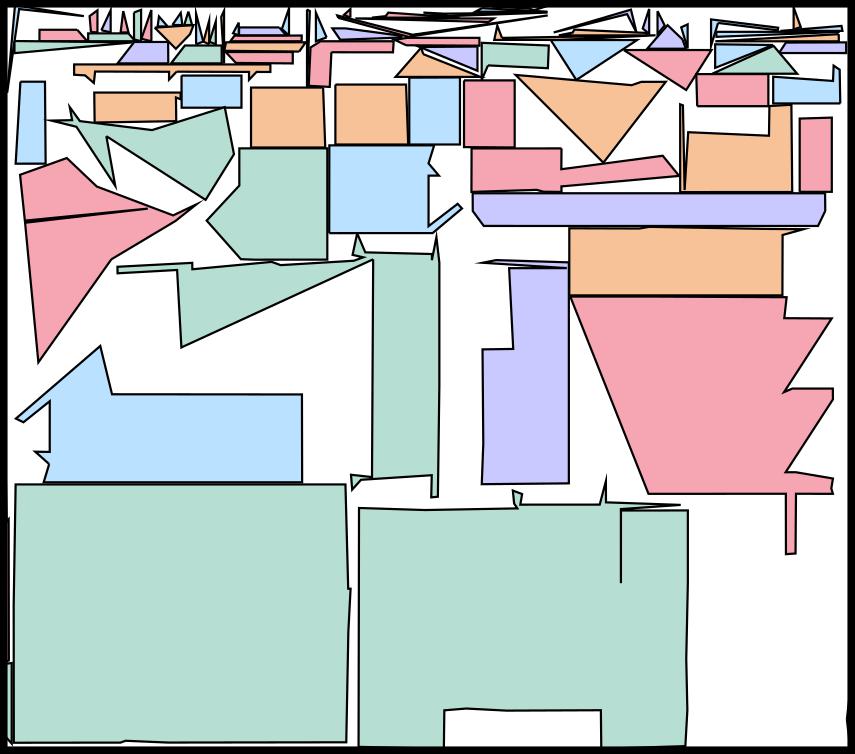}&
\includegraphics[scale=.1]{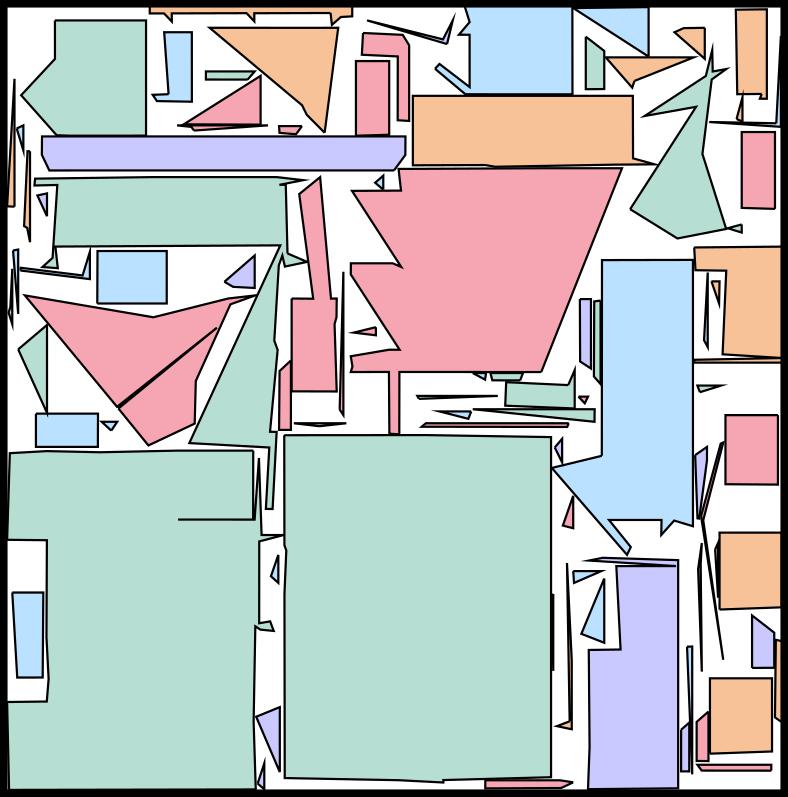}&
\includegraphics[scale=.1]{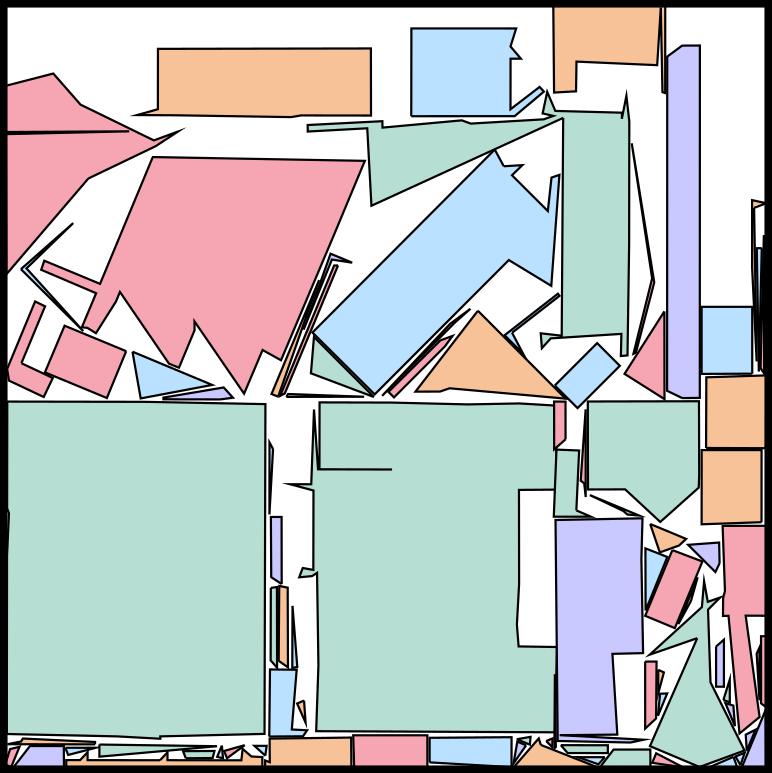}&
\includegraphics[scale=.1]{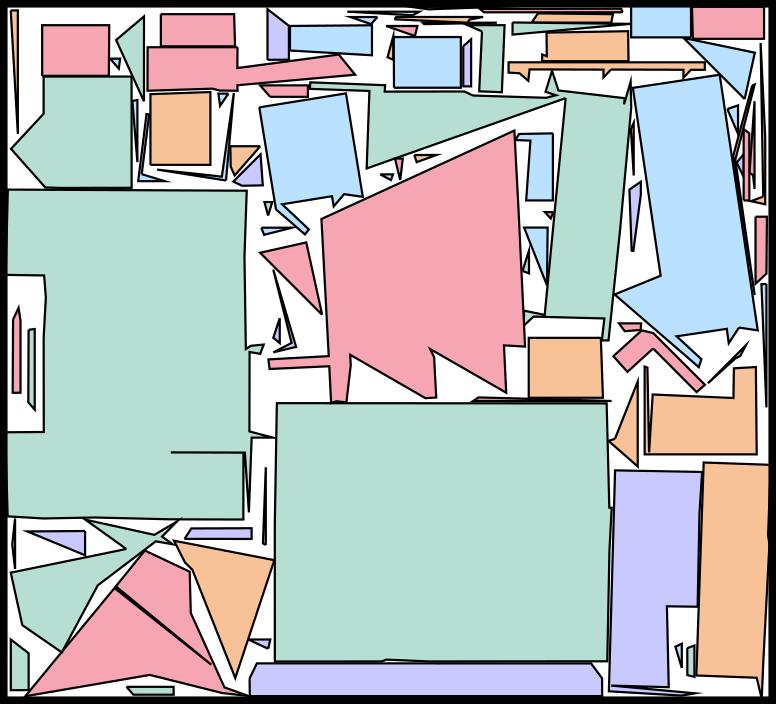}\\
 & \scalebox{0.7}{pr: 69.5\%} &\scalebox{0.7}{pr: 70.1\%} & \scalebox{0.7}{pr: 74.5\%} & \scalebox{0.7}{pr: 82.4\%}\\
 \midrule
\includegraphics[scale=.1]{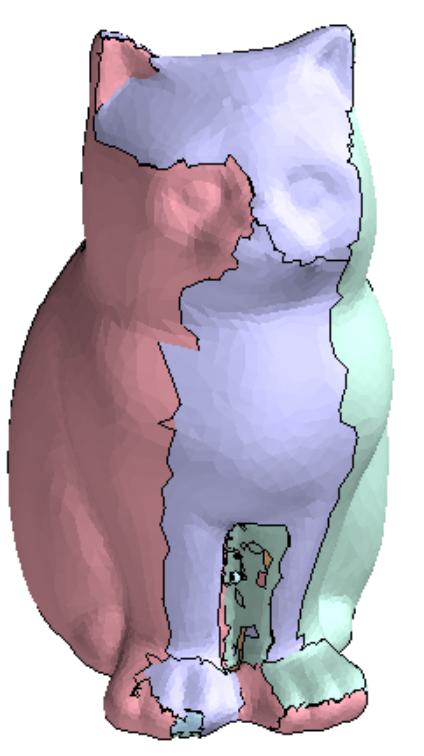}&
\includegraphics[scale=.1]{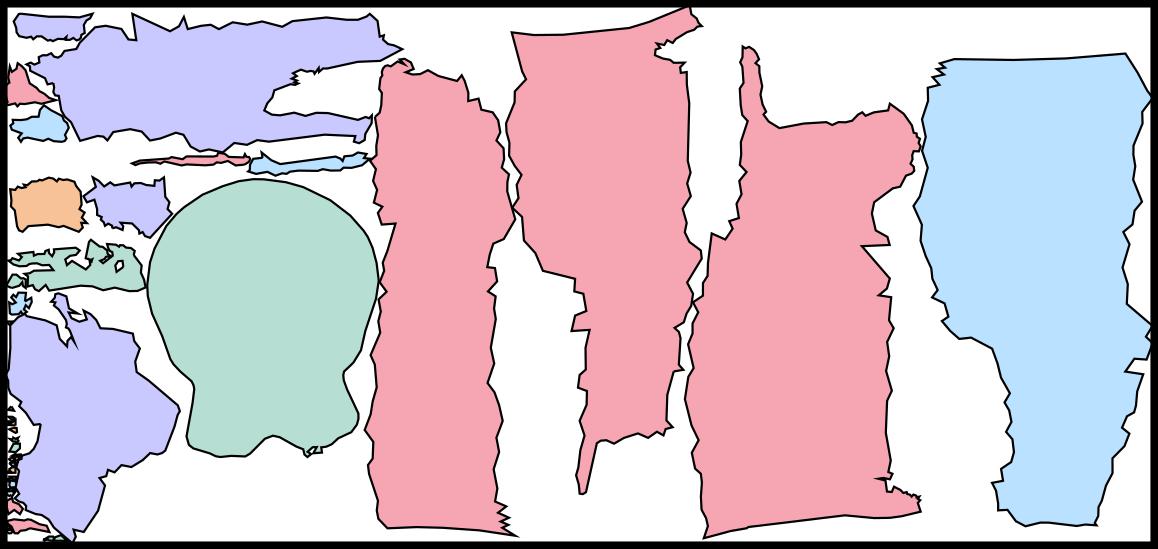}&
\includegraphics[scale=.1]{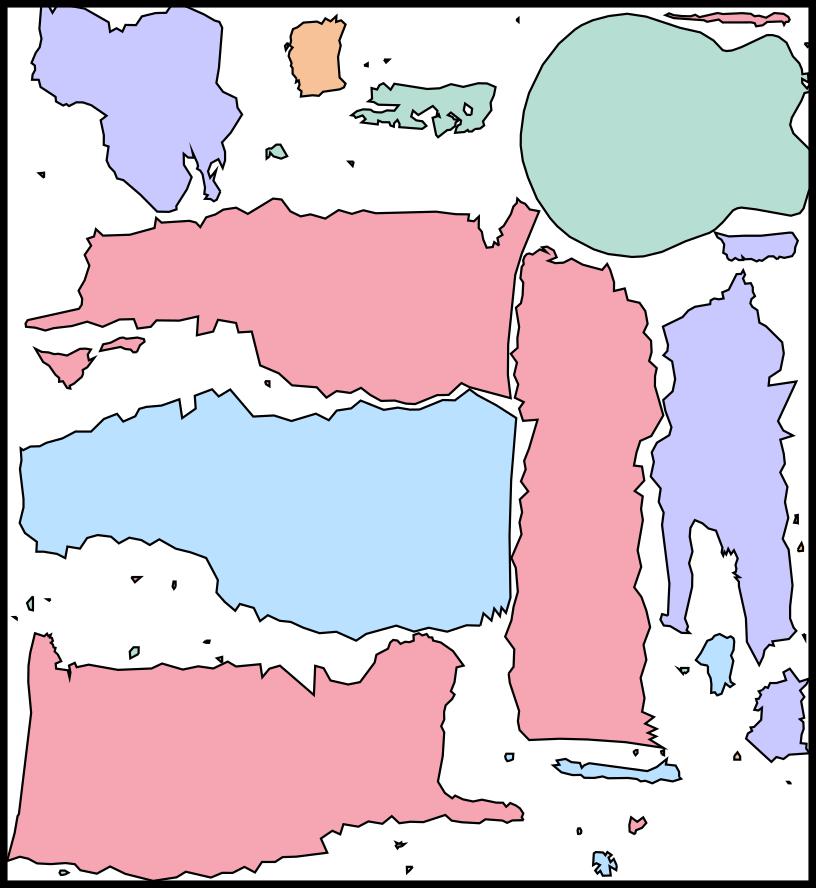}&
\includegraphics[scale=.1]{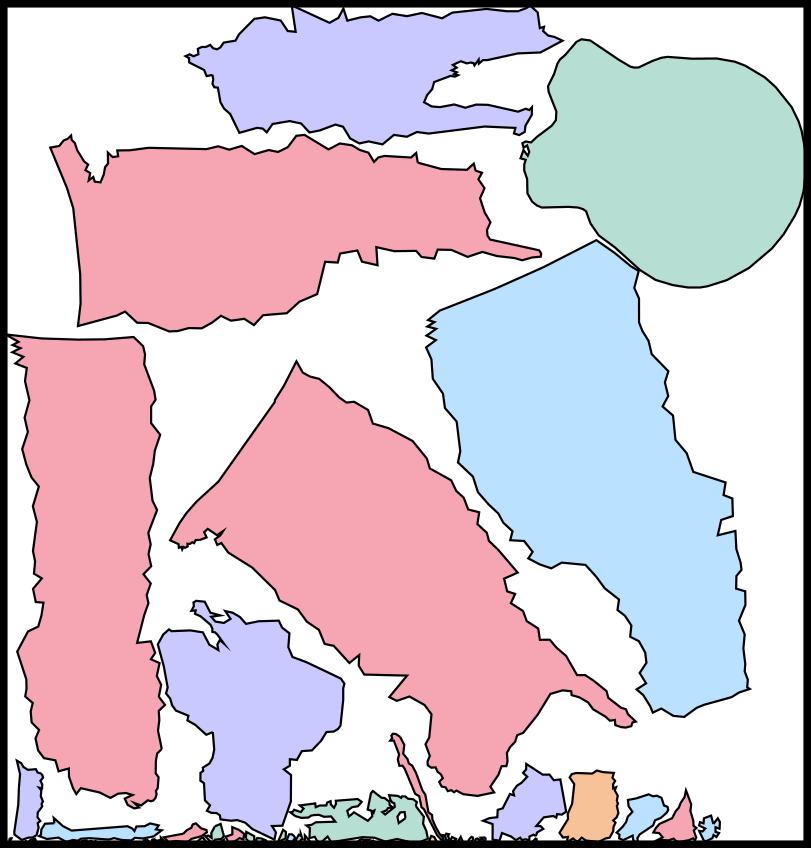}&
\includegraphics[scale=.1]{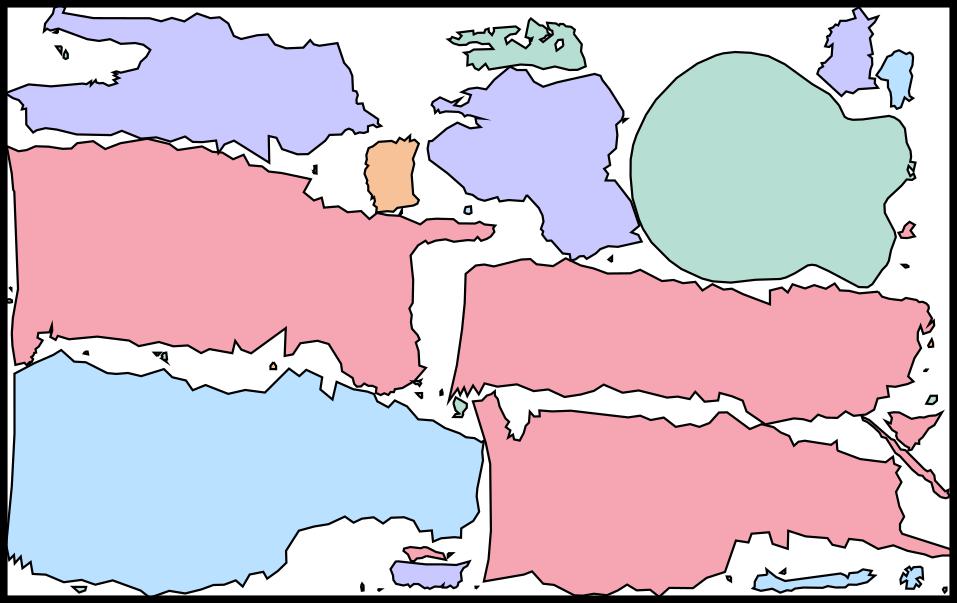}\\
 & \scalebox{0.7}{pr: 68.9\%} &\scalebox{0.7}{pr: 65.4\%} & \scalebox{0.7}{pr: 64.7\%} & \scalebox{0.7}{pr: 77.2\%}\\
 \midrule
\includegraphics[scale=.1]{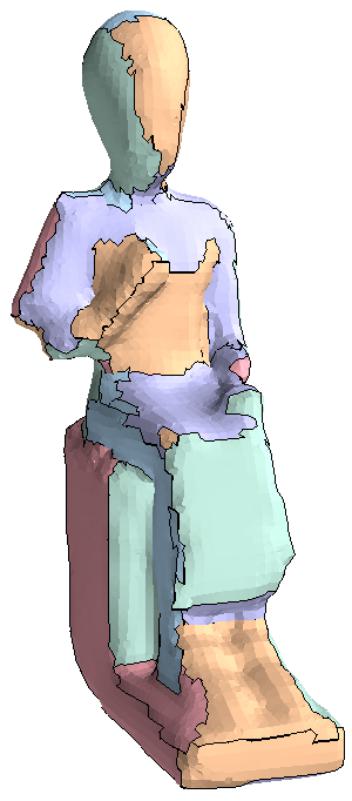}&
\includegraphics[scale=.1]{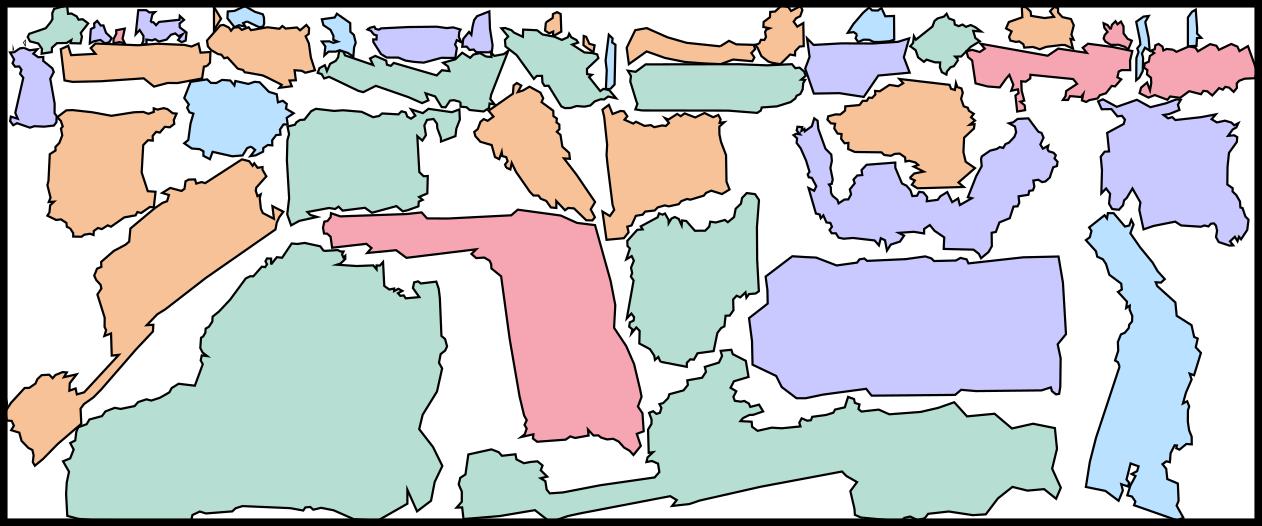}&
\includegraphics[scale=.1]{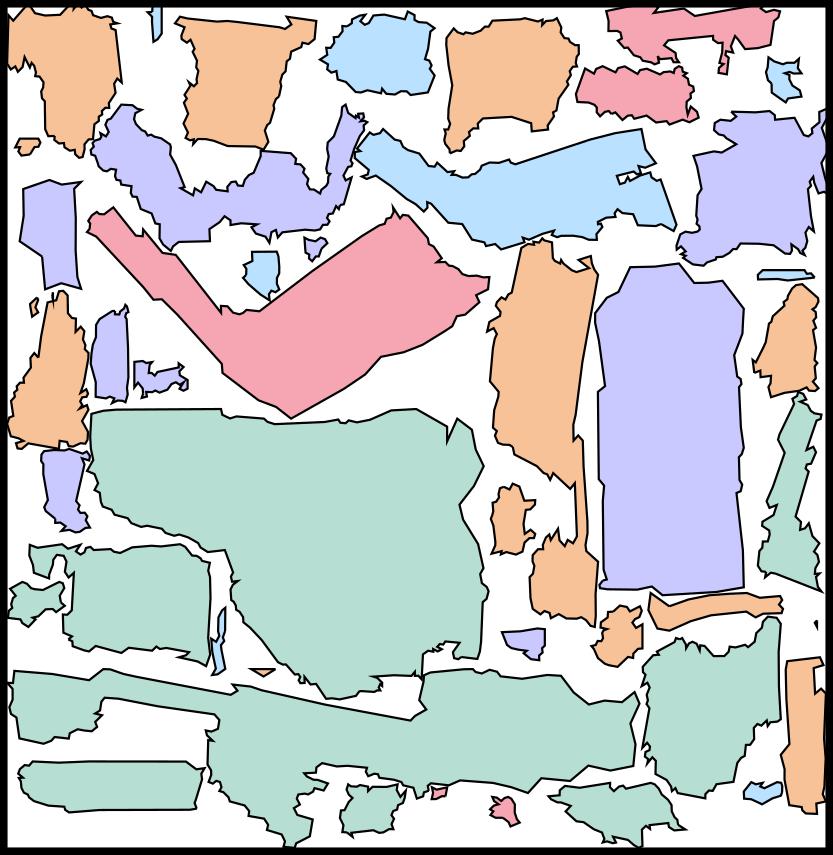}&
\includegraphics[scale=.1]{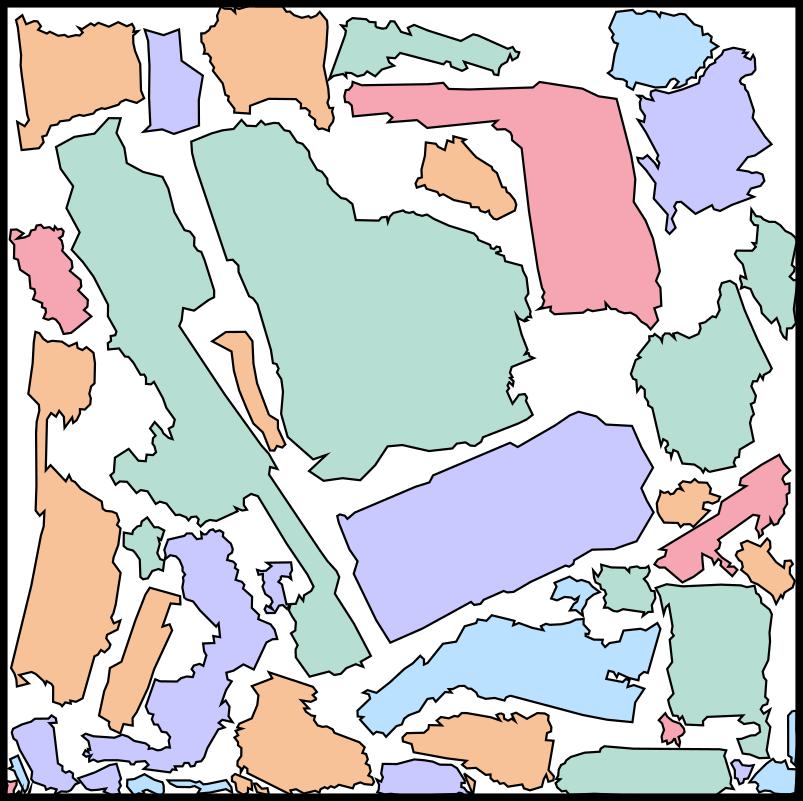}&
\includegraphics[scale=.1]{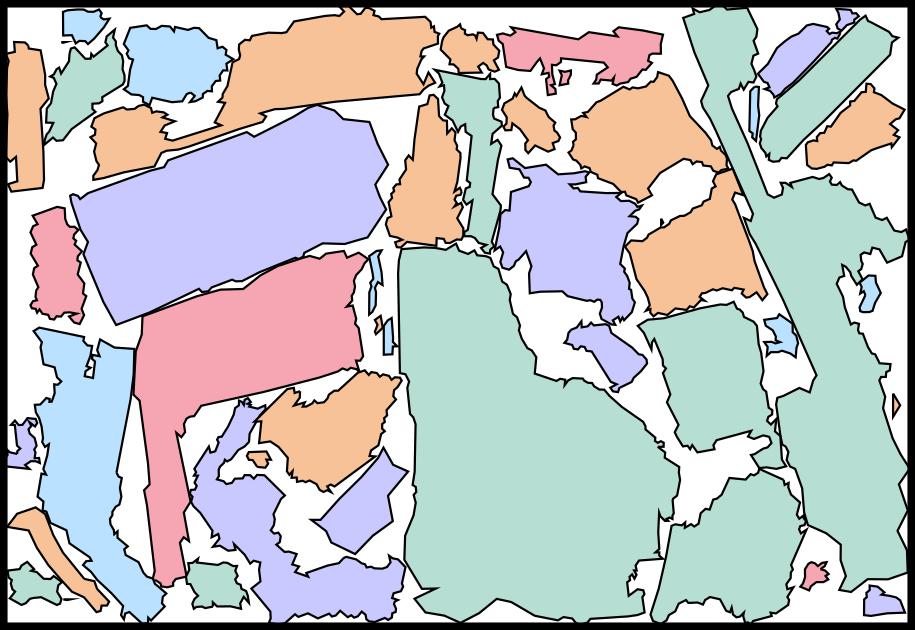}\\
 & \scalebox{0.7}{pr: 67.0\%} &\scalebox{0.7}{pr: 67.9\%} & \scalebox{0.7}{pr: 69.0\%} & \scalebox{0.7}{pr: 76.8\%}\\
 \bottomrule
\end{tabular}}
\egroup
\caption{\label{fig:more_results}Three packing instances from the building, object and general dataset.}
\end{figure*}

\begin{figure*}[ht]
\centering
\bgroup
\def\arraystretch{.5}
\resizebox{\linewidth}{!}{
\begin{tabular}{c|c|c}
\scalebox{0.5}{Building} & \scalebox{0.5}{Object} & \scalebox{0.5}{General}\\
\includegraphics[scale=.1]{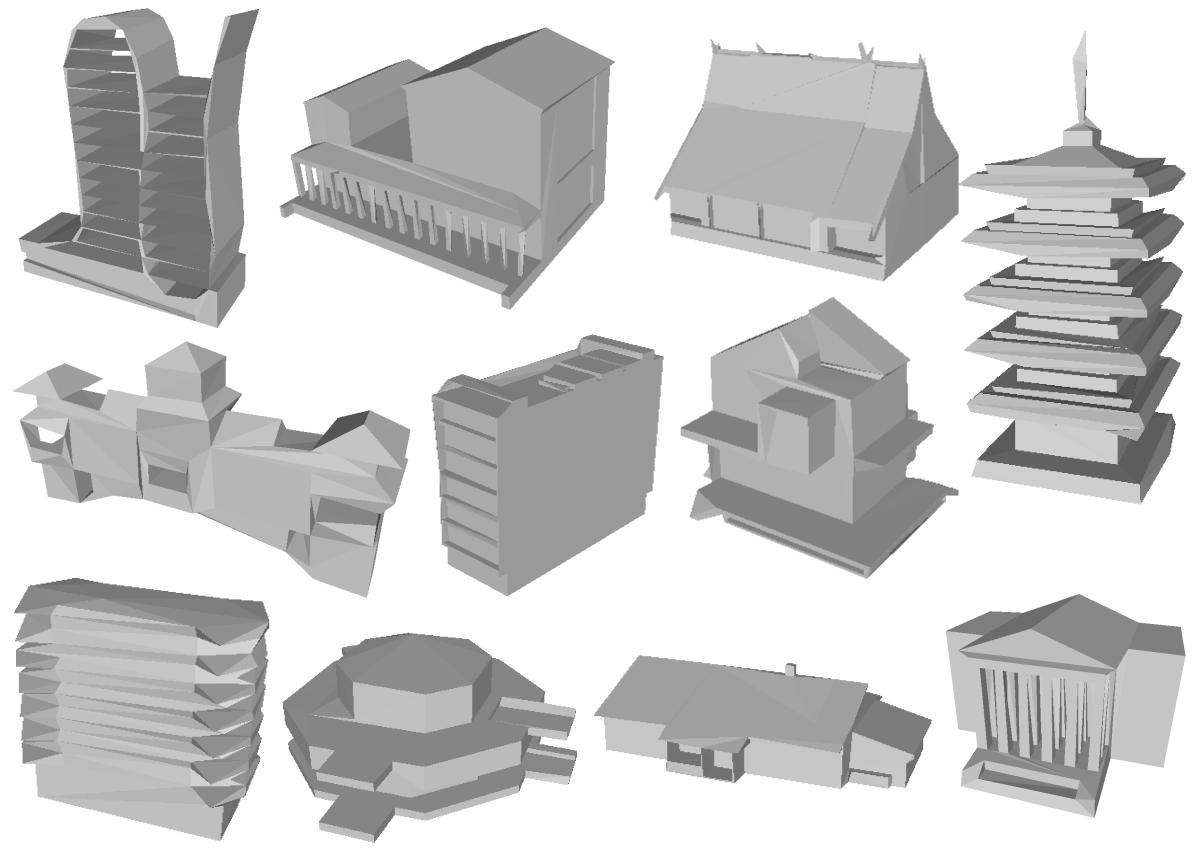}&
\includegraphics[scale=.1]{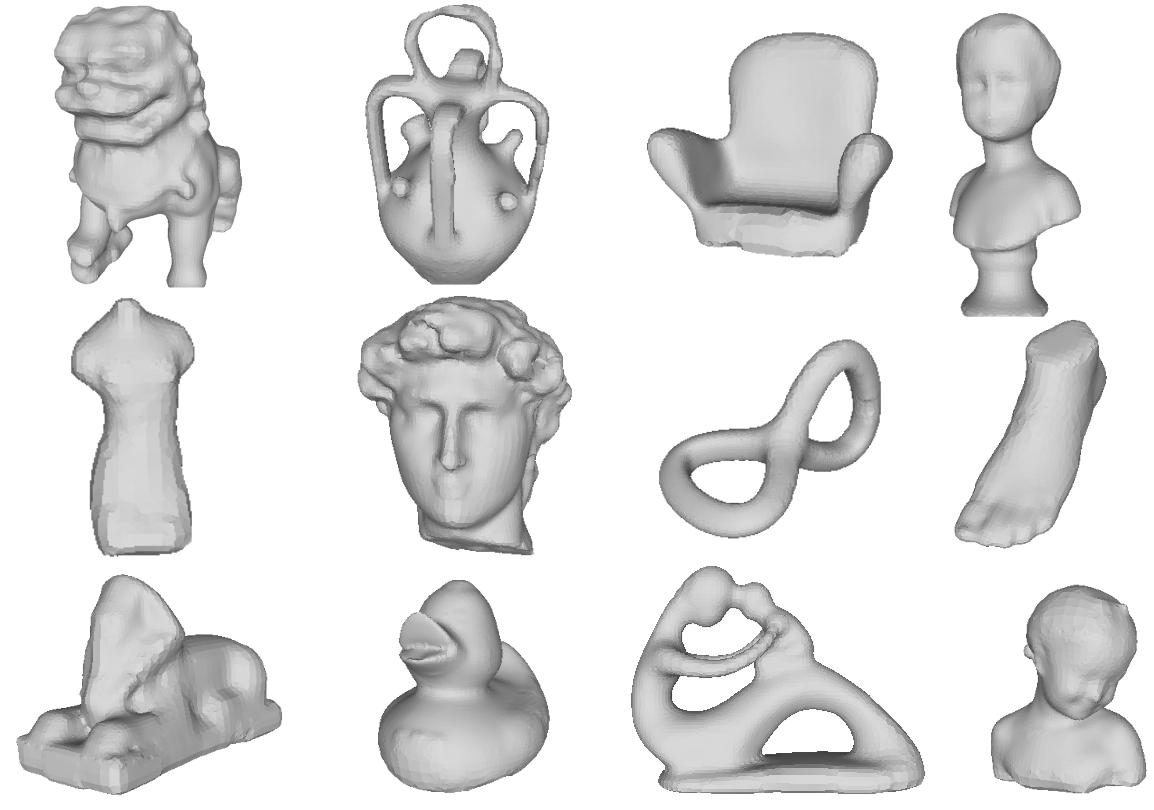}&
\includegraphics[scale=.1]{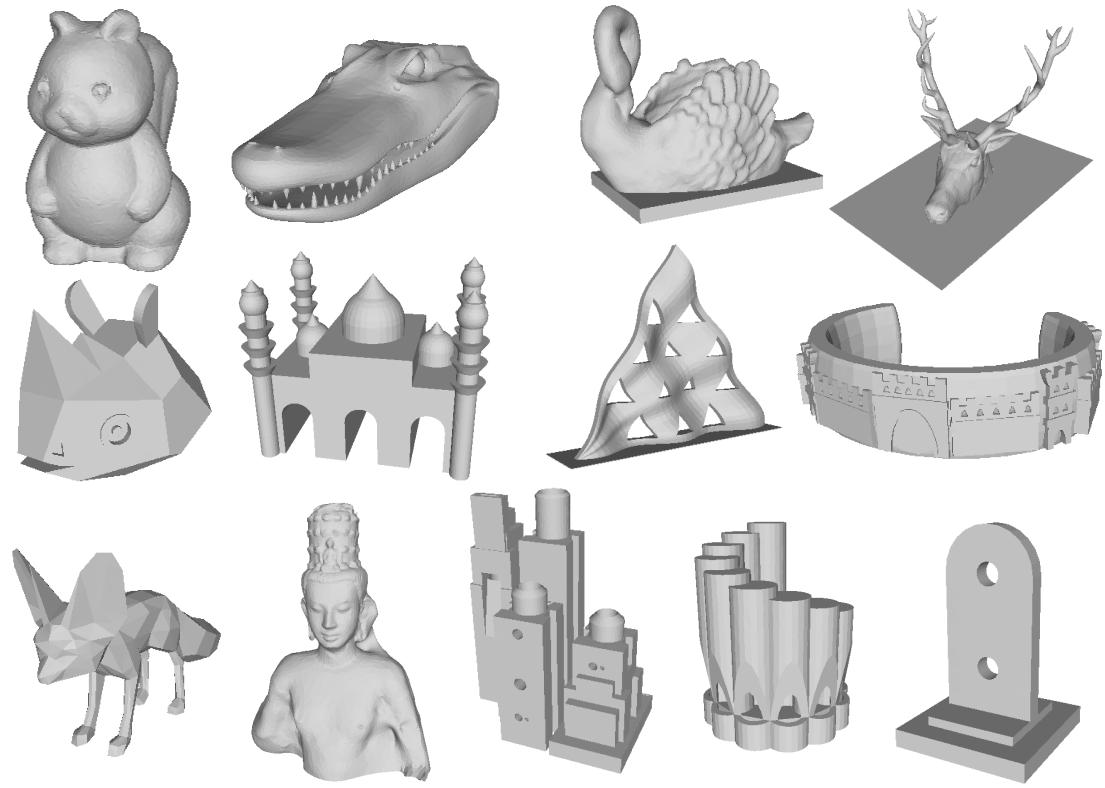}\\
\end{tabular}}
\egroup
\caption{\label{fig:dataset}A set of 3D model samples from our three datasets.}
\end{figure*}

\begin{figure*}[ht]
\centering
\bgroup
\def\arraystretch{.5}
\resizebox{\linewidth}{!}{
\begin{tabular}{ccccc}
\toprule
 & \scalebox{0.7}{\cite{sander2003multi}} & \scalebox{0.7}{XAtlas} & \scalebox{0.7}{NFP} & \scalebox{0.7}{Ours}\\
 \midrule
 \includegraphics[scale=.1]{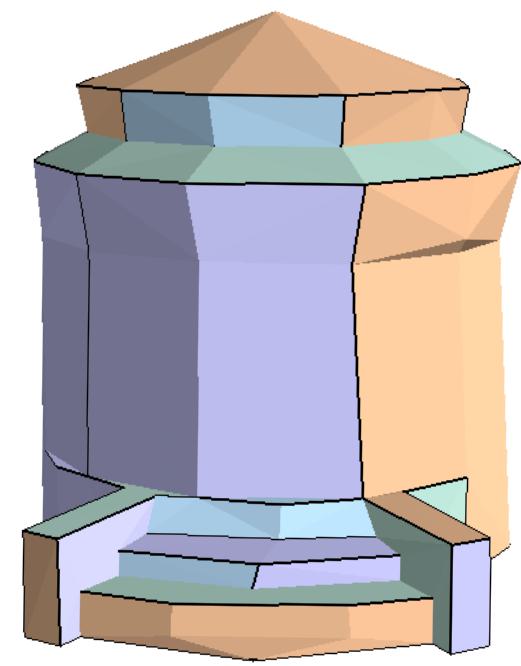}&
\includegraphics[scale=.1]{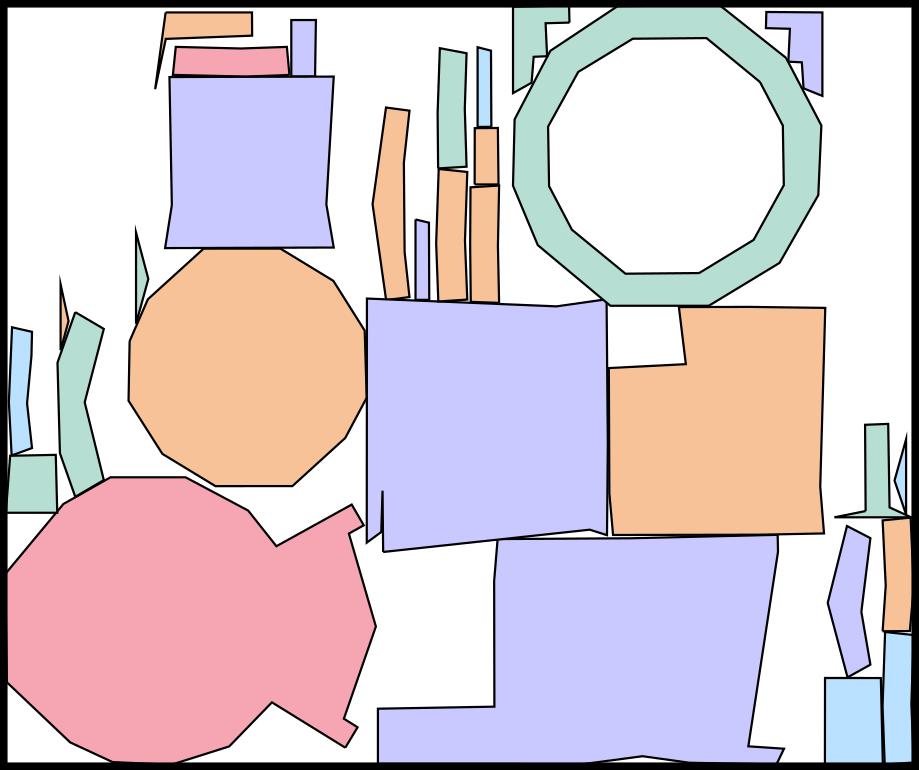}&
\includegraphics[scale=.1]{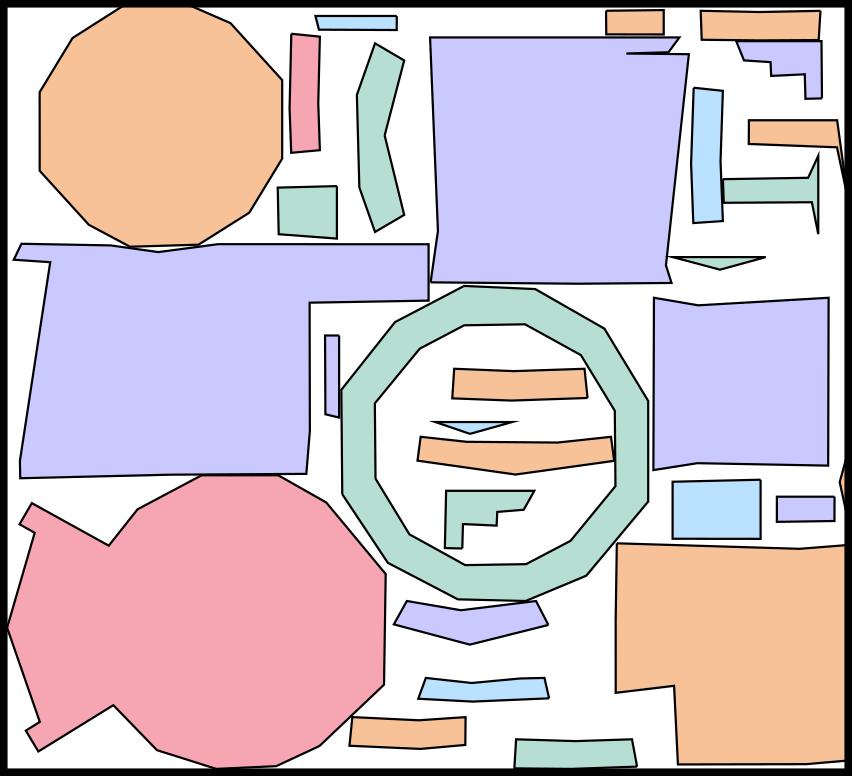}&
\includegraphics[scale=.1]{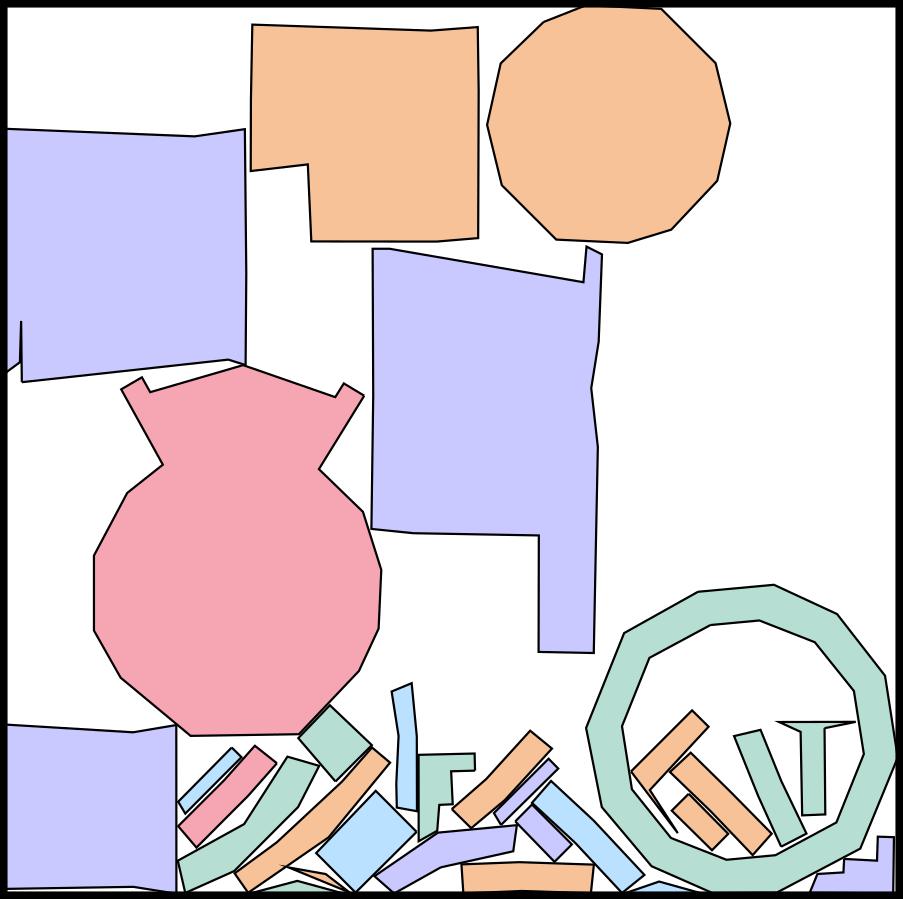}&
\includegraphics[scale=.1]{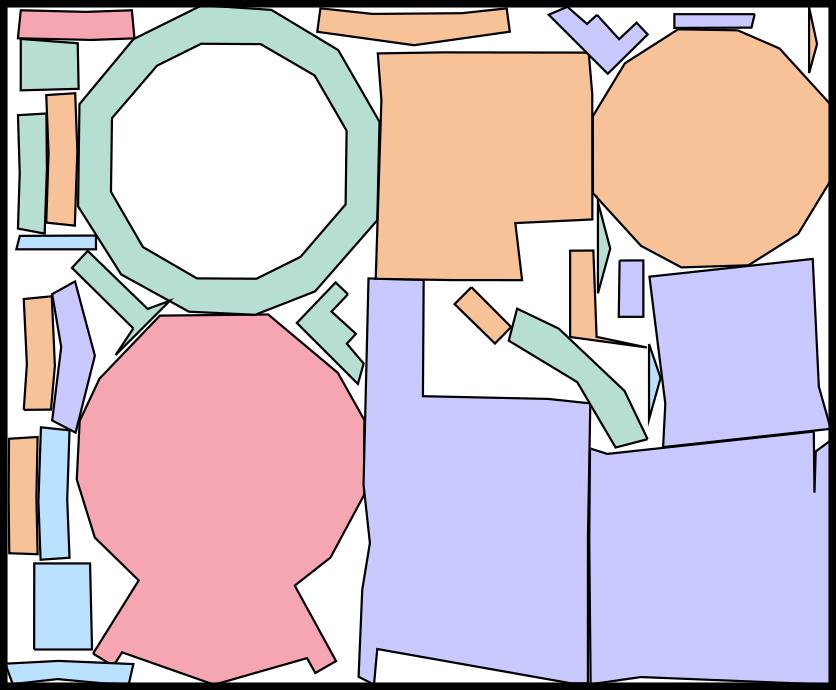}\\
 & \scalebox{0.7}{pr: 59.2\%} &\scalebox{0.7}{pr: 66.4\%} & \scalebox{0.7}{pr: 53.0\%} & \scalebox{0.7}{pr: 73.7\%}\\
 \midrule
\includegraphics[scale=.1]{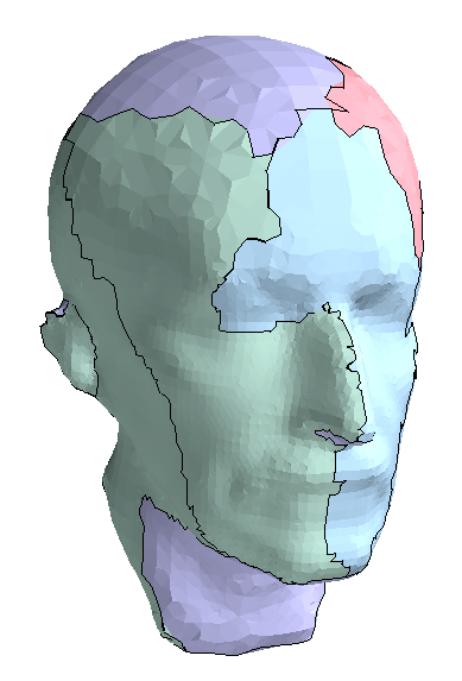}&
\includegraphics[scale=.1]{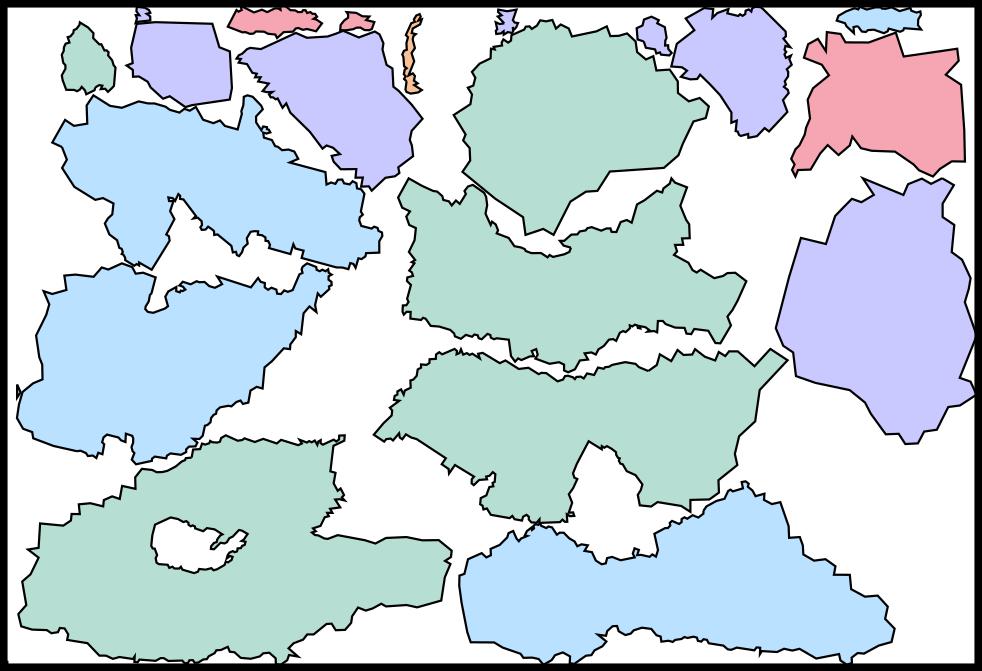}&
\includegraphics[scale=.1]{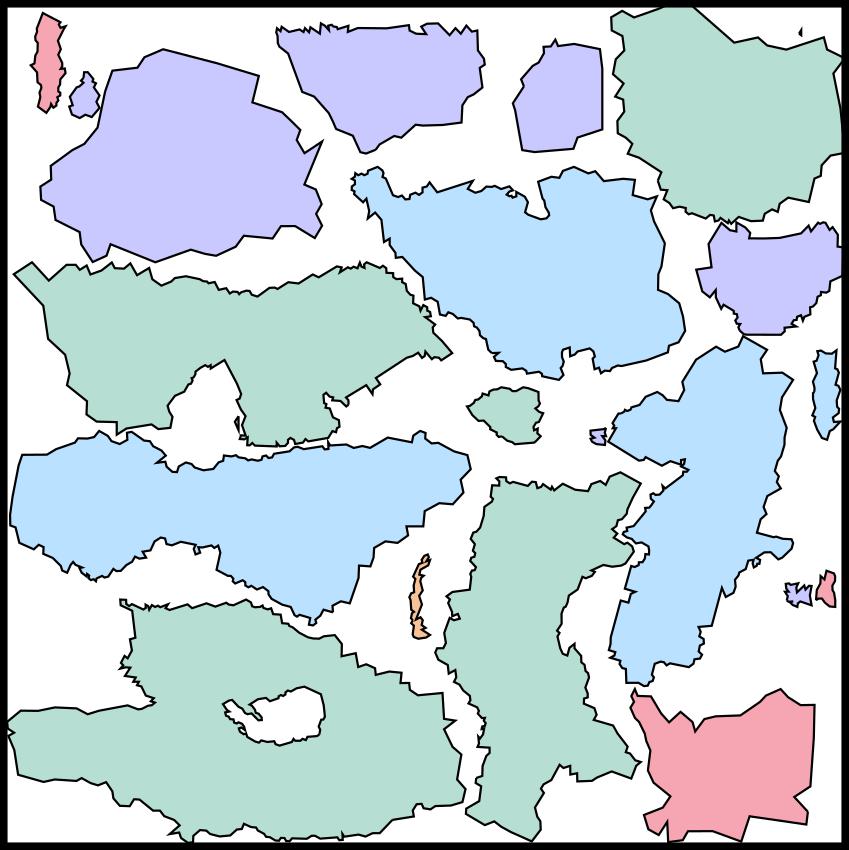}&
\includegraphics[scale=.1]{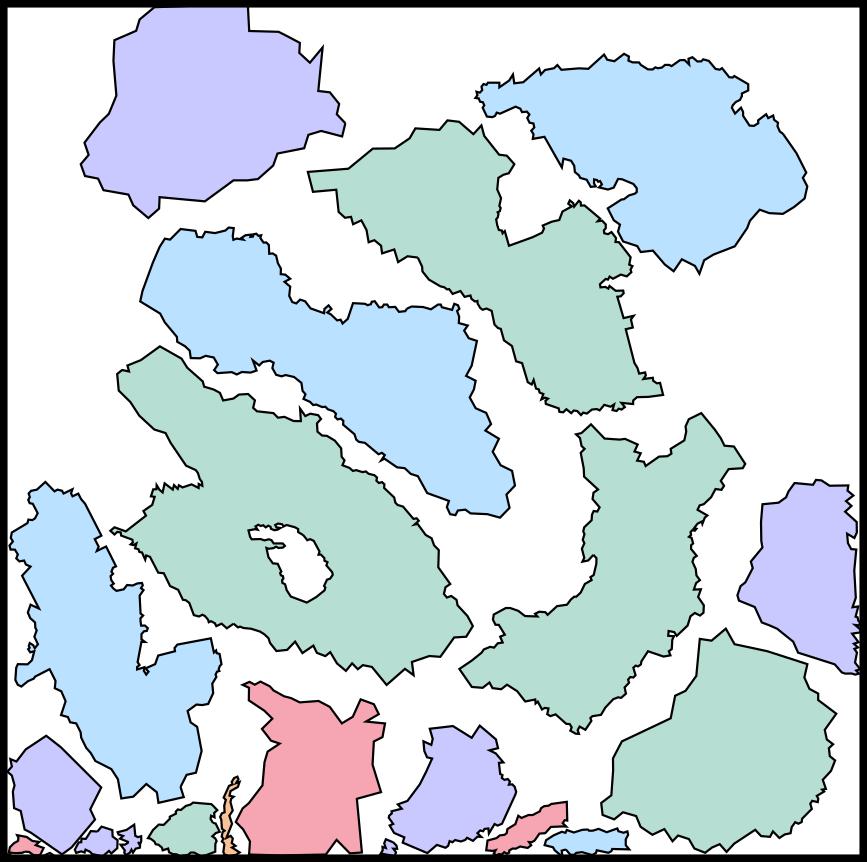}&
\includegraphics[scale=.1]{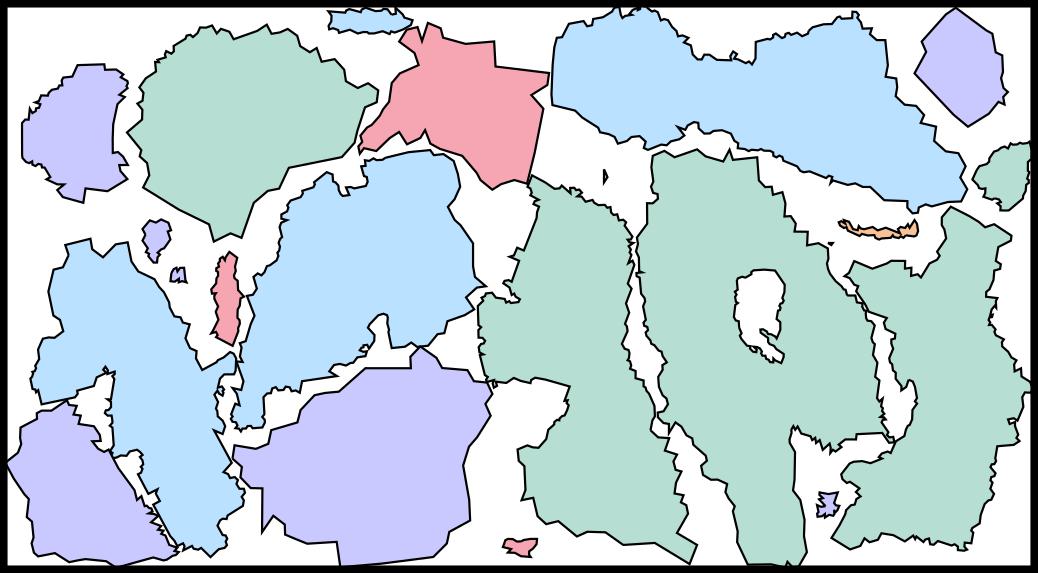}\\
 & \scalebox{0.7}{pr: 66.9\%} &\scalebox{0.7}{pr: 63.9\%} & \scalebox{0.7}{pr: 59.2\%} & \scalebox{0.7}{pr: 74.6\%}\\
 \midrule
\includegraphics[scale=.1]{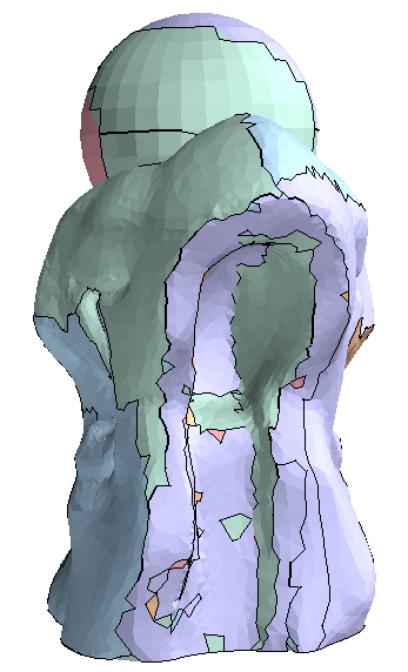}&
\includegraphics[scale=.1]{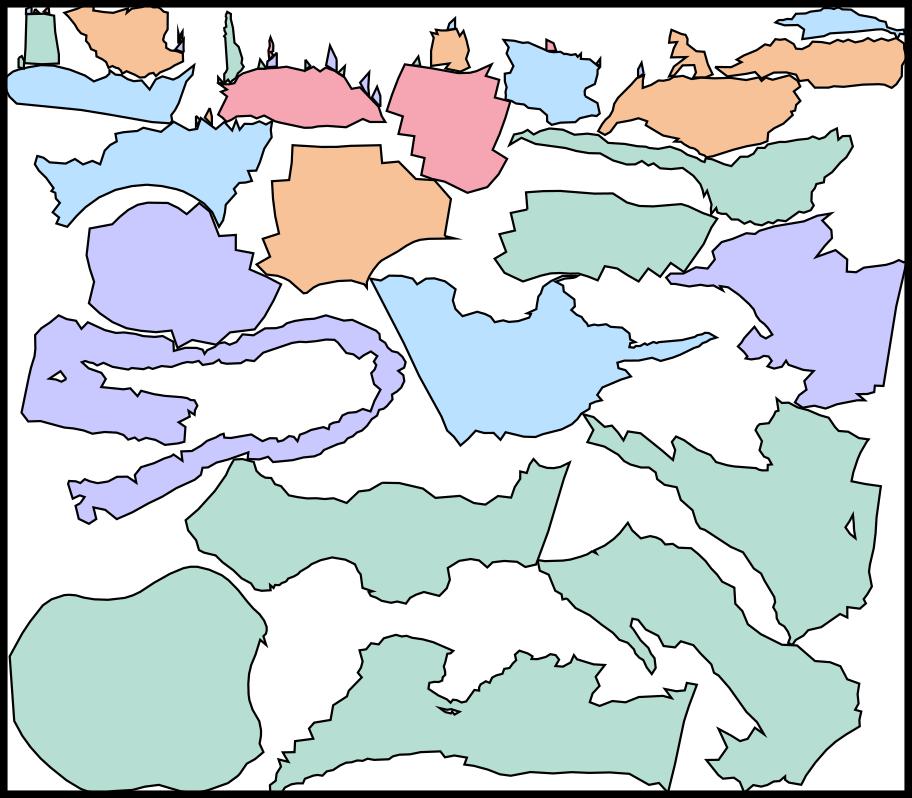}&
\includegraphics[scale=.1]{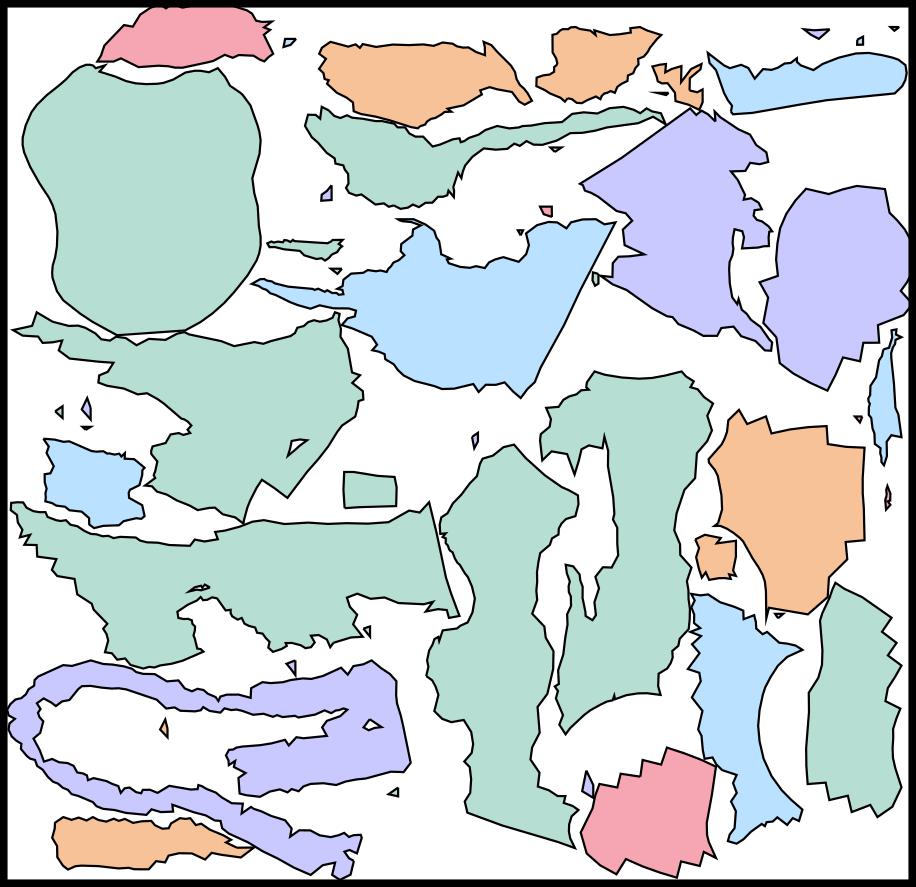}&
\includegraphics[scale=.1]{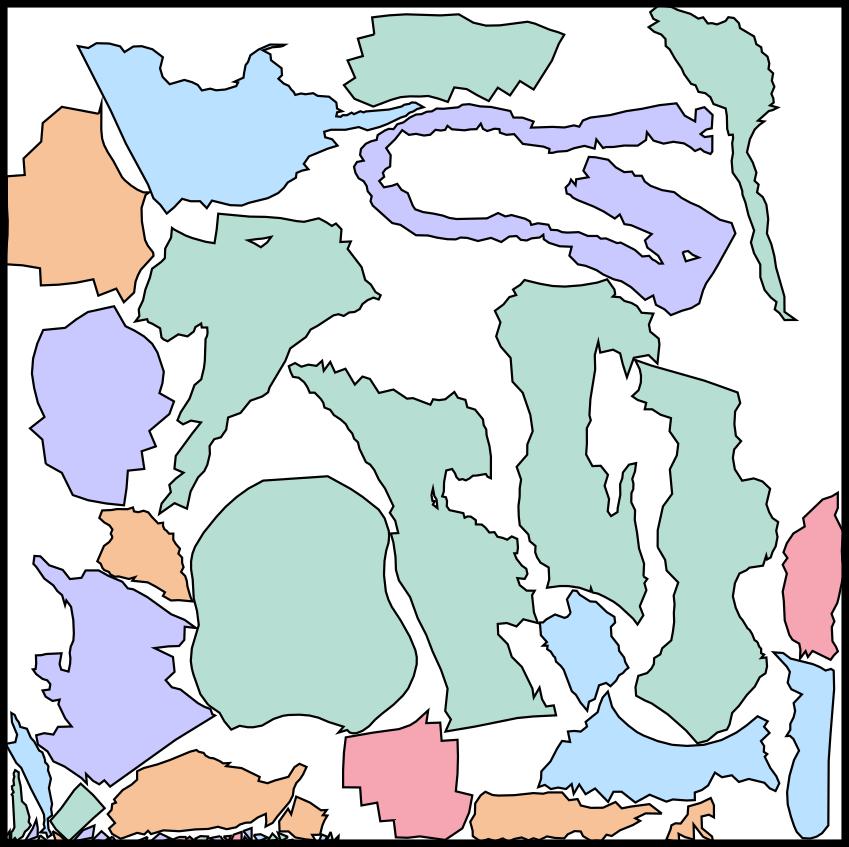}&
\includegraphics[scale=.1]{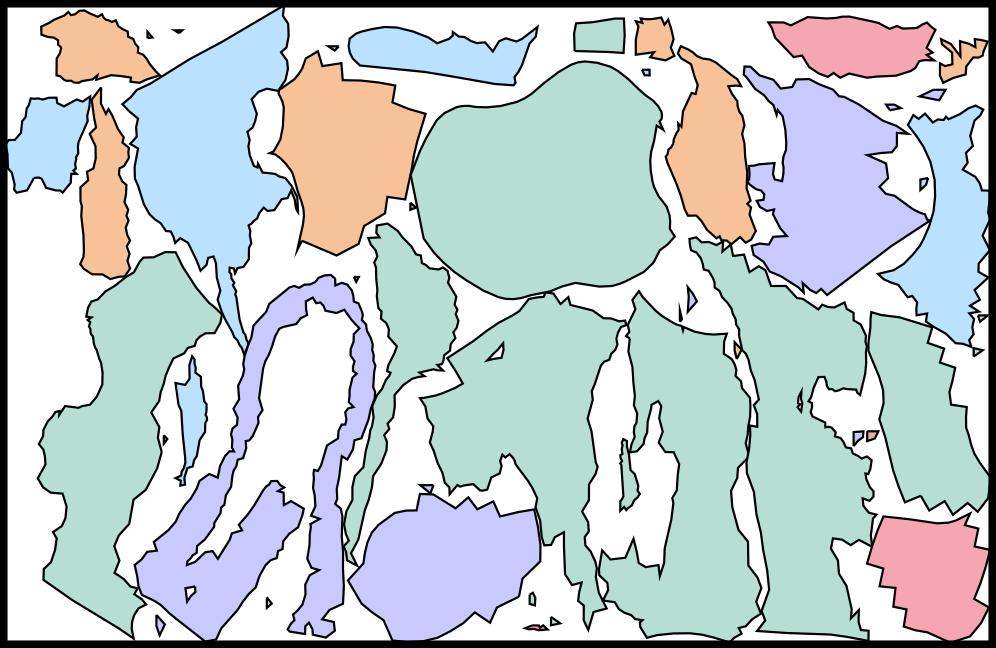}\\
 & \scalebox{0.7}{pr: 54.8\%} &\scalebox{0.7}{pr: 54.2\%} & \scalebox{0.7}{pr: 55.5\%} & \scalebox{0.7}{pr: 62.1\%}\\
\bottomrule
\end{tabular}}
\egroup
\caption{\label{fig:more_results_2}Three packing instances from the building, object, and general dataset.}
\end{figure*}

\paragraph{Datasets} We evaluate our method on three datasets of 2D UV patches obtained from UV unwrapping of 3D models using XAtlas. This software can sometimes generate degenerate patches with zero or negative areas, which are removed from our dataset. As illustrated in~\prettyref{fig:dataset},~\prettyref{fig:more_results}, and~\prettyref{fig:more_results_2}, our first dataset, ``Building'' contains 86 man-made building models with mostly sharp features, where the resulting packing problems have 5 to 131 patches. Our second dataset, ``Object'', contains 81 3D models representing daily objects with few sharp features, where the packing problems have 9 to 200 patches. Our third dataset, ``General'', contains 221 3D models from Thingi10k~\cite{zhou2016thingi10k}, where the packing problems have 4 to 200 patches. \zeshi{Note that although the size of our datasets is not comparable to the sizes of datasets commonly used for computer vision tasks, our learned packing policy demonstrates pretty good generalization ability, as shown in the results section. This is because the training is conducted on patch-level instead of model level. We train our models on randomly sampled patches from these 3D models, leading to millions of combinatorial packing instances for training.}

\paragraph{Baselines} We identify three baselines to compare against. Our first baseline is the exact NFP-based packing approach combining two heuristic methods: maximal packing ratio~\cite{burke2006new} and lowest gravity center~\cite{liu2006algorithm}, which is denoted as NFP-Heuristic. Given a list of patches, NFP-Heuristic first sorts all patches in the area-descending order and then sequentially packs each patch. For a new patch, NFP-Heuristic considers its 16 rotations and computes NFP for each rotation using Minkowski sum~\cite{bennell2008comprehensive} to find collision-free translations. Finally, the pose leading to the highest packing ratio is selected and, if two poses lead to the same packing ratio, we select the one with the lower gravity center position. We compute the NFP using a highly optimized reduced convolution algorithm implemented in CGAL~\cite{fabri2009cgal}. Our second baseline is the packer algorithm implemented in the open-source software: XAtlas~\cite{Xatlas}, which implements aggressive simplification and acceleration techniques, allowing the packing algorithm to scale to problems with hundreds or thousands of patches. For example, it uses voxelized patches instead of piecewise linear ones, so that they could use a scanline algorithm instead of exact NFP computation. Our third baseline re-implements~\cite{sander2003multi} using Python. The major difference between XAtlas and \cite{sander2003multi} lies in their heuristics, where XAtlas maximizes packing ratio and~\citet{sander2003multi} minimizes wasted area.

\begin{table*}[ht]
\centering
\captionof{table}{\label{table:ratio}\small{(Min|Max|Avg) packing ratios of all algorithms on the test datasets; Ours$^\star$ means our algorithm run with a fixed aspect ratio of $1$; $^\dagger$ means our algorithm trained on the general dataset.}}
\vspace{-5px}
\setlength{\tabcolsep}{3px}
\begin{tabular}{lcccccc}
\hline
 Test-set & \cite{sander2003multi} & XAtlas & NFP & Ours & Ours$^\star$ & Our$^\dagger$\\
\hline
 Building & 0.470|0.830|0.675 & 0.525|0.835|0.670 & 0.499|0.907|0.707 & \textbf{0.683|0.980|0.827}  & 0.683|0.980|0.801 & 0.683|0.980|0.805\\
 Object  & 0.290|0.733|0.609 & 0.385|0.788|0.588 & 0.439|0.805|0.630 & \textbf{0.377|0.862|0.687} & 0.377|0.843|0.680 & 0.377|0.862|0.682\\
 General  & 0.455|0.883|0.652 & 0.449|0.886|0.688 & 0.405|0.886|0.690 & \textbf{0.540|0.937|0.776} & 0.509|0.937|0.757 & -\\
 \hline
\end{tabular}

\end{table*} 

\paragraph{Results} For each dataset, we first profile the packing ratio of all the algorithms on the testing problems. As summarized in~\prettyref{table:ratio} (Ours), our algorithm consistently outperforms the other baselines by $5\%-10\%$. To further justify the generality of our method, we train our networks on the General dataset (Ours$^\dagger$) but test it on the other two datasets. In this case, our method suffers from a marginal loss in packing ratio, but still outperforms all the baselines. This result justifies that our method has a reasonable ability of domain transfer and can be ported to pack patches for different classes of 3D models in a zero-shot fashion. More results of our method are visualized in~\prettyref{appen:more_results}.

\begin{table}[ht]
\centering
\caption{\label{table:LSNAccuracy} Average packing ratio over 2500 random problems with $H$ patches.}

\begin{tabular}{lcccc}
\toprule
  & \cite{sander2003multi} & XAtlas & NFP & Ours ($\text{LL}$) \\
\midrule
$\text{pr}$ & 0.618 & 0.651& 0.582 & 0.686 \\
 \bottomrule
\end{tabular}
\end{table}

\paragraph{Ablation Study} We further analyze aspects of our learning-assisted technique. First, we profile the accuracy of HSN, which is measured by the fraction of patch pairs that are correctly ranked. Our HSN achieves an accuracy of $90.8\%$, $86.9\%$, and $84.6\%$ on the Building, Object, and General test sets, respectively. Our trained HSN achieves a high ranking accuracy for the Building dataset and the accuracy degrades for the Object and General datasets, in which the patch shapes are more complex than those in the Building dataset. Next, we highlight the packing ratio of our low-level $\pi^\text{LPN}$ and $\pi^\text{LSN}$ alone. To this end, we sample a random subset of $H$ patches and use our $\text{LL}$ to perform packing. We compare $\text{LL}$ with the baselines and summarize the results averaged over $2500$ random problems in~\prettyref{table:LSNAccuracy}. It is shown that our DRL-based packing policy still outperforms other baselines for smaller packing problems with $H$ patches, which validates the necessity of using a learned packing policy as the low-level packer. Next, we compare the packing ratio of $\text{LL}$ under different horizons $H$. We train four low-level algorithms with $H=2,\cdots,5$ and our resulting packing ratios over the $2500$ random problems are show in~\prettyref{table:H_ablation}. Our approach performs the worst when $H=2$, where our low-level policies become myopic, but the quality varies only slightly when $H\geq2$. Therefore, we choose $H=4$ for the highest quality. 

\begin{table}[ht]
\centering
\caption{\label{table:H_ablation}Packing ratio comparison with different H.}
\begin{tabular}{lcccc}
\toprule
 &H = 2 & H = 3 & H = 4 & H = 5 \\
\midrule
\text{pr} & 0.741 & 0.771 & 0.776 & 0.770\\
\bottomrule
\end{tabular}
\end{table}

\zeshi{Finally, to validate the design choices of our pipeline, we analyze several variants of our method. The corresponding results are summarized in~\prettyref{table:Ablation}. In LPN+LSN, we remove our hierarchical grouping procedure, and train LPN and LSN with a larger horizon $H = 10$ to directly pack a sequence of patches sorted in area-descending order. In NFP+HSN, we replace $\text{LL}(\bullet)$ with NFP and area-descending ordering, but still use our HSN for hierarchical grouping. In LPN(c)+LSN+HSN, we train the LPN using continuous action space, where we adopt PPO algorithm~\cite{schulman2017proximal} to optimize the policy. The learning process of LSN and HSN remains the same as our method. In LPN+LSN+HSN, we use our low- and high-level algorithms but without hole-filling, i.e., we consider all the patches in the high-level algorithm and do not filter out tiny patches. As shown in~\prettyref{table:Ablation}, LSN+LPN performs extremely poorly. We find that the policy learned with large horizons fails to converge due to the huge search space and extremely high complexity involved in irregular shape packing problems, which validates the necessity of our hierarchical grouping procedure. We can see that LPN(c)+LSN+HSN leads to better performances than NFP+HSN, verifying the effectiveness of our learning-based packing policy. By the packing ratio comparisons among LPN(c)+LSN+HSN, LPN+LSN+HSN, and ours, we justify the necessity of a discrete action space and the hole-filling strategy.}

\begin{table}[ht]
\centering
\caption{\label{table:Ablation}Packing ratio comparison of algorithm variants on General dataset.}
\begin{tabular}{lc}
\hline
Methods            & pr    \\ \hline
LSN + LPN          & 0.448     \\
NFP                & 0.690 \\
NFP + HSN          & 0.700 \\
LPN(c) + LSN + HSN & 0.725    \\
LPN + LSN + HSN    & 0.744 \\
Ours               & 0.776 \\ \hline
\end{tabular}
\end{table}

\begin{figure}[ht]
\centering
\includegraphics[trim=0 0 1.2cm 1.2cm,clip,width=.99\linewidth]{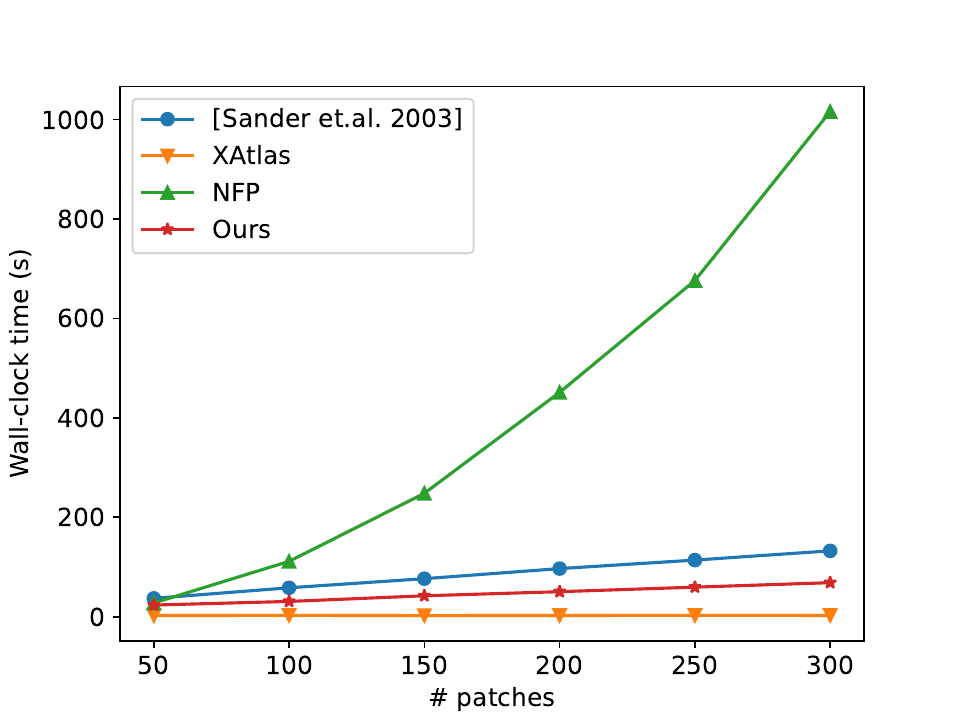}
\caption{\label{fig:scalability}The average packing time of various algorithms plotted against the number of patches.}
\end{figure}

\begin{figure*}[ht]
\centering
\bgroup
\def\arraystretch{.5}
\resizebox{\linewidth}{!}{
\begin{tabular}{ccccc}
\toprule
 & \scalebox{0.7}{\cite{sander2003multi}} & \scalebox{0.7}{XAtlas} & \scalebox{0.7}{NFP} & \scalebox{0.7}{Ours}\\
 \midrule
\includegraphics[scale=.1]{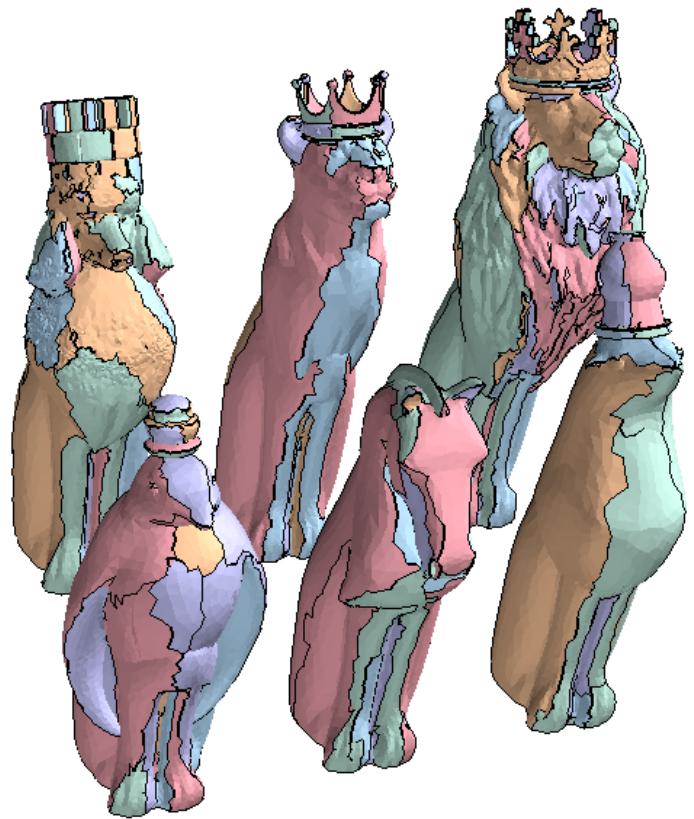}&
\includegraphics[scale=.1]{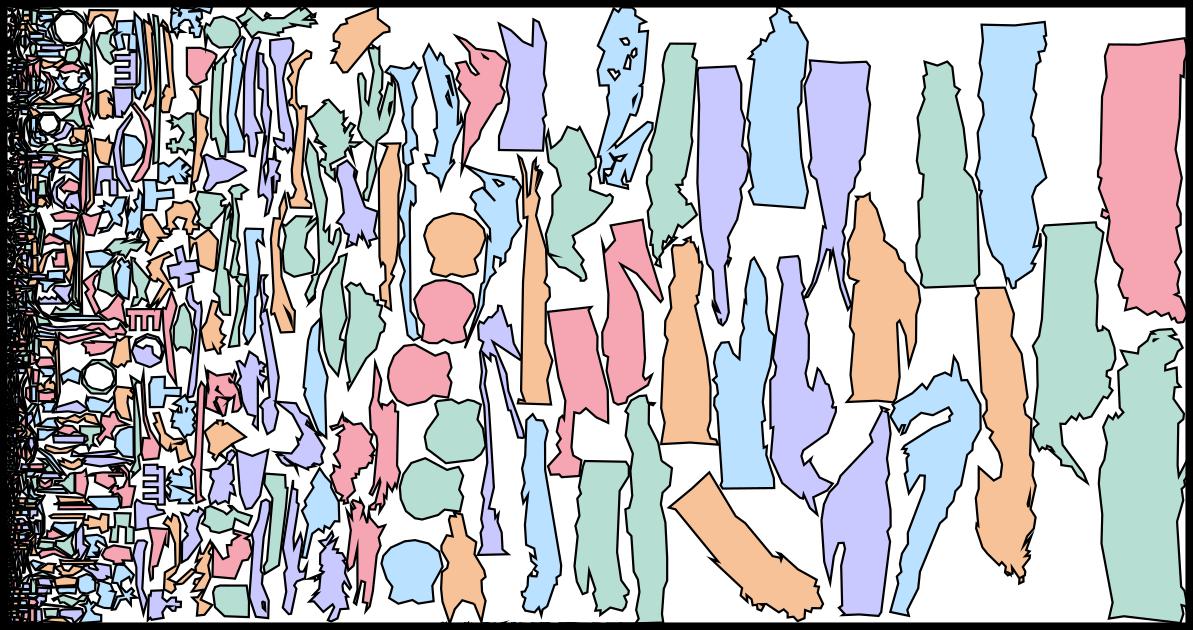}&
\includegraphics[scale=.1]{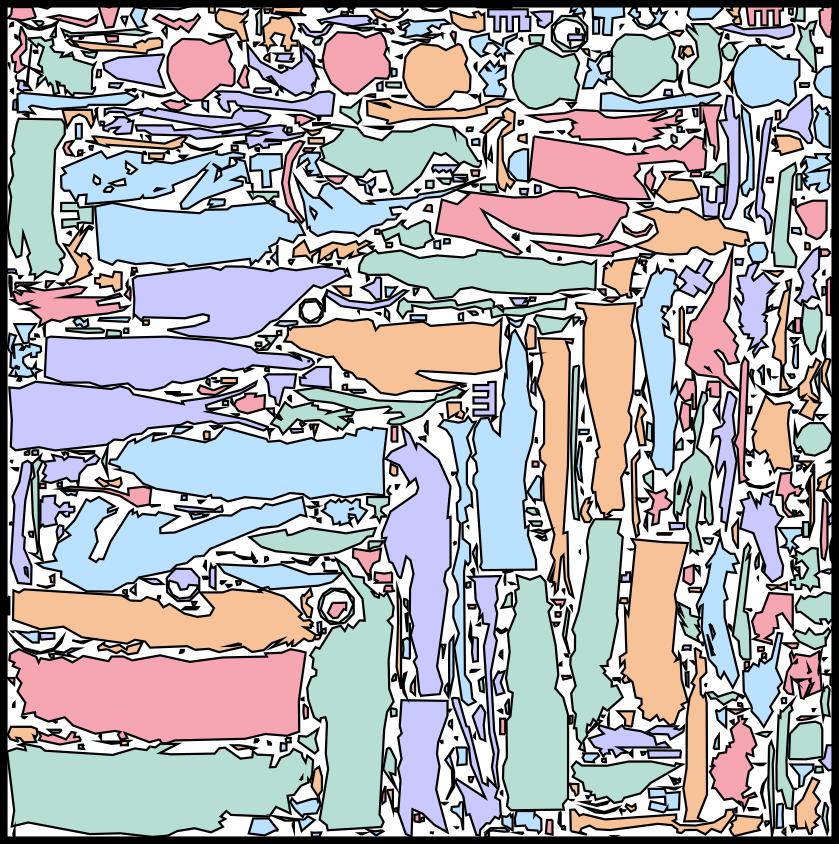}&
\includegraphics[scale=.1]{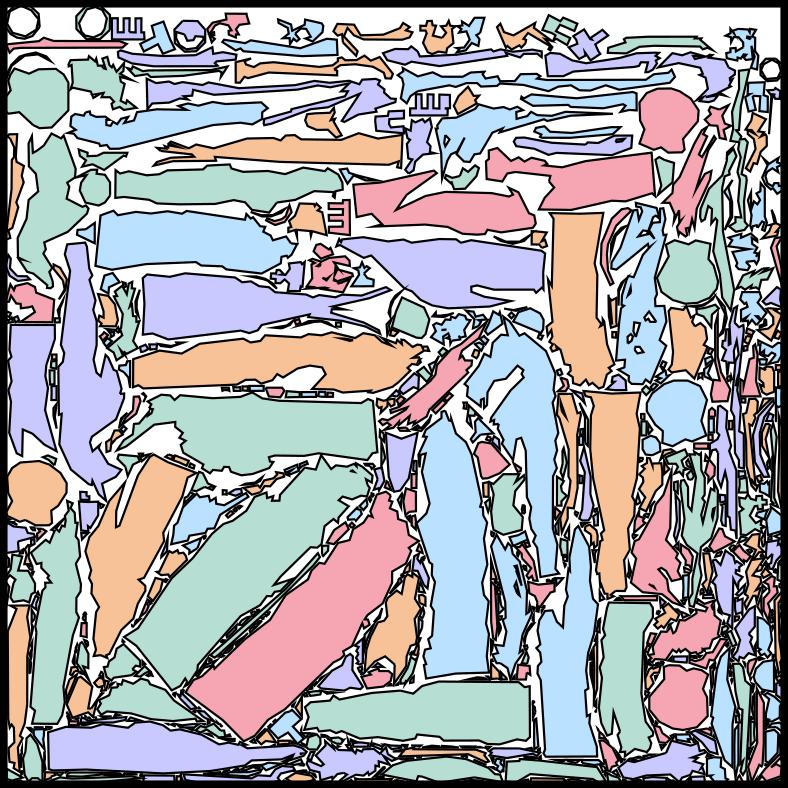}&
\includegraphics[scale=.1]{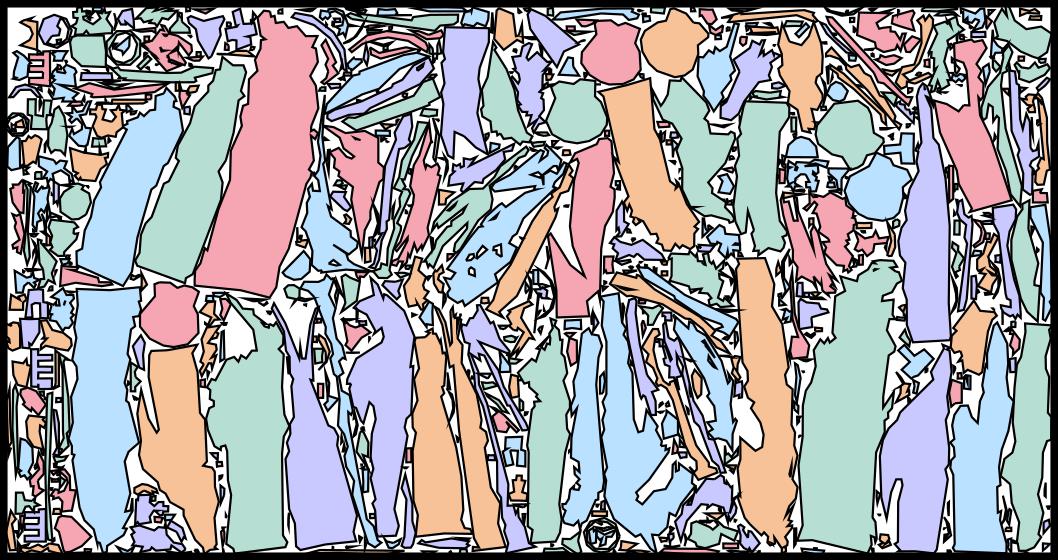}\\
 & \scalebox{0.7}{pr: 59.6\%} &\scalebox{0.7}{pr: 67.4\%} & \scalebox{0.7}{pr: 72.3\%} & \scalebox{0.7}{pr: 75.9\%}\\
\bottomrule
\end{tabular}}
\egroup
\caption{\label{fig:large_example}A large problem instance with $784$ charts.}
\end{figure*}

\paragraph{Computational Cost} Although our method achieves better packing ratios, our computational efficacy is inferior to XAtlas, due to the repeated network evaluation. For the general dataset, the average packing time of XAtlas~\cite{Xatlas}, \cite{sander2003multi}, NFP, and our method are $1.81s$, $33.52s$, $93.62s$, and $37.76s$, respectively. The performance breakdown for our packing method on the general dataset is summarized in~\prettyref{table:breakdown}. It is shown that the computational bottleneck lies in the scanline-based hole filling, which involves nested for loops and is implemented in Python. Our method can be accelerated if we switch to a scanline algorithm implemented in native-C++. We further study the scalability of our method in dealing with large UV packing problems. To this end, we combine patches from several 3D models and use each algorithm to pack all the patches into a single texture. We create a dataset including packing instances with $50,100,150,200,250,300$ patches, on which we run our method and other baselines. The computational overhead is plotted against the number of patches in~\prettyref{fig:scalability}. The cost of NFP is much higher than other algorithms due to the superlinear increase of computational complexity in computing the Minkowski sum. By design of our high-level selection policy, our method scales linearly against the number of patches, although our method is slower than XAtlas. In~\prettyref{fig:large_example}, we show an example in which $784$ charts segmented from $6$ animal chesses are packed into a single Atlas. The packing time of \cite{sander2003multi}, XAtlas, NFP and our method are $180.68s$, $4.68s$, $2966.63s$ and $278.87s$, respectively. Our method achieves better packing ratio than NFP while requiring significantly fewer computational resources.

\begin{table}[ht]
\centering
\caption{\label{table:breakdown}The performance breakdown for packing one patch on average.}
\begin{tabular}{lc}
\toprule
 Procedure & Cost(\%) \\
\midrule
HSN Evaluation (\prettyref{sec:highLevelSelect}) & 10.1\%\\
LPN+LSN Evaluation (\prettyref{sec:lowLevelPack},\ref{sec:lowLevelOrder}) & 18.4\%\\
Bin-packing (\prettyref{sec:global}) & 3.9\%\\
Joint optimization (\prettyref{eq:jointOpt}) & 12.9\%\\
Scanline-based hole filling & 52.9\%\\
\bottomrule
\end{tabular}
\end{table}

\begin{figure*}[ht]
\centering
\bgroup
\def\arraystretch{.5}
\resizebox{\linewidth}{!}{
\begin{tabular}{ccccc}
\toprule
\scalebox{0.7}{\rotatebox{90}{Unfixed aspect ratios}}&
\includegraphics[scale=.1]{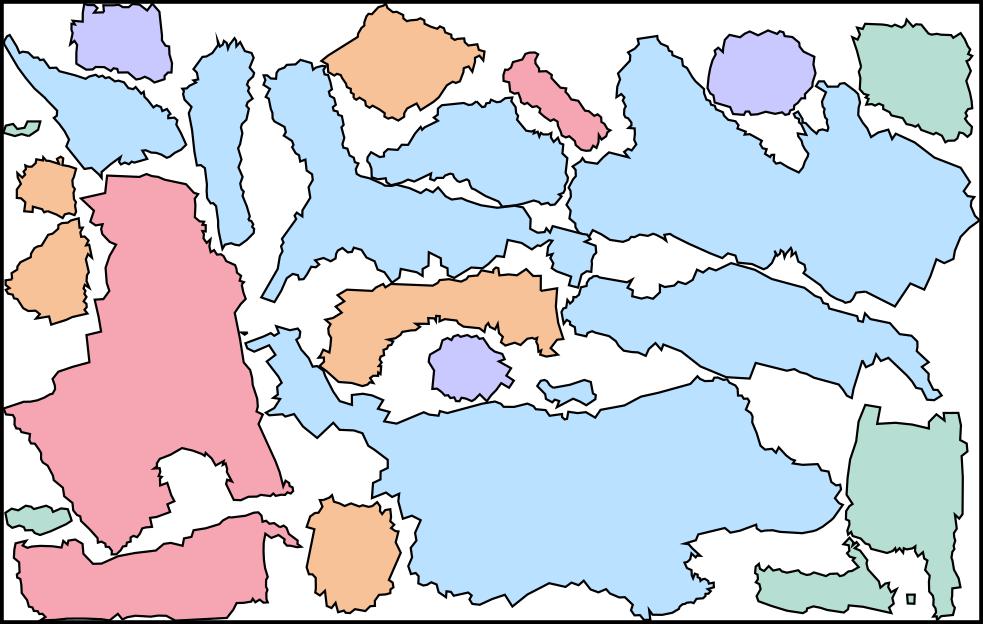}&
\includegraphics[scale=.1]{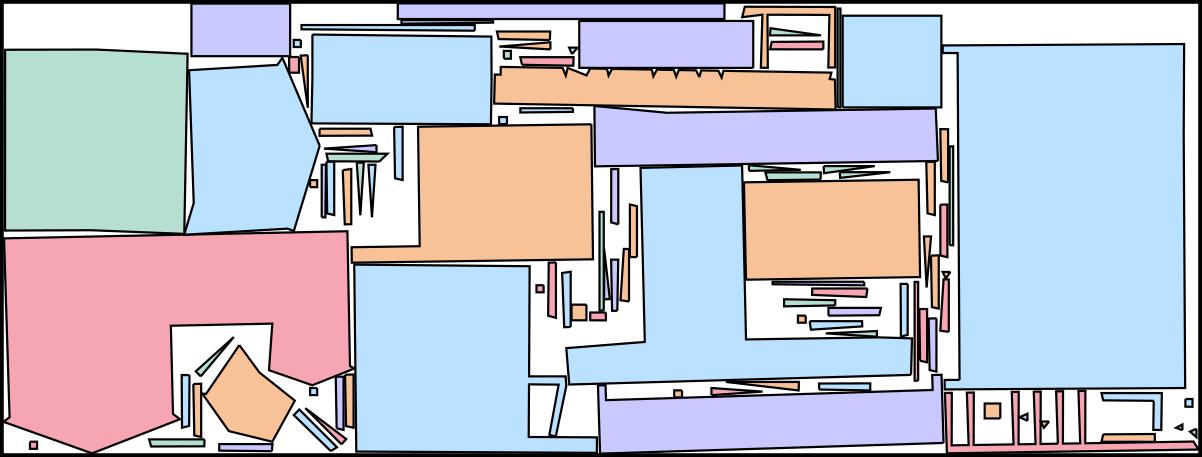}&
\includegraphics[scale=.1]{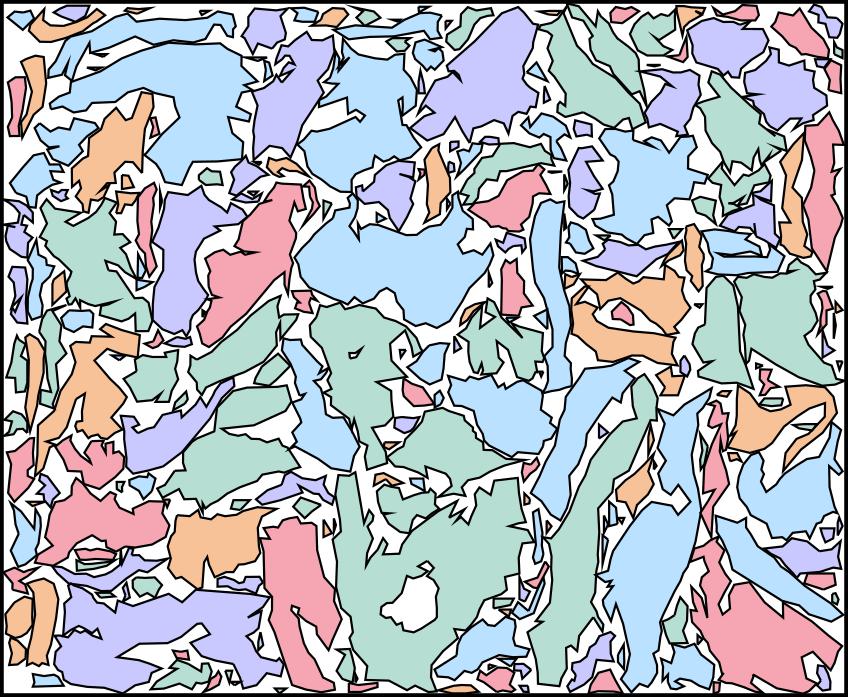}&
\includegraphics[scale=.1]{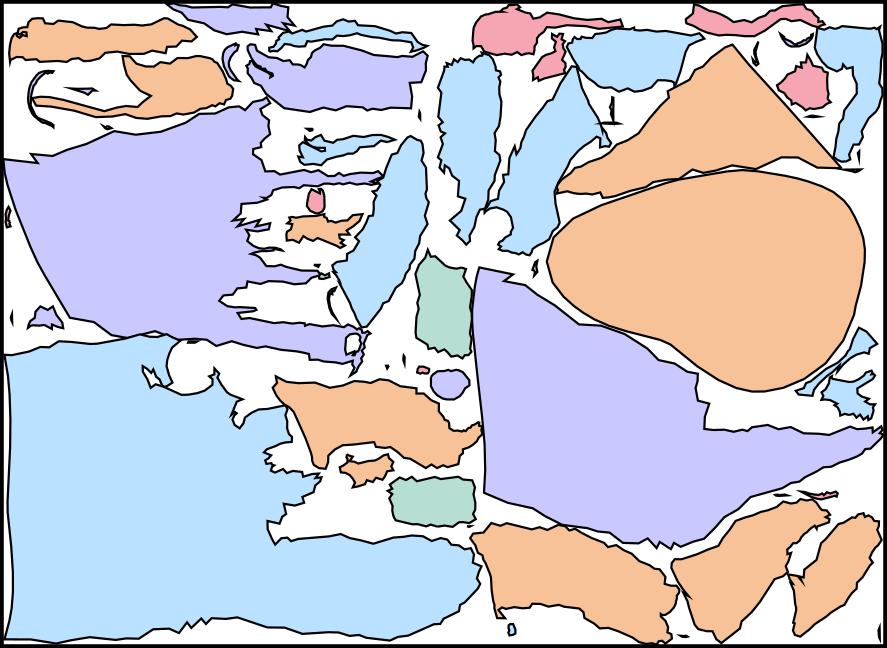}\\
&\scalebox{0.7}{pr: 63.7\%} & \scalebox{0.7}{pr: 77.0\%} &\scalebox{0.7}{pr: 67.2\%} & \scalebox{0.7}{pr: 69.7\%}\\
\midrule
\scalebox{0.7}{\rotatebox{90}{Fixed aspect ratios}}&
\includegraphics[scale=.1]{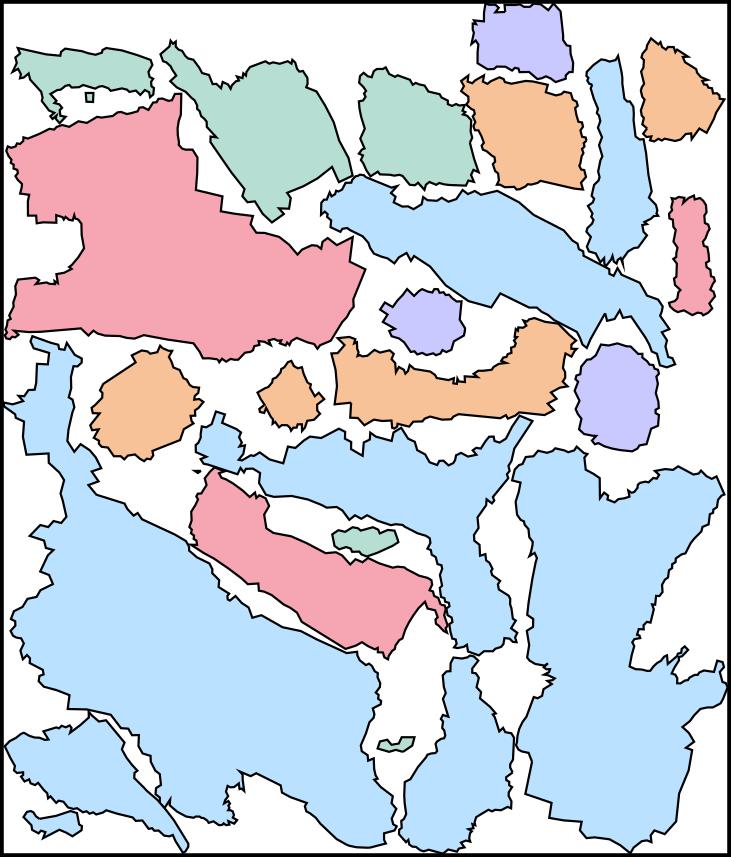}&
\includegraphics[scale=.1]{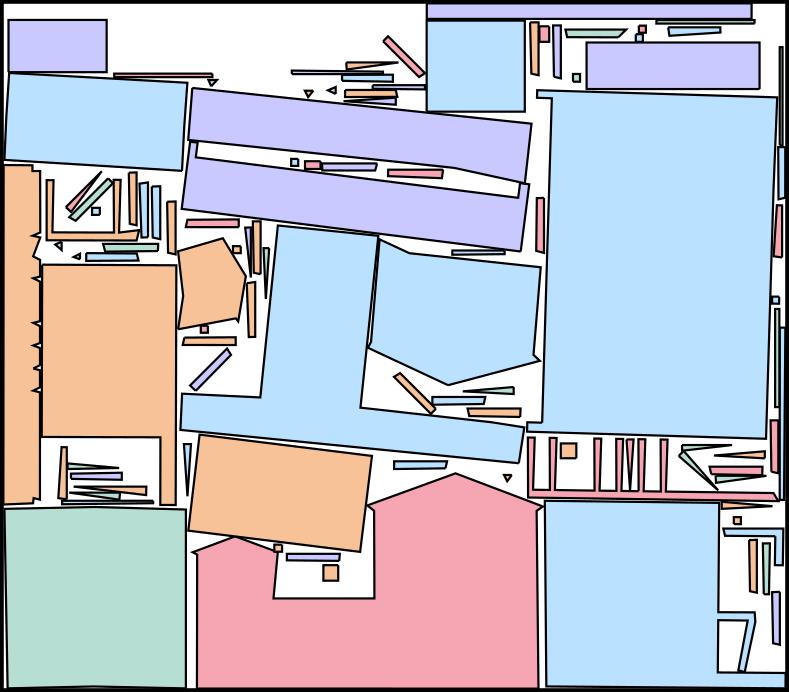}&
\includegraphics[scale=.1]{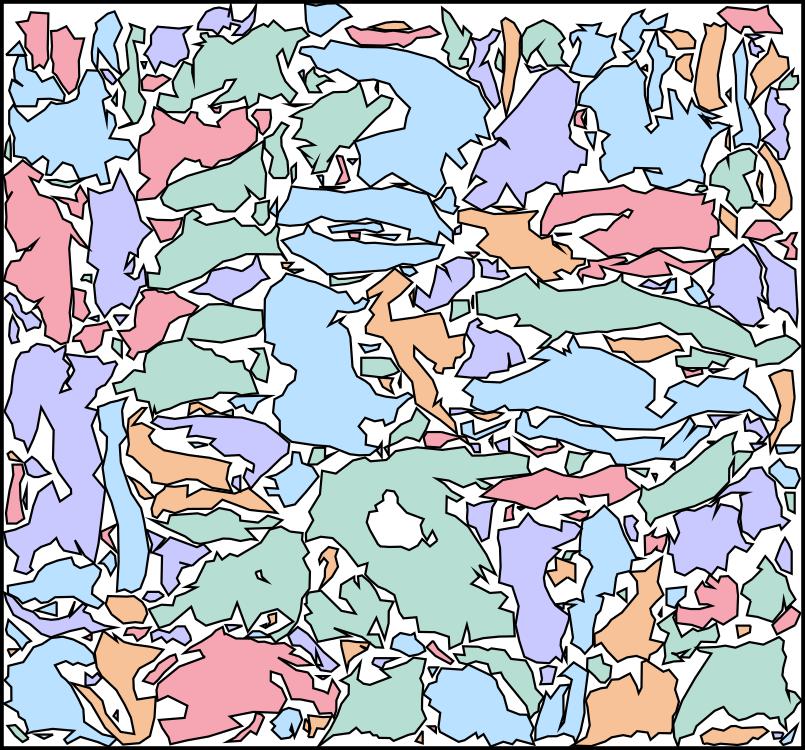}&
\includegraphics[scale=.1]{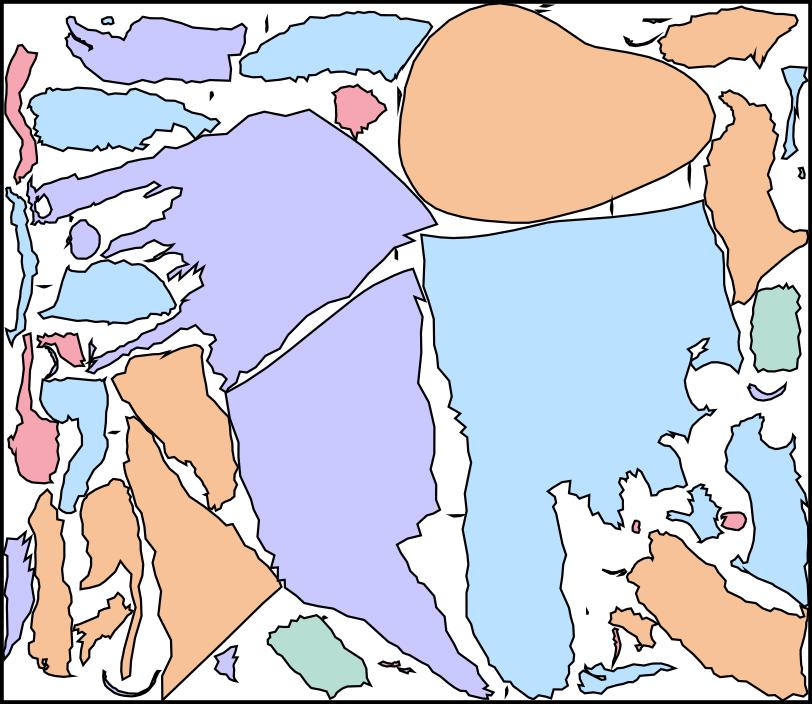}\\
&\scalebox{0.7}{pr: 62.1\%} & \scalebox{0.7}{pr: 76.0\%} &\scalebox{0.7}{pr: 66.3\%} & \scalebox{0.7}{pr: 69.7\%}\\
\bottomrule
\end{tabular}}
\egroup
\caption{\label{fig:ratio1} \zeshi{Packing results with and without fixed aspect ratios.}}
\end{figure*}

\paragraph{User-controlled Aspect Ratio}
By default, we search for the optimal aspect ratio of the texture image that maximizes the packing ratio. However, our method can be easily adapted to support a user-specified packing ratio, by forwarding it to the bin-packing procedure. To compare our method with XAtlas, we conduct experiments where we specify the aspect ratio of $1$. The results of these experiments are also summarized in~\prettyref{table:ratio} (Ours$^\star$), where our approach still generates the best results compared to baselines, although with an expected degradation compared to our results with the default setting. \zeshi{\prettyref{fig:ratio1} visualizes some of the packing results with fixed and adjustable aspect ratios.}

\begin{figure}[ht]
\centering
\setlength{\tabcolsep}{1px}
\begin{tabular}{rrll}
\rotatebox{90}{\quad\quad After First Stage}&
\includegraphics[scale=.15]{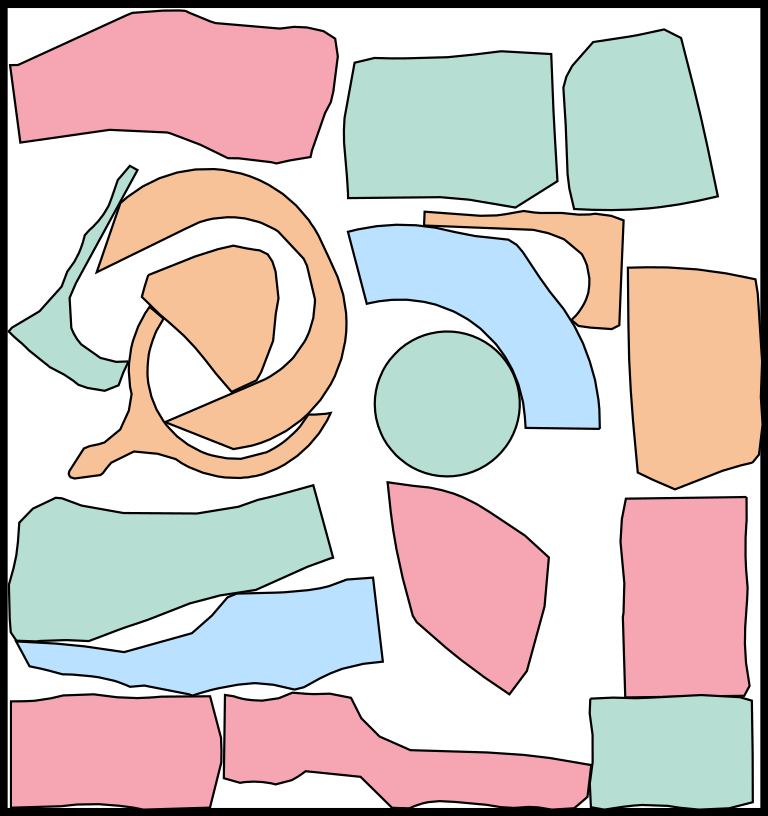}&
\includegraphics[scale=.15]{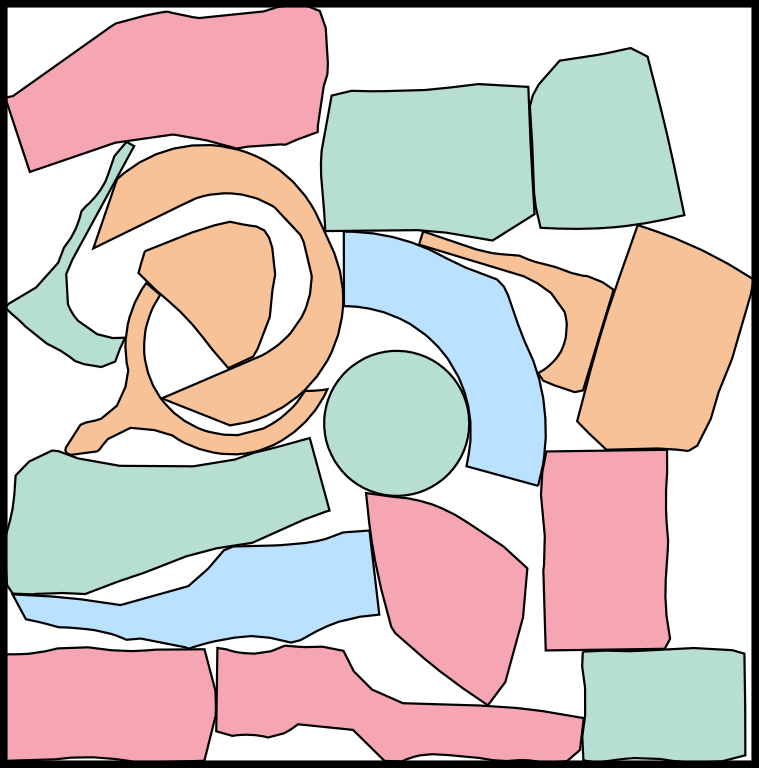}&
\rotatebox{90}{\quad\quad After Joint Opt.}\\
\rotatebox{90}{\;\TWORCell{No-filtering baseline}{pr: 71.9\%}}&
\includegraphics[scale=.15]{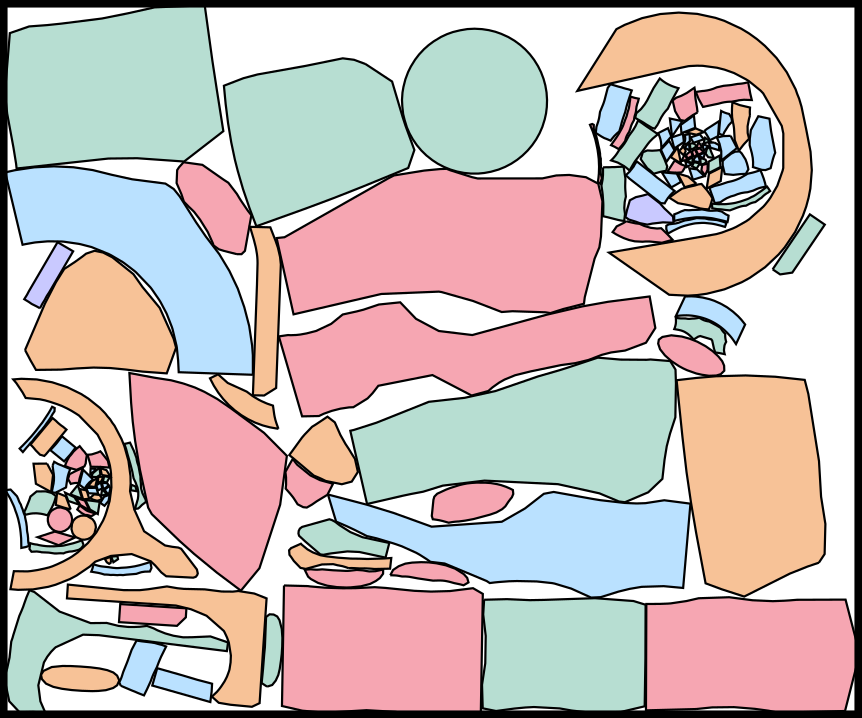}&
\includegraphics[scale=.15]{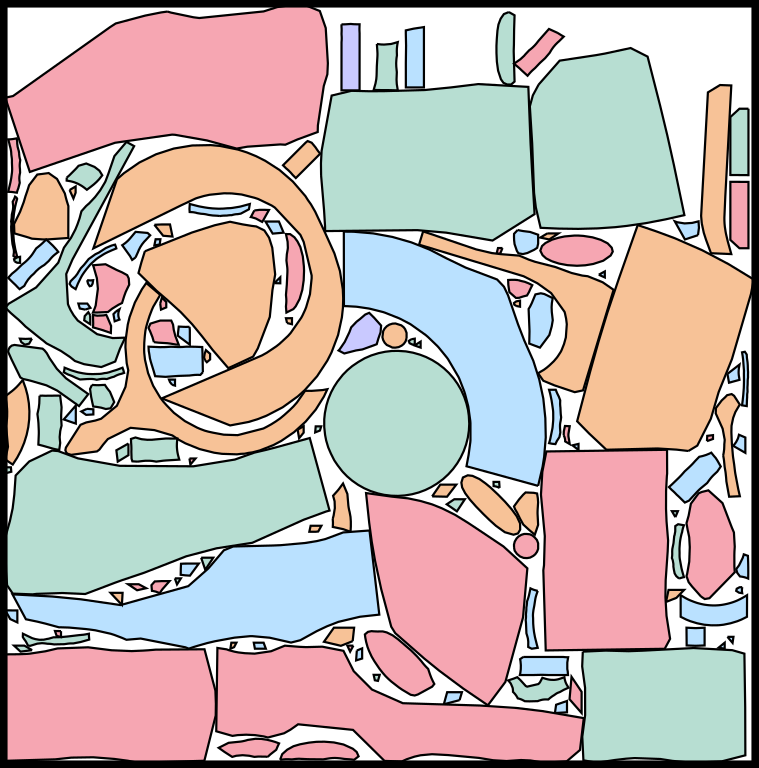}&
\rotatebox{90}{\quad\TWORCell{After Hole-filling}{pr: 76.5\%}}\\
\end{tabular}
\caption{\label{fig:stageTwo} After the first stage, there are gaps between non-tiny patches (top left). We perform a joint optimization to squeeze patches and remove gaps (top right). After hole-filing, we achieve a packing ratio of $76.5\%$ (bottom right). If we do not filter out the tiny patches and forward all patches to HSN, we can only achieve a ratio of $71.9\%$ (bottom left).}
\end{figure}


\section{Conclusions and Future Work}
We present a learning-assisted irregular shape packing algorithm for UV patches. On three datasets with various topology and geometry properties, we achieve $5\%-10\%$ packing ratio improvement over XAtlas, NFP, and \cite{sander2003multi} baseline algorithms. Our algorithm can deal with problem instances with up to hundreds of patches within tolerable computational overhead for offline packing. By optimizing only the rigid transformations for the patches, our approach respects the input UV patch shapes and parameterizations, which can be immediately incorporated into existing UV unwrapping pipelines.

\begin{figure}[ht]
\centering
\bgroup
\def\arraystretch{.5}
\resizebox{\linewidth}{!}{
\begin{tabular}{cc}
\toprule
\scalebox{0.5}{XAtlas} & \scalebox{0.5}{Ours}\\
\midrule
\includegraphics[scale=.1]{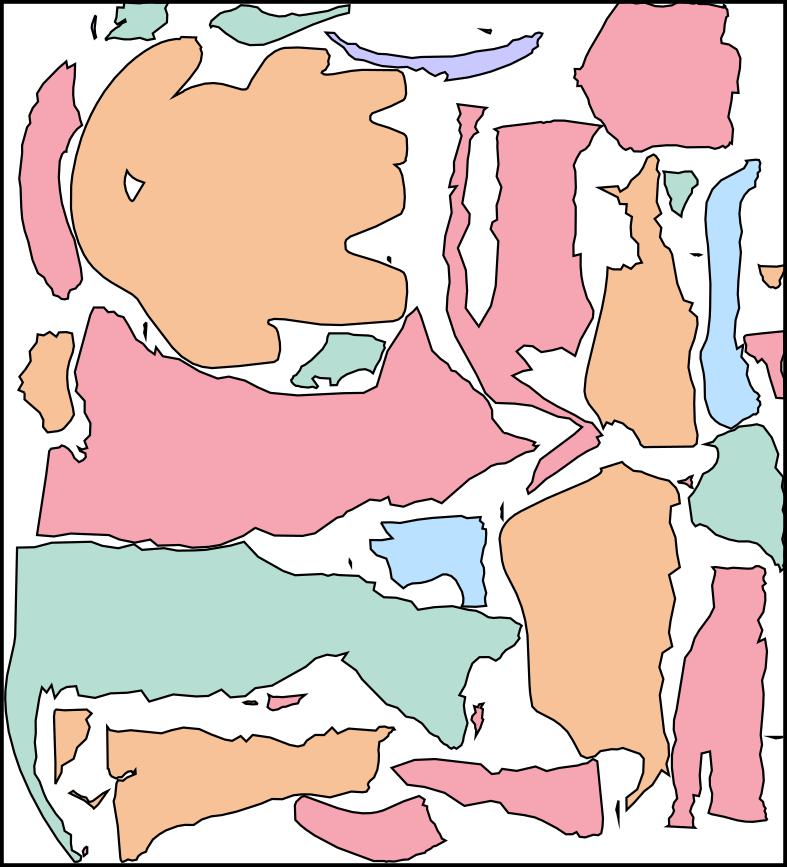}&
\includegraphics[scale=.1]{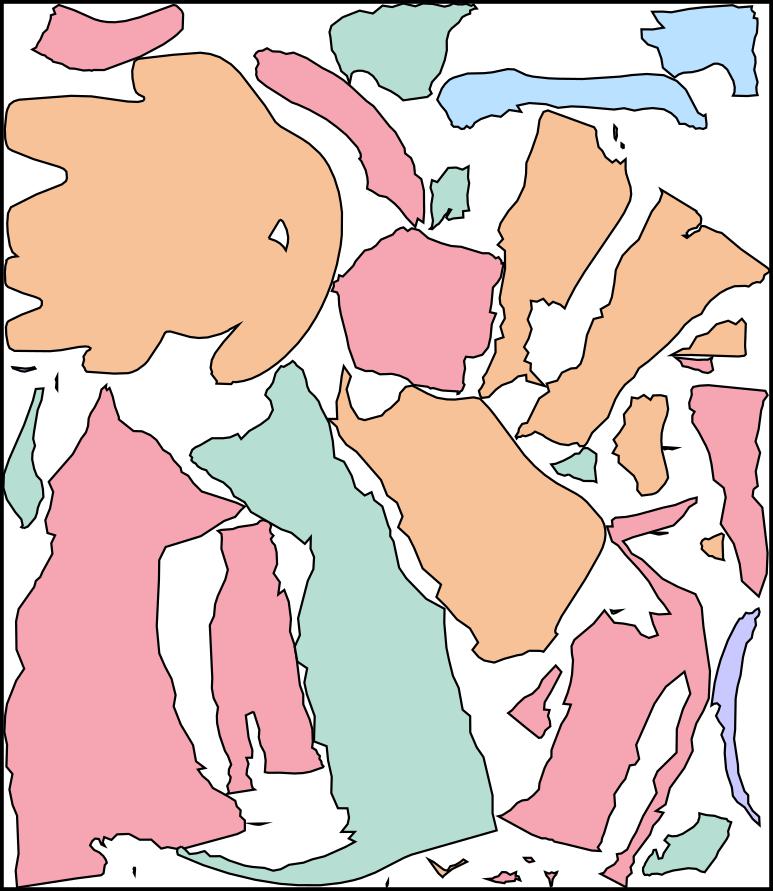}\\
\scalebox{0.5}{pr: 64.0\%} & \scalebox{0.5}{pr: 63.7\%}\\
\midrule
\includegraphics[scale=.1]{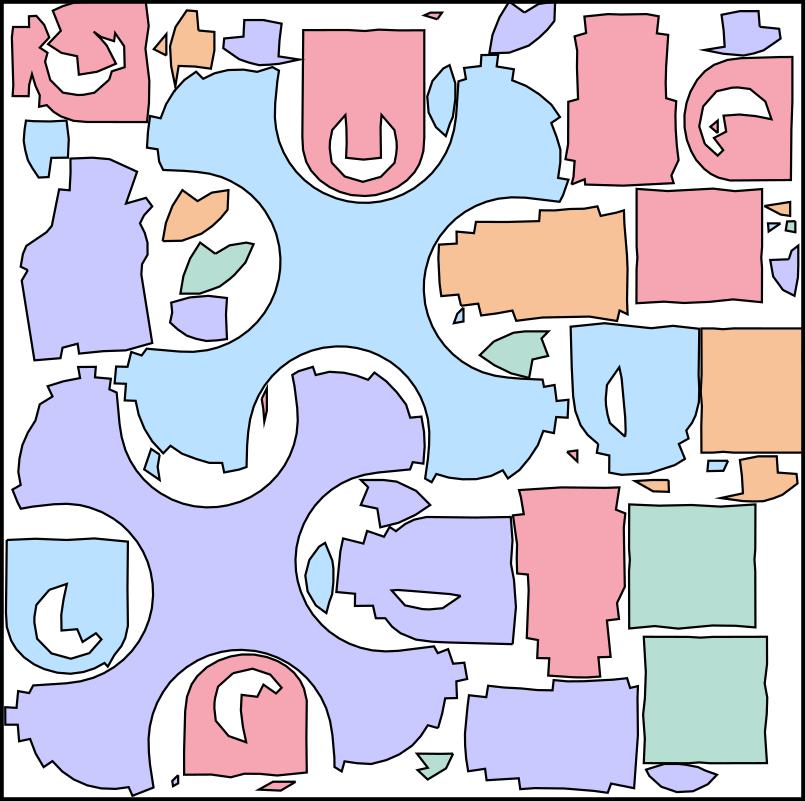}&
\includegraphics[scale=.1]{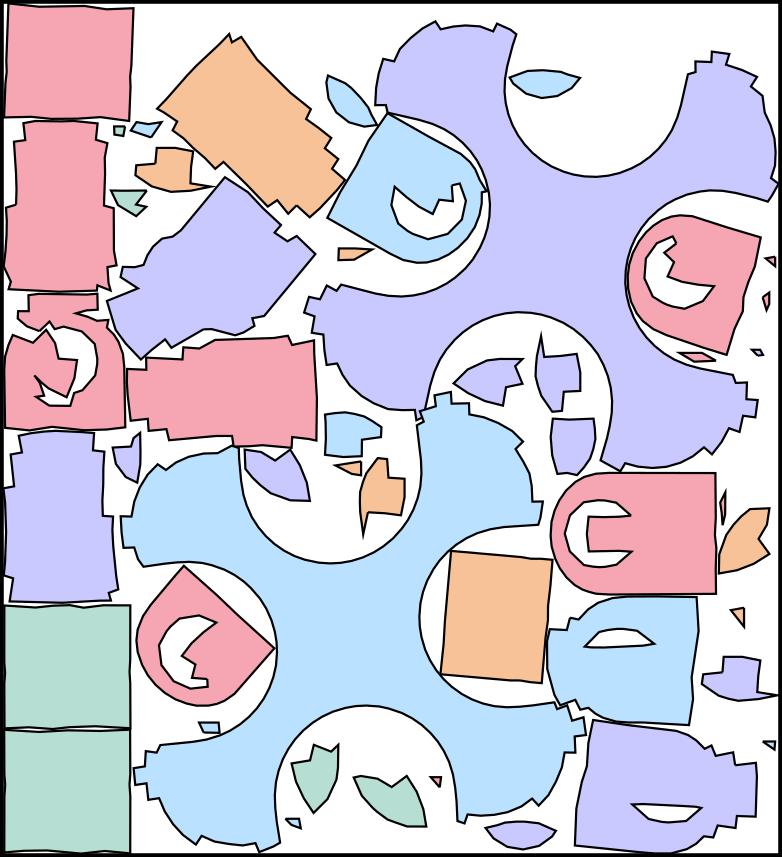}\\
\scalebox{0.5}{pr: 66.9\%} & \scalebox{0.5}{pr: 64.9\%}\\
\bottomrule
\end{tabular}}
\egroup
\caption{\label{fig:suboptimal} \zeshi{Sub-optimal packing instances generated by our pipeline.}}
\end{figure}

\zeshi{Our method leads to several noteworthy future directions. First, although our method achieves averaged best performance, there can be sub-optimal solution instances as illustrated in~\prettyref{fig:suboptimal}. Second, our action space does not inherently allow patches to be put into gaps between other patches, which requires a more flexible action space design. Next, our existing hierarchical grouping approach, while efficient, operates in a greedy manner and may be confined to local optima. An alternative solution is to reframe the grouping process as a sequential decision-making problem that can be solved by advanced policy optimization algorithms. Finally, we are also looking into more efficient policy parameterizing utilizing diffusion models and transformers.}


\begin{acks}
    We would like to thank the anonymous reviewers for their constructive suggestions and feedback. We also thank Yujie Wang for her valuable discussions and help. Her expertise in solving NP-hard problems using deep neural networks has been instrumental in shaping the direction of this paper, allowing us to refine our methods and focus our efforts in the most promising directions. Additionally, Manyi Li is supported by the Excellent Young Scientists Fund Program (Overseas) of Shandong Province (Grant No.2023HWYQ-034).
\end{acks}

\bibliographystyle{ACM-Reference-Format}
\bibliography{main}
\appendix
\section{\label{appen:architecture}Network Architecture}
We summarized the detailed architecture of our three network in~\prettyref{table:LPN},~\prettyref{table:LSN}, and~\prettyref{table:HSN}. The full network pipeline of LPN is defined as:
\begin{align*}
\underset{\triangleq\bar{s}_i}{(\bar{P}_{i-1},\bar{p}_i,\cdots,\bar{p}_{i+H-1})}\gets&
\underset{\triangleq\text{FCN}(s_i)}{(\text{FCN}(P_{i-1}),\text{FCN}(p_i),\cdots,\text{FCN}(p_{i+H-1}))}\\
\text{Q}_{256}\gets&\pi^\text{LPN}(\bar{s}_i),
\end{align*}
where the output is $256$-dimensional Q-values over the action space. The full network pipeline of LSN is defined as:
\begin{align*}
\bar{s}_i\gets&\text{FCN}(s_i)\\
(\text{Q}_{p_i},\cdots,\text{Q}_{p_{i+H-1}})\gets&
\pi^\text{LSN}\left[
\left(\setlength{\arraycolsep}{1pt}\begin{array}{c}\bar{P}_{i-1} \\ \bar{p}_i\end{array}\right),
\cdots,
\left(\setlength{\arraycolsep}{1pt}\begin{array}{c}\bar{P}_{i-1} \\ \bar{p}_{i+H-1}\end{array}\right)
\right],
\end{align*}
where the output is the $H$-dimensional Q-value of each candidate next patch. Finally, the full network pipeline of HSN is defined as:
\begin{align*}
\text{pr}(\mathcal{S}')\gets&\text{HSN}(\bar{p}_1,\cdots,\bar{p}_{H}),
\end{align*}
which outputs the predicted weighted average packing ratio.

\begin{table}[ht]
\centering
\caption{The architecture of FCN.}
\begin{tabular}{ccc}
\hline
Layer Type  & Input Size               & Output Size              \\ \hline
Convolution & (batch size, 1, 50, 50)  & (batch size, 6, 24, 24)  \\ \hline
ReLU        & (batch size, 6, 24, 24)  & (batch size, 6, 24, 24)  \\ \hline
Convolution & (batch size, 6, 24, 24)  & (batch size, 12, 12, 12) \\ \hline
ReLU        & (batch size, 12, 12, 12) & (batch size, 12, 12, 12) \\ \hline
Convolution & (batch size, 12, 12, 12) & (batch size, 12, 6, 6)   \\ \hline
ReLU        & (batch size, 12, 6, 6)   & (batch size, 12, 6, 6)   \\ \hline
Flatten     & (batch size, 12, 6, 6)   & (batch size, 432)        \\ \hline
\end{tabular}
\end{table}
\begin{table}[ht]
\centering
\caption{\label{table:LPN} The architecture of LPN.}
\begin{tabular}{ccc}
\hline
Layer Type & Input Size        & Output Size       \\ \hline
Linear     & (batch size, 432 $\times$ (1 + H)) & (batch size, 1024) \\ \hline
ReLU       & (batch size, 1024) & (batch size, 1024) \\ \hline
Linear     & (batch size, 1024) & (batch size, 512) \\ \hline
ReLU       & (batch size, 512) & (batch size, 512) \\ \hline
Linear     & (batch size, 512) & (batch size, 256) \\ \hline
ReLU       & (batch size, 256) & (batch size, 256) \\ \hline
Linear     & (batch size, 256) & (batch size, 256) \\ \hline
\end{tabular}
\end{table}
\begin{table}[ht]
\centering
\caption{\label{table:LSN} The architecture of LSN.}
\begin{tabular}{ccc}
\hline
Layer Type & Input Size                & Output Size           \\ \hline
GAT        & (batch size, H, 432 $\times$ 2 ) & (batch size, H,  512) \\ \hline
ELU        & (batch size, H,  512)     & (batch size, H,  512) \\ \hline
GAT        & (batch size, H,  512)     & (batch size, H,  256) \\ \hline
ELU        & (batch size, H,  256)     & (batch size, H,  256) \\ \hline
Linear     & (batch size, H,  256)     & (batch size, H,  128) \\ \hline
ReLU       & (batch size, H,  128)     & (batch size, H,  128) \\ \hline
Linear     & (batch size, H,  128)     & (batch size, H,  64)  \\ \hline
ReLU       & (batch size, H,  64)      & (batch size, H,  64)  \\ \hline
Linear     & (batch size, H,  64)      & (batch size, H,  32)  \\ \hline
ReLU       & (batch size, H,  32)      & (batch size, H,  32)  \\ \hline
Linear     & (batch size, H,  32)      & (batch size, H,  1)   \\ \hline
\end{tabular}
\end{table}
\begin{table}[ht]
\centering
\caption{\label{table:HSN} The architecture of HSN.}
\begin{tabular}{ccc}
\hline
Layer Type  & Input Size            & Output Size          \\ \hline
GAT         & (batch size, H, 432 ) & (batch size, H, 256) \\ \hline
ELU         & (batch size, H, 256)  & (batch size, H, 256) \\ \hline
GAT         & (batch size, H, 256)  & (batch size, H, 64)  \\ \hline
ELU         & (batch size, H, 64)   & (batch size, H, 64)  \\ \hline
Max Pooling & (batch size, H, 64)   & (batch size, 64)     \\ \hline
ReLU        & (batch size, 64)      & (batch size, 32)     \\ \hline
Linear      & (batch size, 32)      & (batch size, 1)      \\ \hline
Sigmoid     & (batch size, 1)       & (batch size, 1)      \\ \hline
\end{tabular}
\end{table}

\newpage
\section{\label{appen:algorithm}Algorithm Pipeline}
We summarize our training procedure in~\prettyref{alg:training} and runtime pipeline in~\prettyref{alg:testing}.
\begin{algorithm} [H]
\caption{\label{alg:training} Training $\pi^\text{LPN}$, $\pi^\text{LSN}$, and $\text{HSN}$} 
\begin{algorithmic}[1] 
\Require{A dataset of packing problems $\mathbb{S}=\{\mathcal{S}_1,\mathcal{S}_2,\cdots\}$}
\Ensure{$\pi^\text{LPN}$, $\pi^\text{LSN}$, $\text{HSN}$}
\LineComment{Initial training}
\While{not converged}
\State Sample $\mathcal{S}\in\mathbb{S}$
\State Sample \textit{ordered} subset of $H$ patches $\mathcal{S}'\subseteq\mathcal{S}$
\State Use MDP of~\prettyref{sec:lowLevelPack} to populate experience buffer
\State Update $\pi^\text{LPN}$ via DDQN
\EndWhile
\While{not converged}
\State Sample $\mathcal{S}\in\mathbb{S}$
\State Sample \textit{unordered} subset of $H$ patches $\mathcal{S}'\subseteq\mathcal{S}$
\State Use MDP of~\prettyref{sec:lowLevelOrder} and $\pi^\text{LPN}$ to populate experience buffer
\State Update $\pi^\text{LSN}$ via DDQN
\EndWhile
\While{not converged}
\State Sample $\mathcal{S}\in\mathbb{S}$
\State Sample two \textit{unordered} subsets of $H$ patches $\mathcal{S}',\mathcal{S}''\subseteq\mathcal{S}$
\State Use $\pi^\text{LPN}$ and $\pi^\text{LSN}$ to compute groundtruth $\text{pr}'$ and $\text{pr}''$
\State Update $\text{HSN}$ using SGD on loss~\prettyref{eq:metricLoss}
\EndWhile
\end{algorithmic} 
\end{algorithm}
\begin{algorithm} [H]
\caption{\label{alg:testing} Learning-Assisted UV Packing} 
\begin{algorithmic}[1] 
\Require{A packing problem $\mathcal{S}$}
\Ensure{Pose for each $p_i\in\mathcal{S}$}
\LineComment{Exclude small patches}
\State $\mathcal{S}_/\gets\emptyset$
\State Calculate average area of salient patches $\bar{a}$
\For{$p\in\mathcal{S}$}
\If{$\text{area}(p)<\bar{a}/5$}
\State $\mathcal{S}\gets\mathcal{S}-\{p\}$
\State $\mathcal{S}_/\gets\mathcal{S}_/\cup\{p\}$
\EndIf
\EndFor
\LineComment{Main loop}
\State Compute $\text{pr}(\mathcal{S})$
\While{not converged}
\State Sample $400$ random subsets of $H$ patches $\mathcal{S}'_{1,\cdots,400}$
\State Sort $\mathcal{S}_i'$ in $\text{HSN}(\mathcal{S}_i')$-descending order
\For{$i=1,\cdots,10$}
\State Compute groundtruth $\text{pr}_i\gets\text{pr}(\mathcal{S}-\mathcal{S}_i'\cup\{\text{LL}(\mathcal{S}_i')\})$
\EndFor
\State Sort $\mathcal{S}_1',\cdots,\mathcal{S}_{10}'$ in $\text{pr}_i$-descending order
\If{$\text{pr}_1>\text{pr}(\mathcal{S})$}
\State Compute alpha shape for $\text{LL}(\mathcal{S}_1')$
\State $\mathcal{S}\gets\mathcal{S}-\mathcal{S}_1'\cup\{\text{LL}(\mathcal{S}_1')\}$
\State $\text{pr}(\mathcal{S})\gets\text{pr}_1$
\EndIf
\State {\algorithmicelse} Break
\EndWhile
\State Bin-packing all patches in $\mathcal{S}$
\State Locally squeeze patches via \prettyref{eq:jointOpt}
\LineComment{Pack small patches}
\For{$p\in\mathcal{S}_/$ in area-descending order}
\State $\mathcal{S}\gets\text{scanline}(\mathcal{S},p)$
\EndFor
\State Return all poses
\end{algorithmic} 
\end{algorithm}

\section{\label{appen:more_results}More Results}
We show more results packed by various baselines.
\begin{figure*}[ht]
\centering
\bgroup
\def\arraystretch{.5}
\resizebox{\linewidth}{!}{
\begin{tabular}{ccccc}
\toprule
 & \scalebox{0.7}{\cite{sander2003multi}} & \scalebox{0.7}{XAtlas} & \scalebox{0.7}{NFP} & \scalebox{0.7}{Ours}\\
\midrule
\includegraphics[scale=.1]{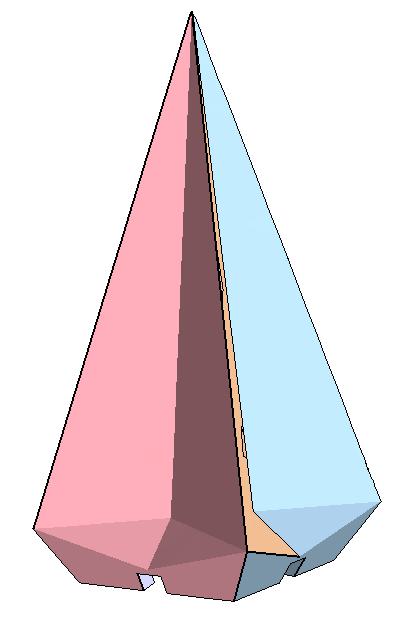}&
\includegraphics[scale=.1]{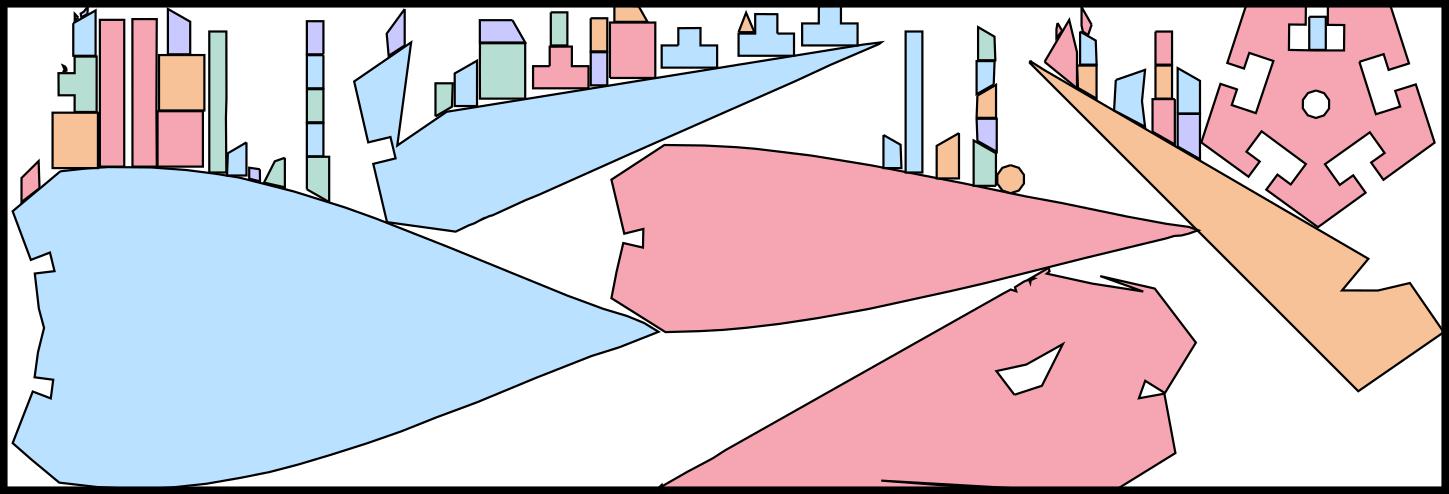}&
\includegraphics[scale=.1]{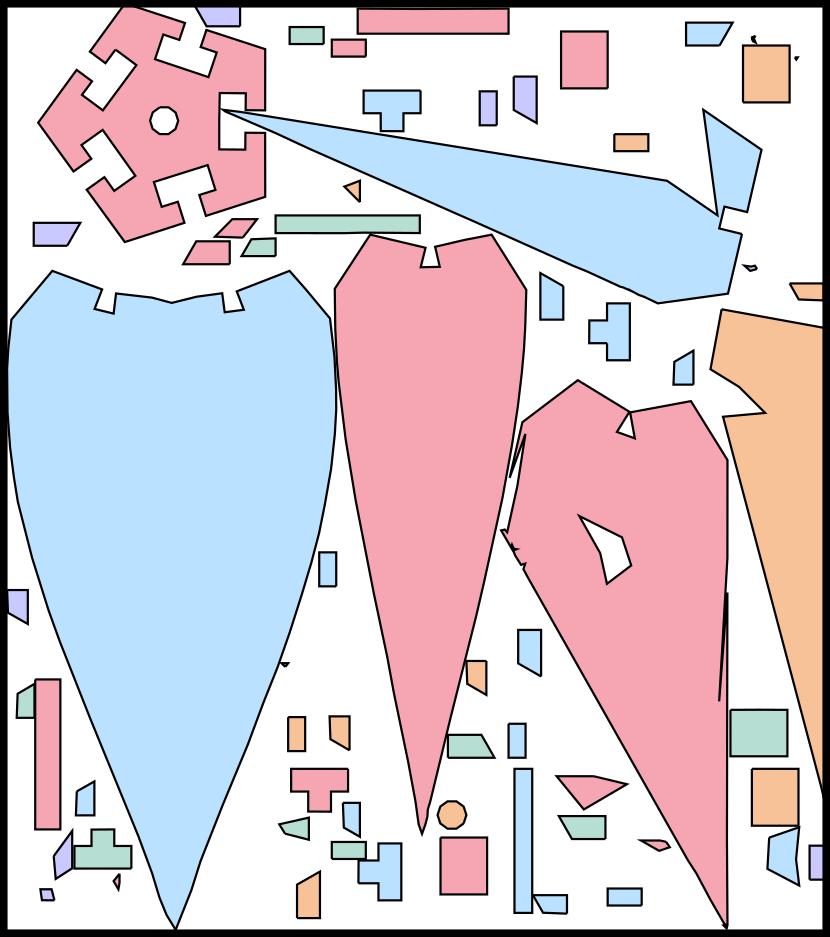}&
\includegraphics[scale=.1]{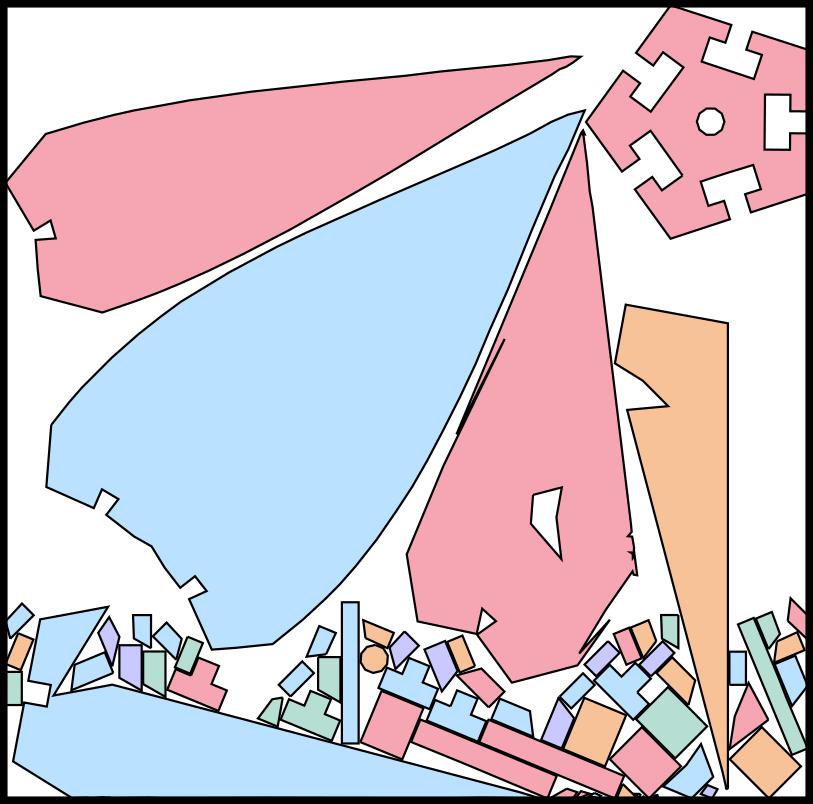}&
\includegraphics[scale=.1]{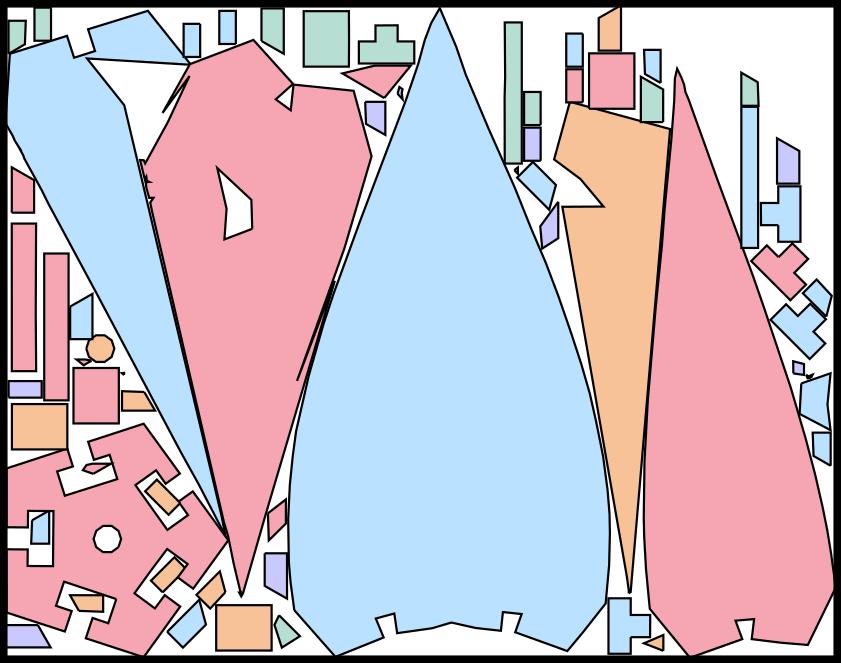}\\
 & \scalebox{0.7}{pr: 59.2\%} &\scalebox{0.7}{pr: 56.7\%} & \scalebox{0.7}{pr: 64.8\%} & \scalebox{0.7}{pr: 76.9\%}\\
\midrule
\includegraphics[scale=.1]{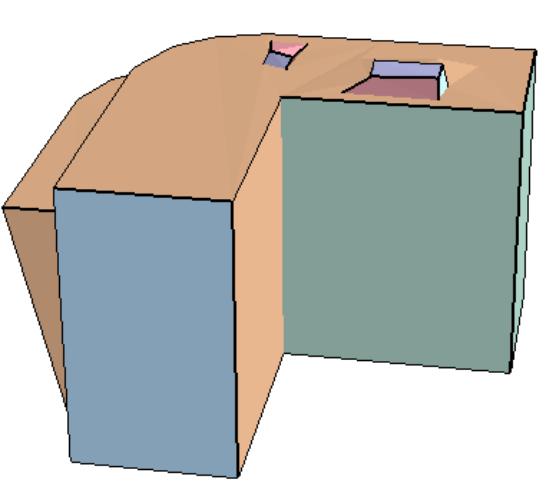}&
\includegraphics[scale=.1]{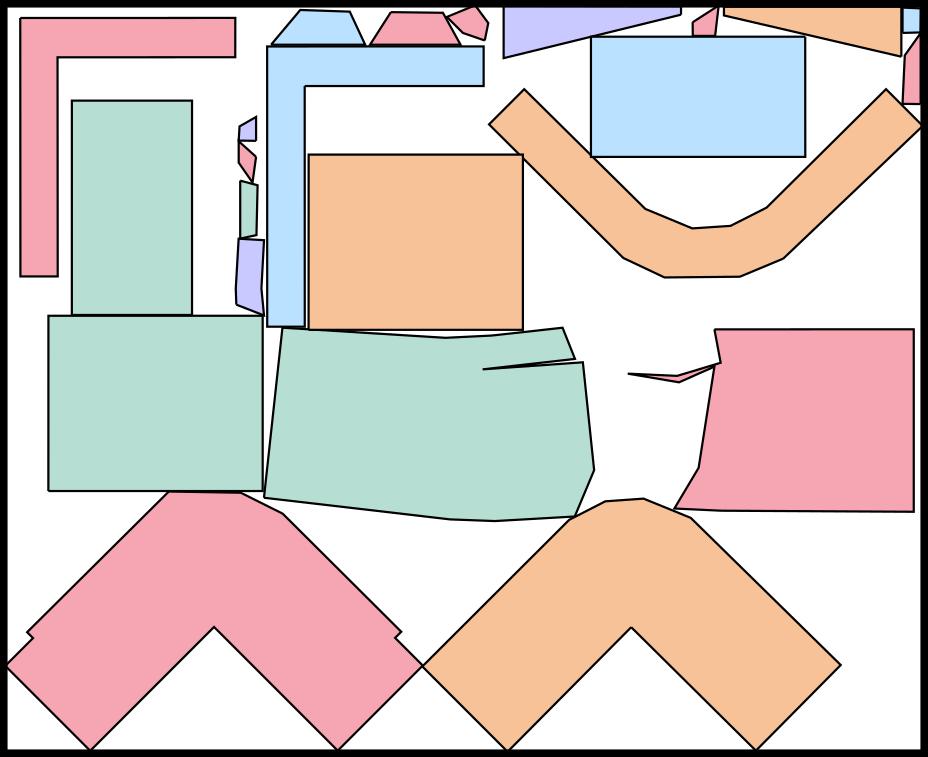}&
\includegraphics[scale=.1]{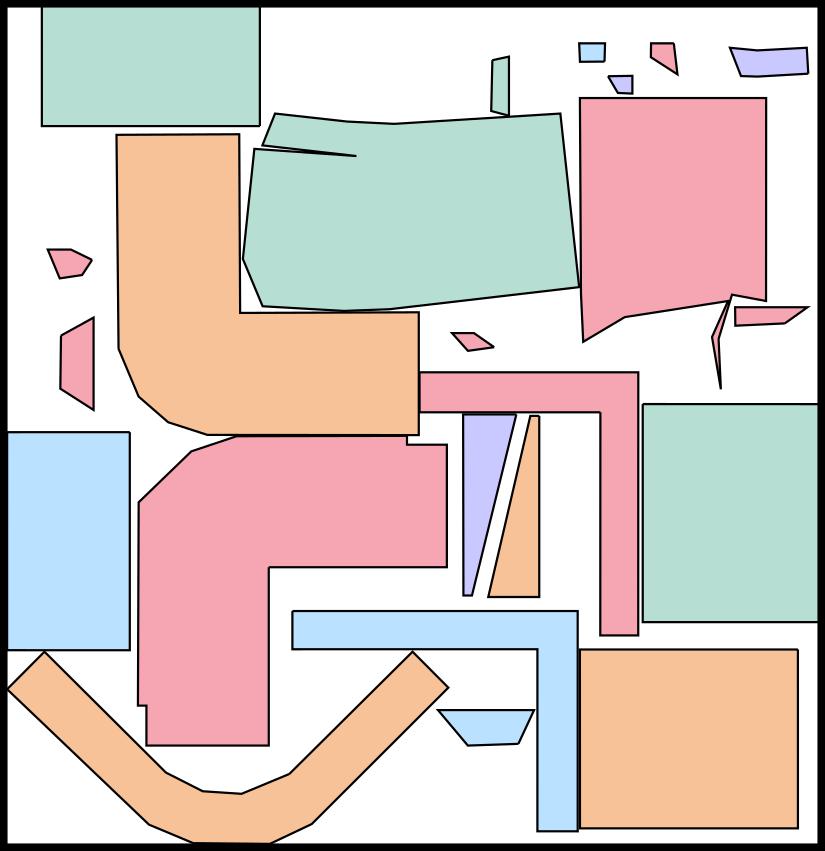}&
\includegraphics[scale=.1]{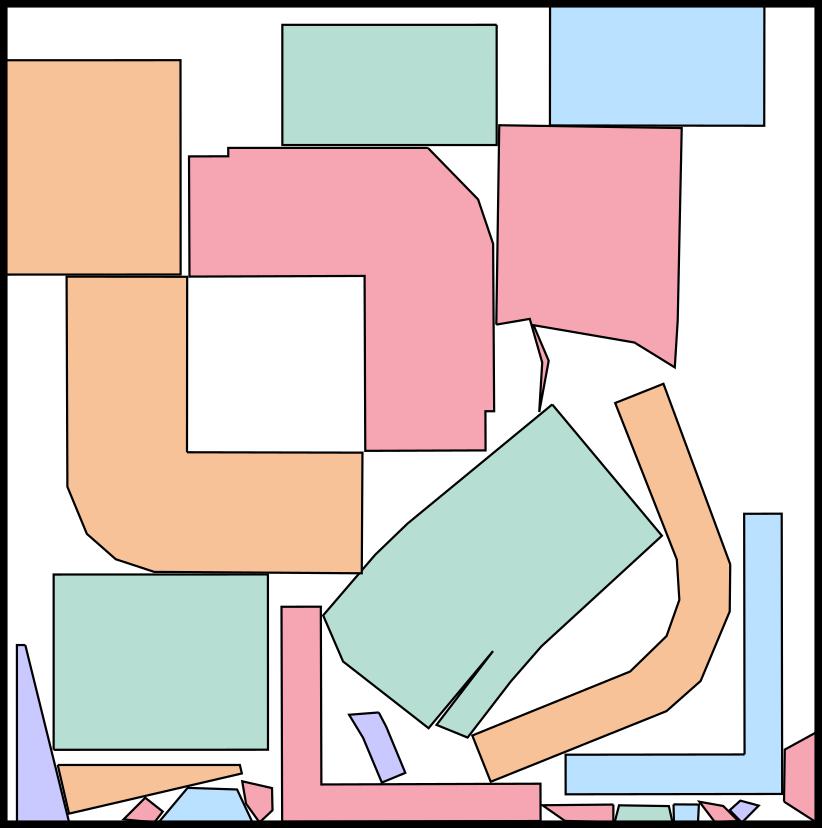}&
\includegraphics[scale=.1]{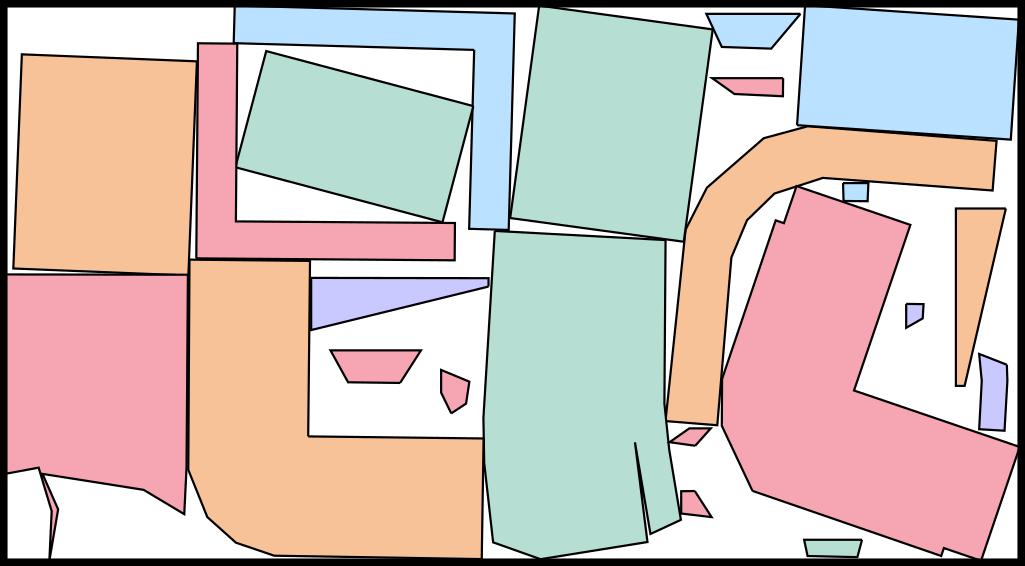}\\
 & \scalebox{0.7}{pr: 63.4\%} &\scalebox{0.7}{pr: 62.9\%} & \scalebox{0.7}{pr: 65.5\%} & \scalebox{0.7}{pr: 76.8\%}\\
\midrule
\includegraphics[scale=.1]{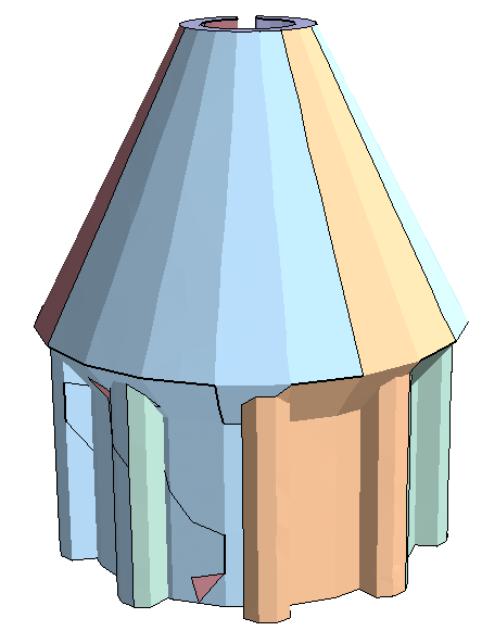}&
\includegraphics[scale=.1]{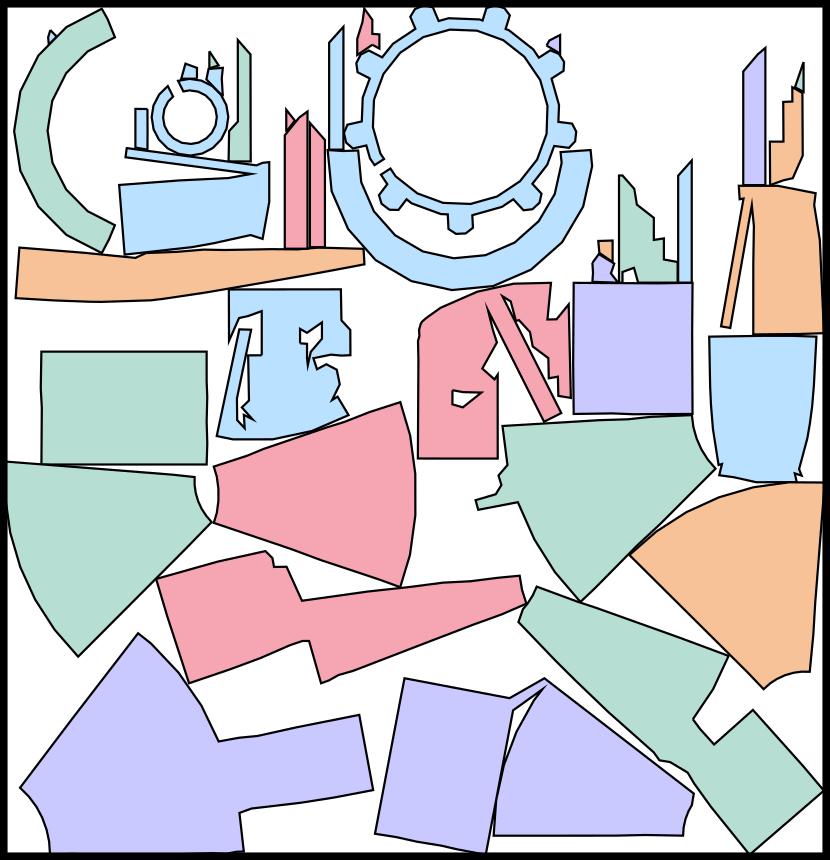}&
\includegraphics[scale=.1]{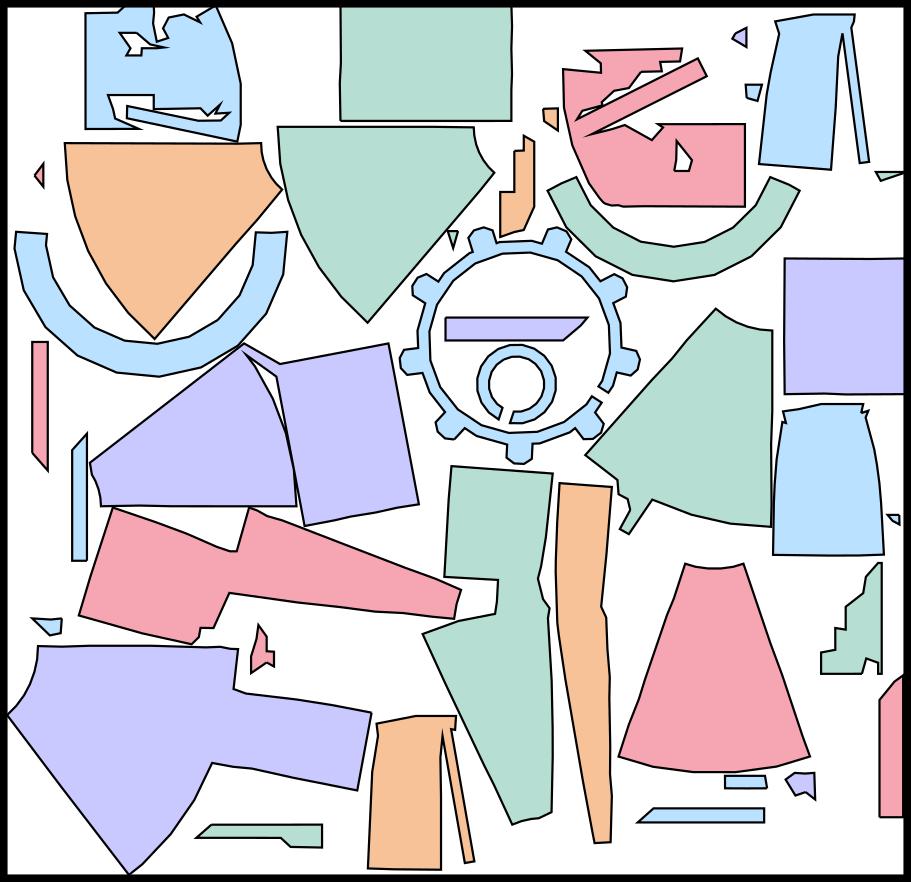}&
\includegraphics[scale=.1]{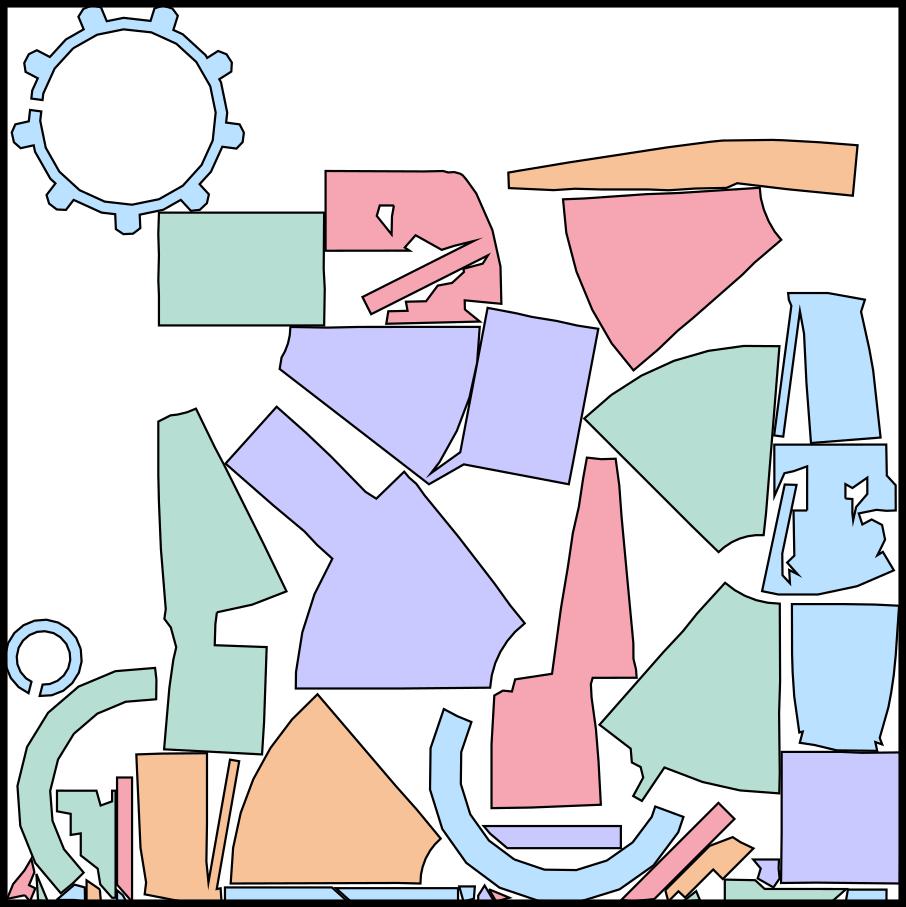}&
\includegraphics[scale=.1]{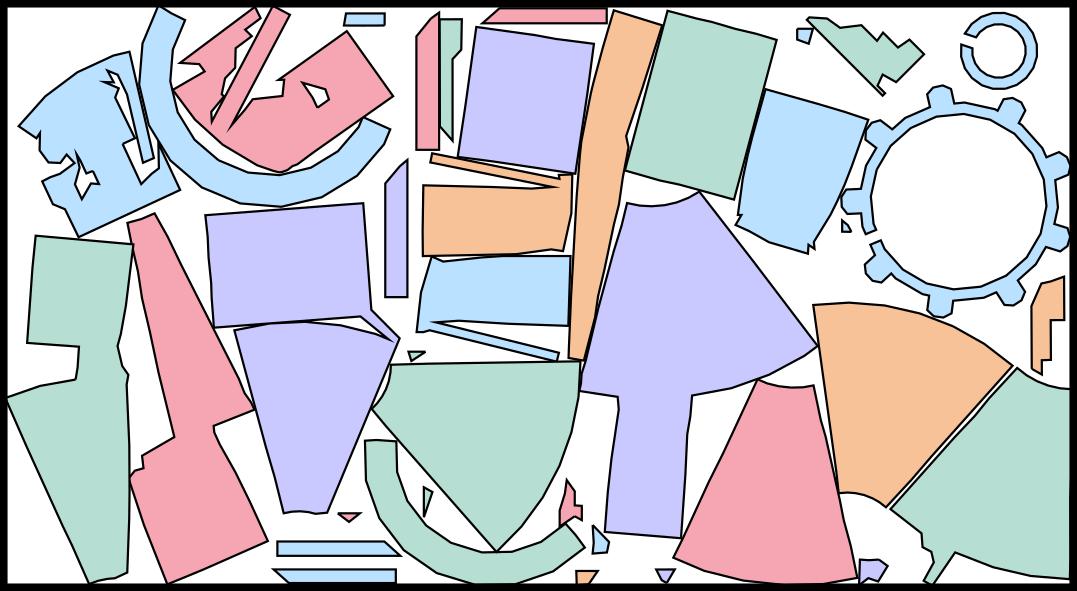}\\
 & \scalebox{0.7}{pr: 57.7\%} &\scalebox{0.7}{pr: 55.0\%} & \scalebox{0.7}{pr: 50.3\%} & \scalebox{0.7}{pr: 65.4\%}\\
\midrule
 \includegraphics[scale=.1]{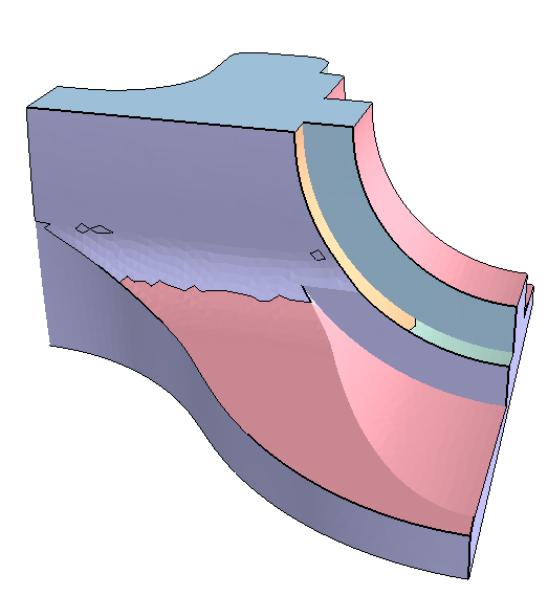}&
\includegraphics[scale=.1]{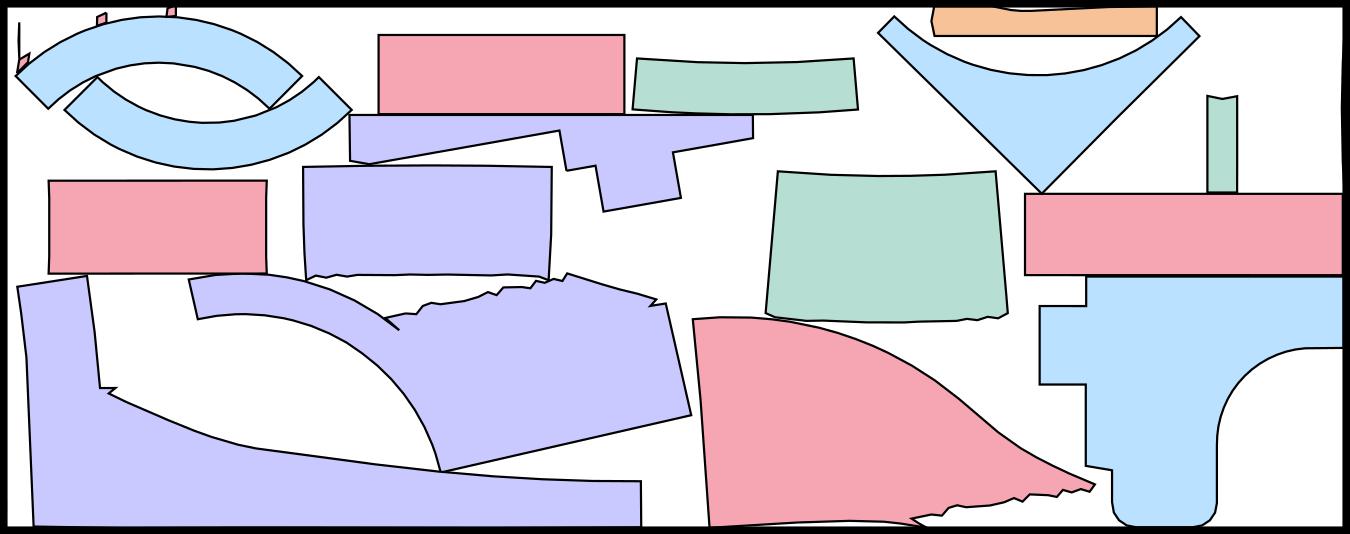}&
\includegraphics[scale=.1]{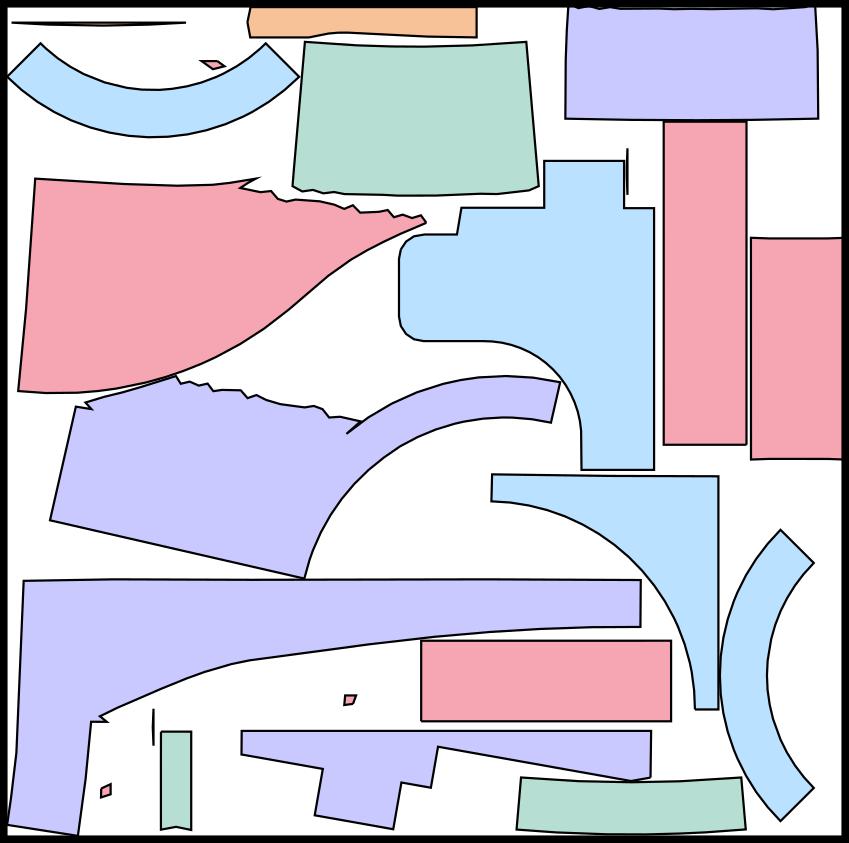}&
\includegraphics[scale=.1]{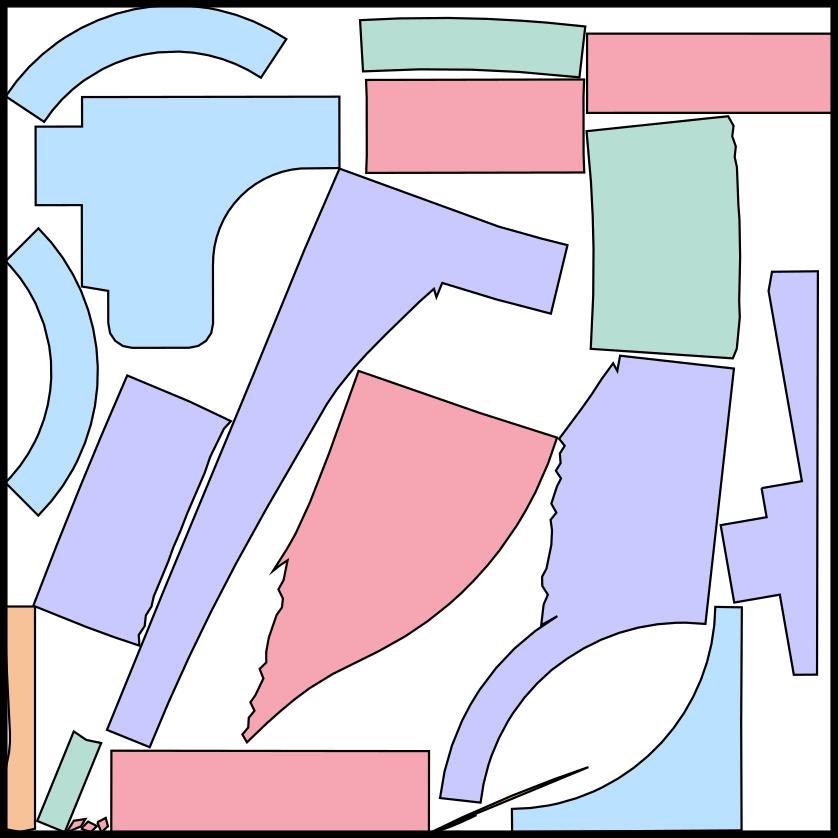}&
\includegraphics[scale=.1]{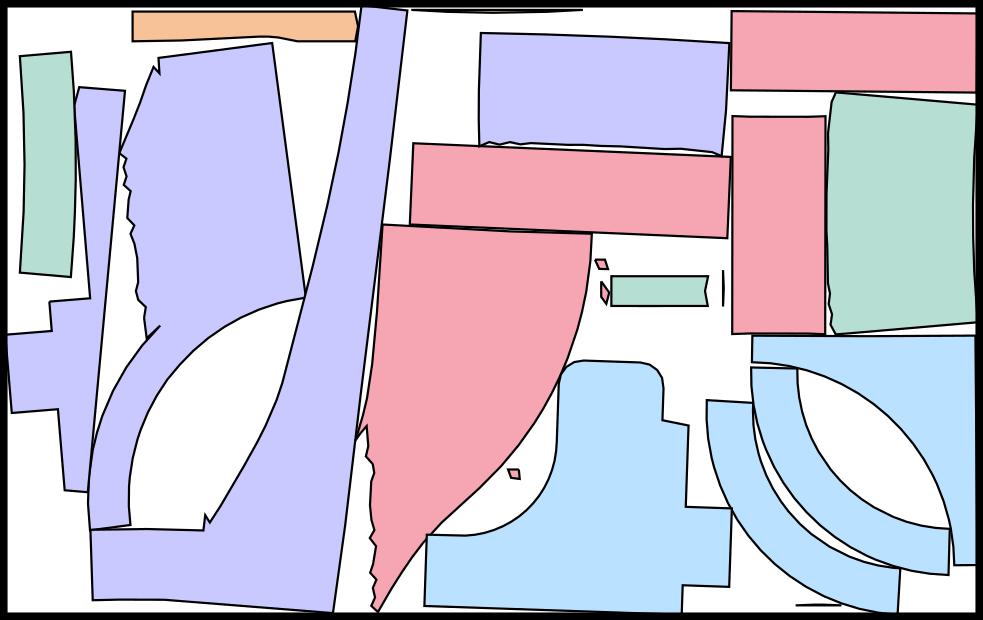}\\
 & \scalebox{0.7}{pr: 61.3\%} &\scalebox{0.7}{pr: 63.7\%} & \scalebox{0.7}{pr: 63.0\%} & \scalebox{0.7}{pr: 72.6\%}\\
\midrule
\includegraphics[scale=.1]{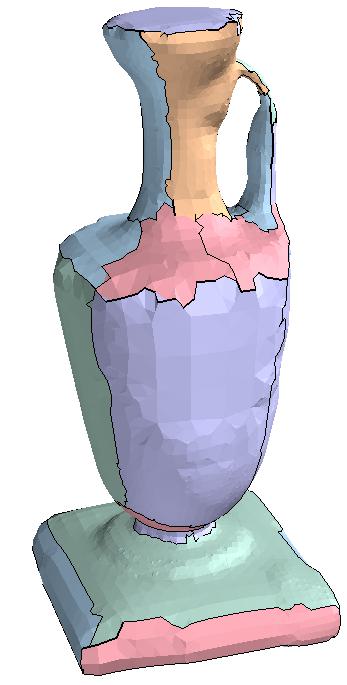}&
\includegraphics[scale=.1]{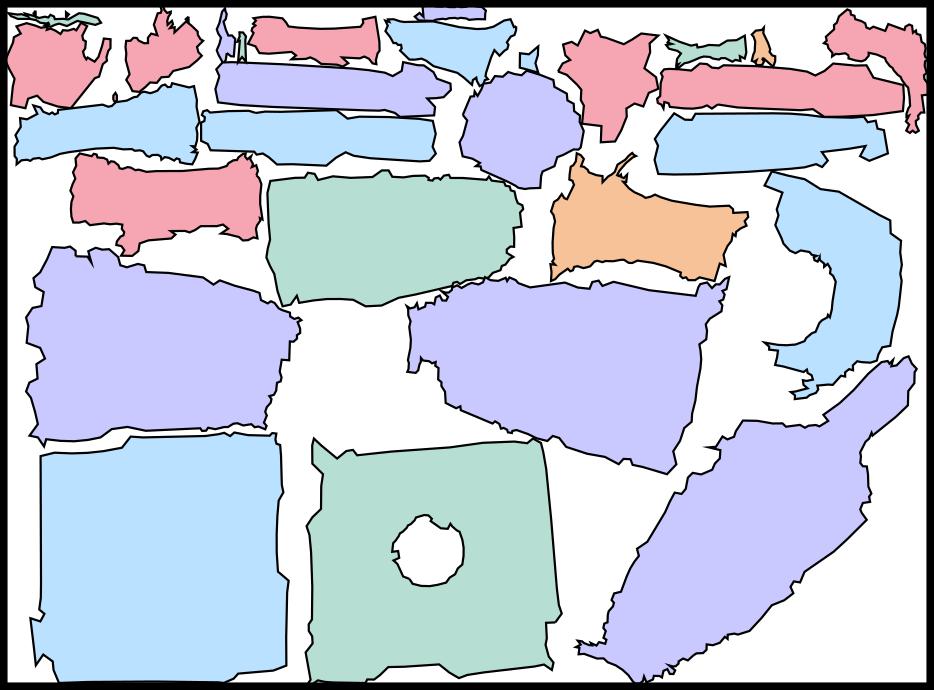}&
\includegraphics[scale=.1]{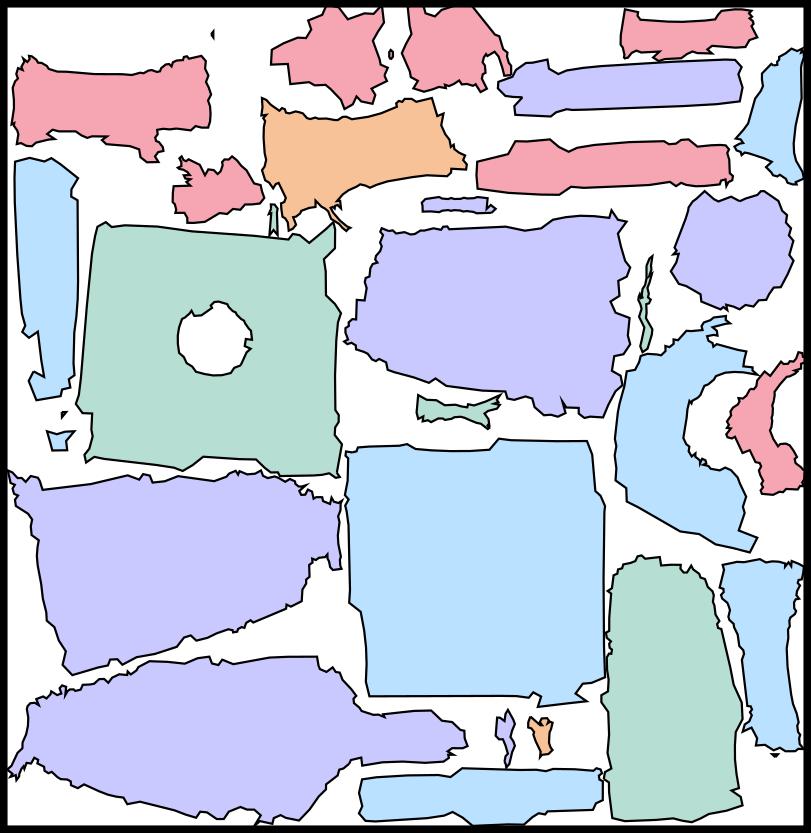}&
\includegraphics[scale=.1]{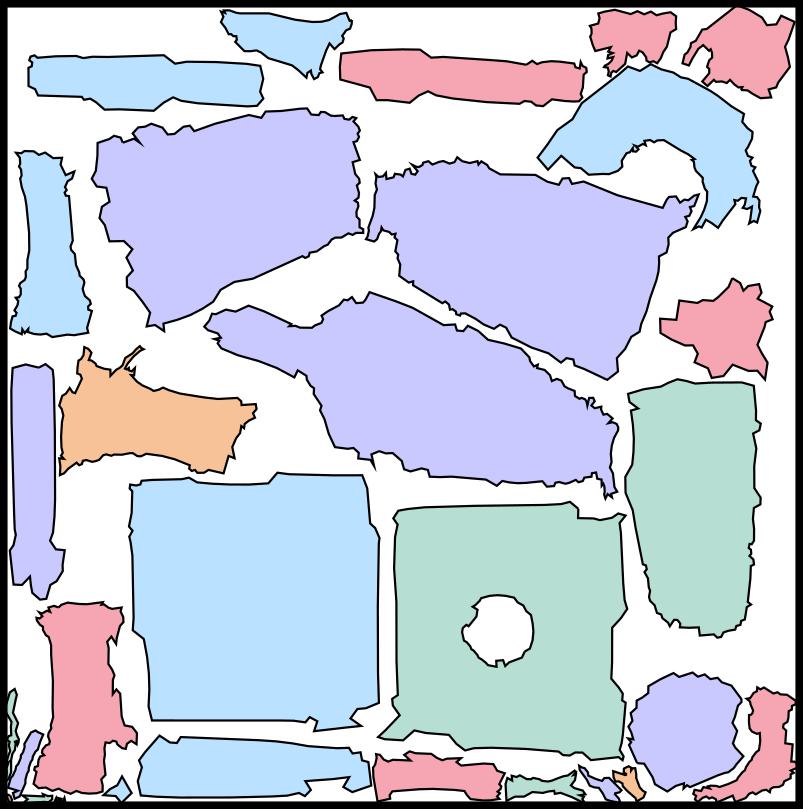}&
\includegraphics[scale=.1]{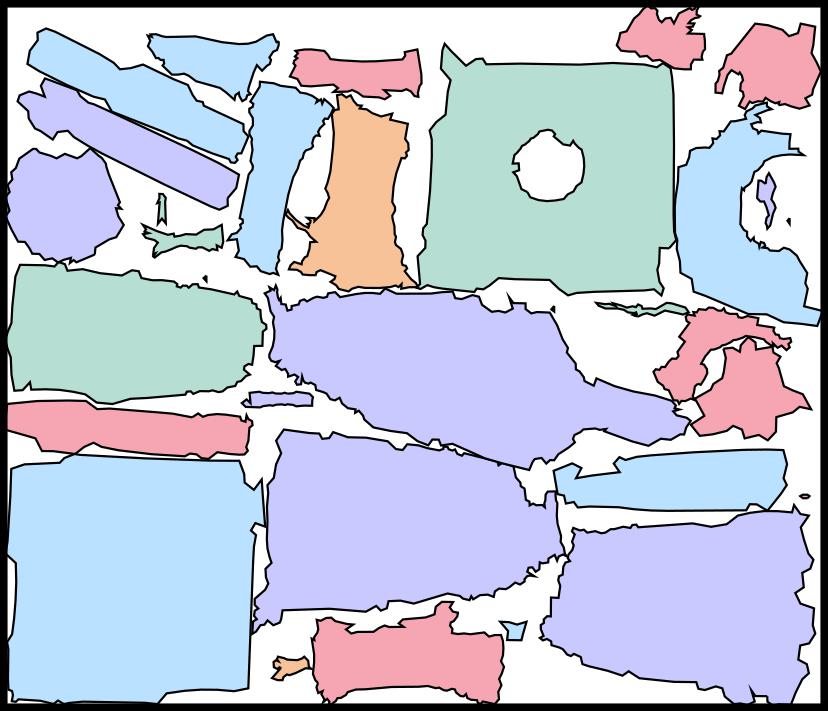}\\
 & \scalebox{0.7}{pr: 67.6\%} &\scalebox{0.7}{pr: 70.5\%} & \scalebox{0.7}{pr: 68.4\%} & \scalebox{0.7}{pr: 75.7\%}\\
\midrule
\includegraphics[scale=.1]{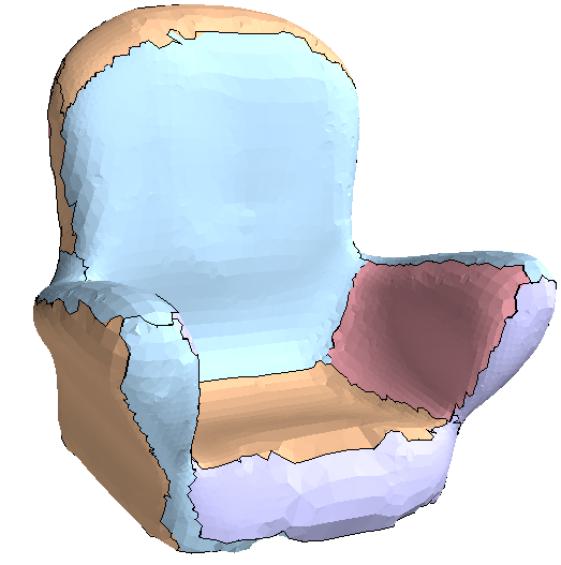}&
\includegraphics[scale=.1]{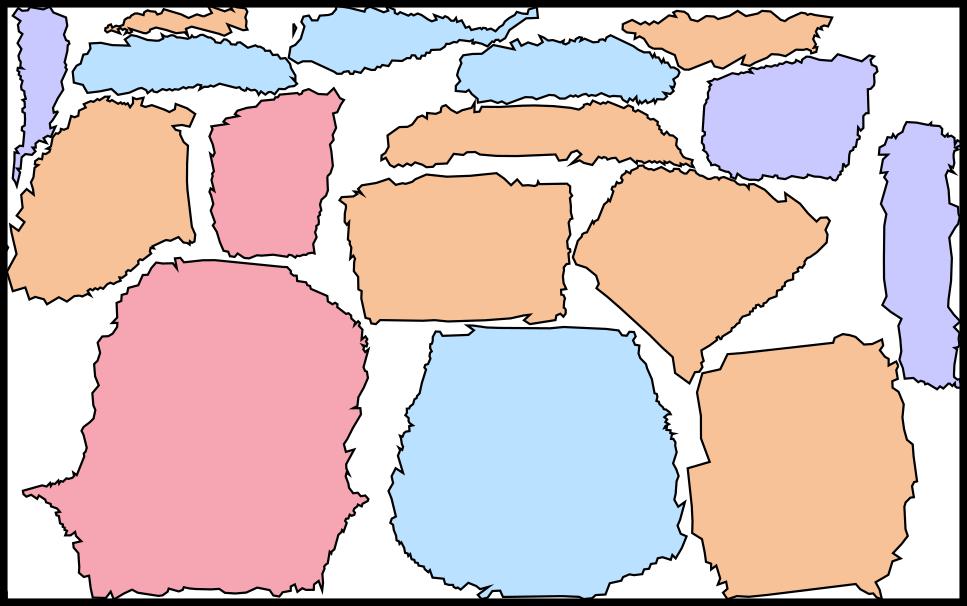}&
\includegraphics[scale=.1]{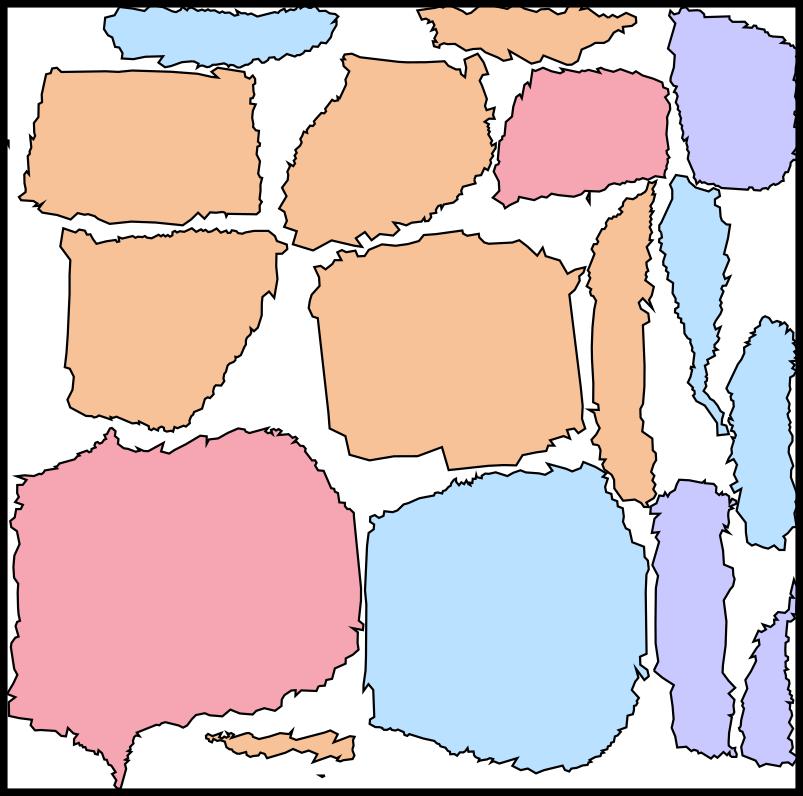}&
\includegraphics[scale=.1]{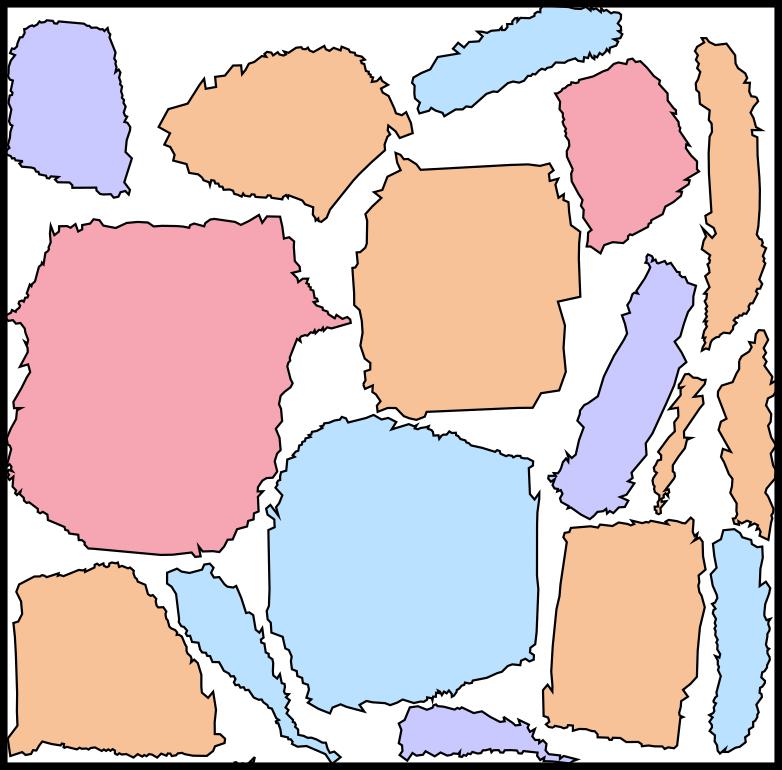}&
\includegraphics[scale=.1]{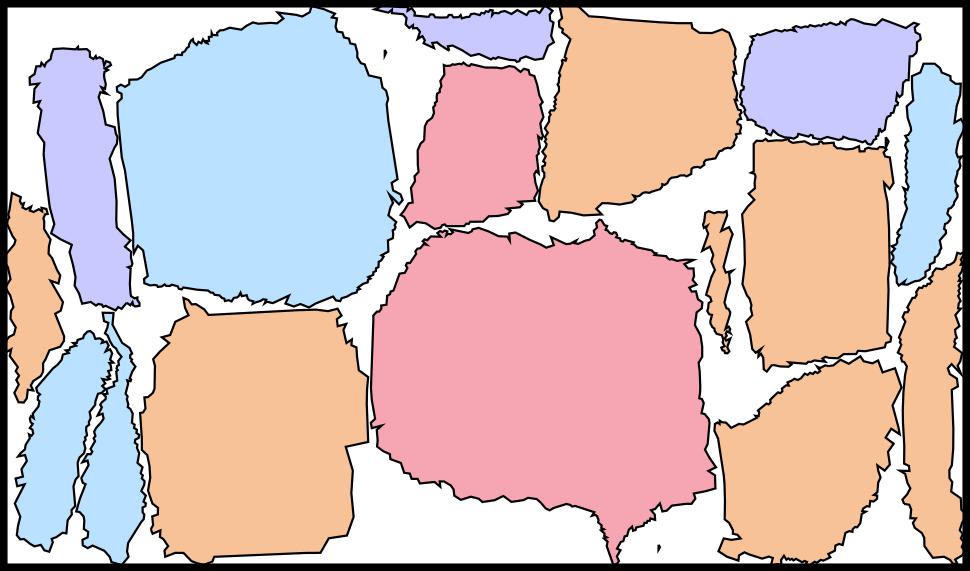}\\
 & \scalebox{0.7}{pr: 74.5\%} &\scalebox{0.7}{pr: 72.2\%} & \scalebox{0.7}{pr: 73.8\%} & \scalebox{0.7}{pr: 80.0\%}\\
 \bottomrule
\end{tabular}}
\egroup
\end{figure*}

\begin{figure*}[ht]
\centering
\bgroup
\def\arraystretch{.5}
\resizebox{\linewidth}{!}{
\begin{tabular}{ccccc}
\toprule
 & \scalebox{0.7}{\cite{sander2003multi}} & \scalebox{0.7}{XAtlas} & \scalebox{0.7}{NFP} & \scalebox{0.7}{Ours}\\
\midrule
\includegraphics[scale=.1]{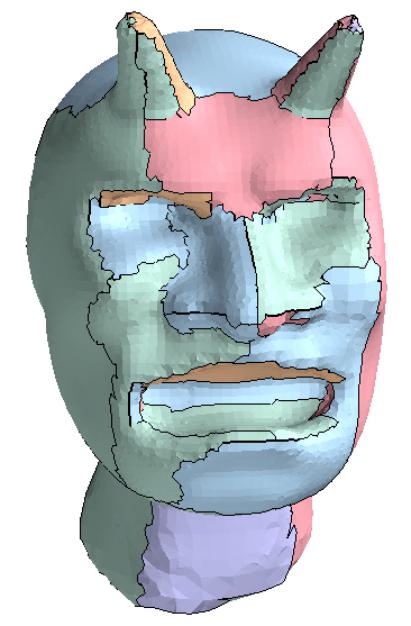}&
\includegraphics[scale=.1]{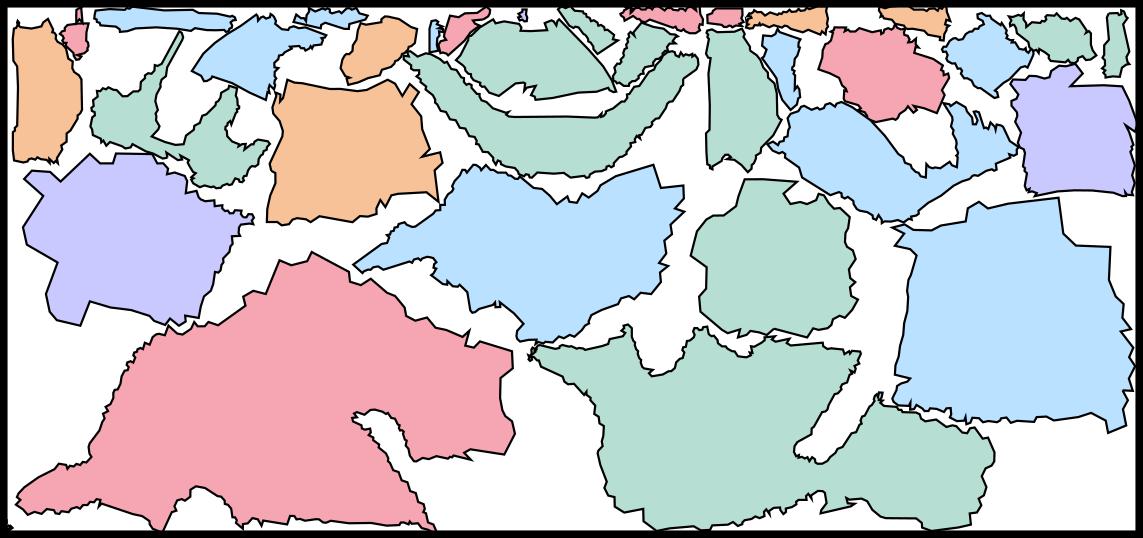}&
\includegraphics[scale=.1]{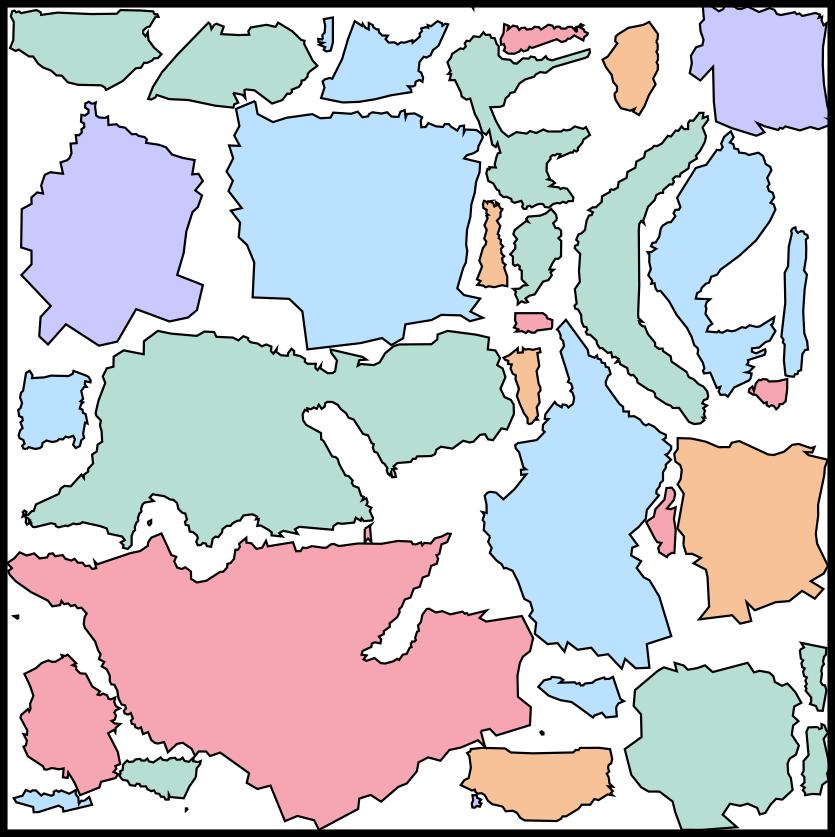}&
\includegraphics[scale=.1]{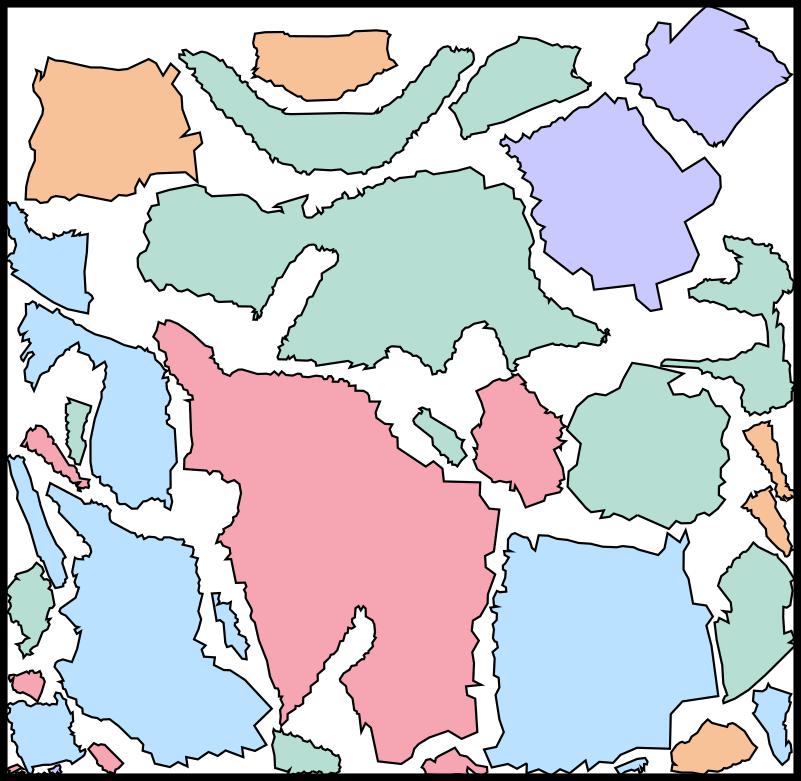}&
\includegraphics[scale=.1]{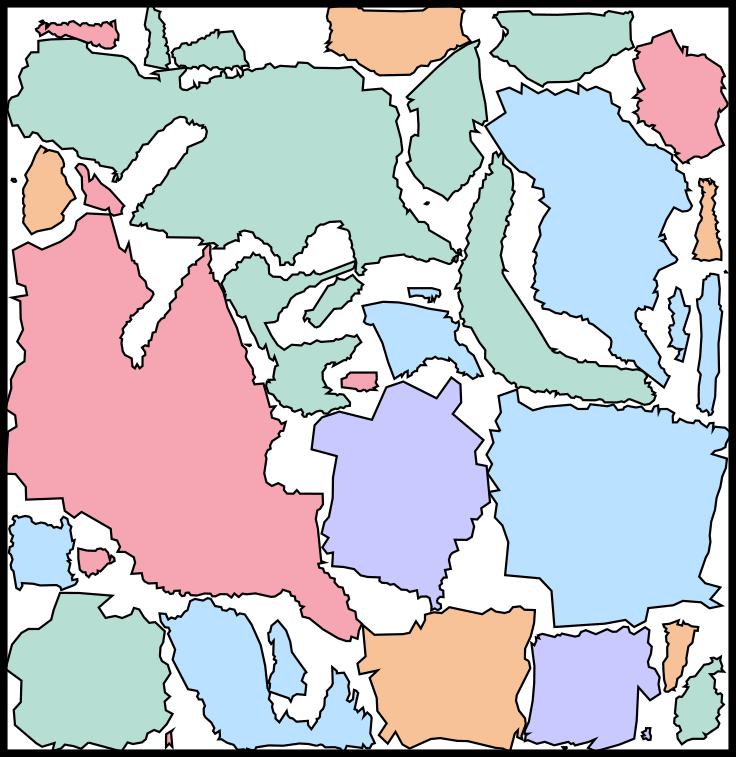}\\
 & \scalebox{0.7}{pr: 72.5\%} &\scalebox{0.7}{pr: 63.2\%} & \scalebox{0.7}{pr: 70.8\%} & \scalebox{0.7}{pr: 79.7\%}\\
\midrule
\includegraphics[scale=.1]{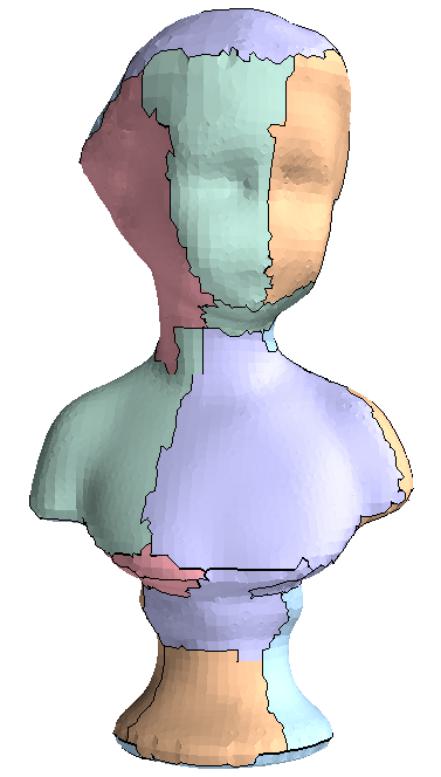}&
\includegraphics[scale=.1]{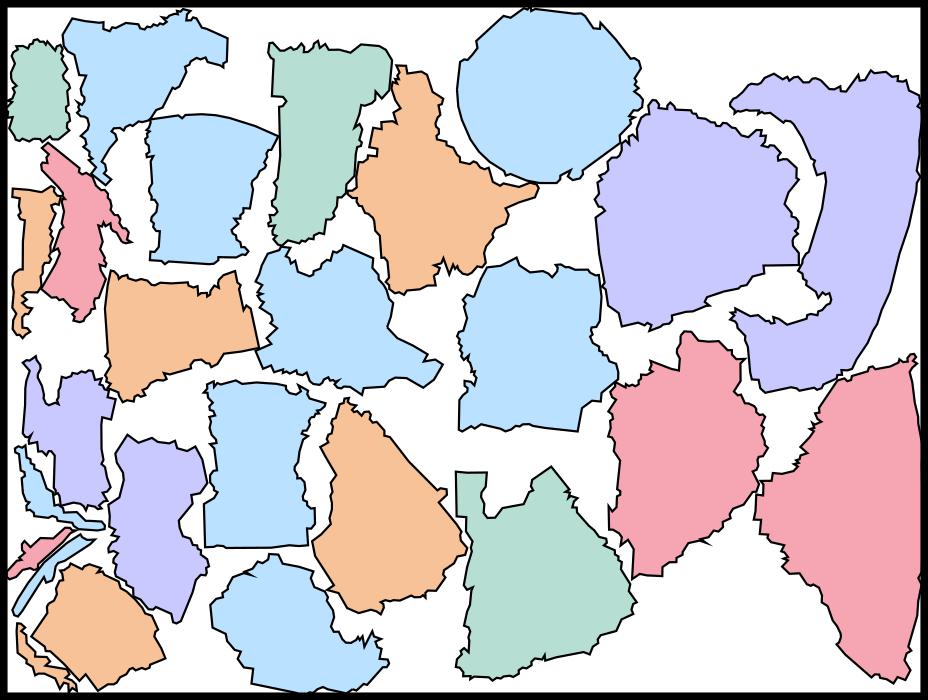}&
\includegraphics[scale=.1]{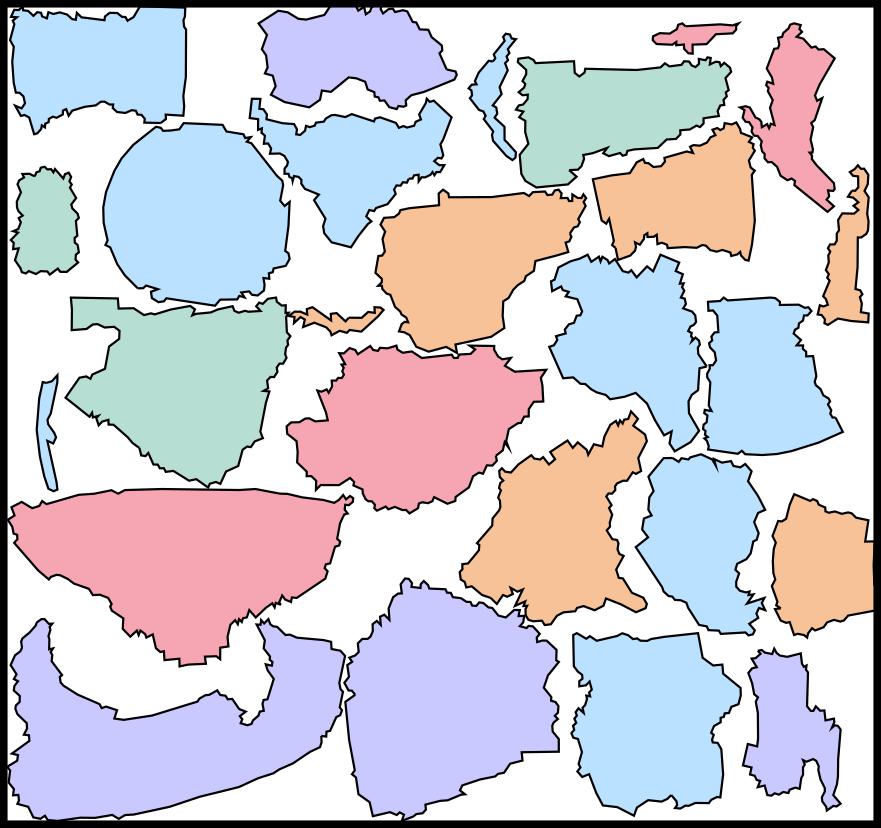}&
\includegraphics[scale=.1]{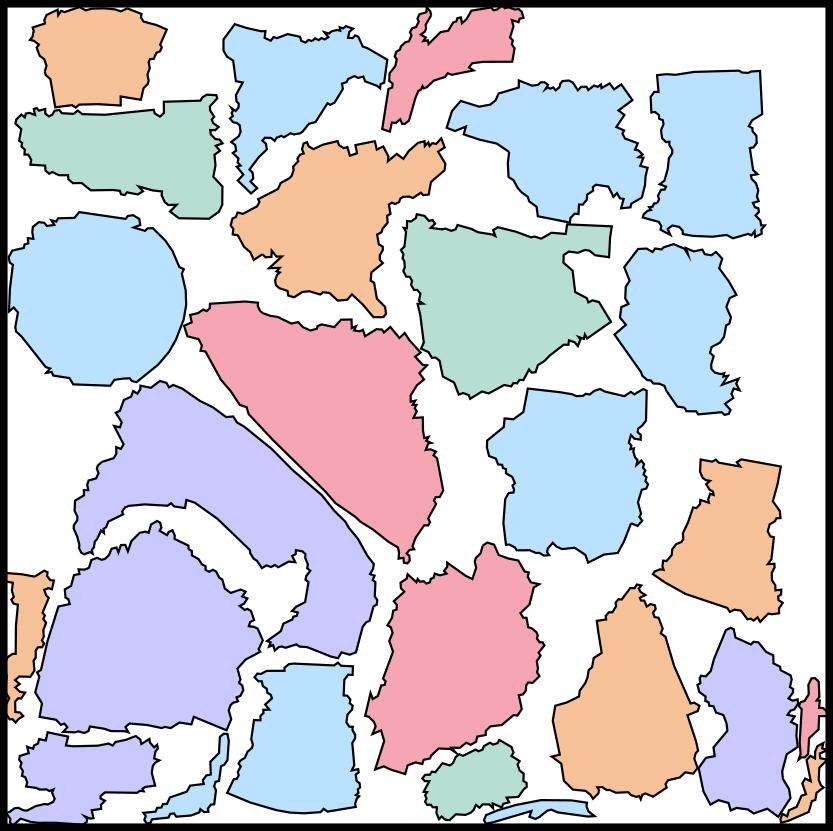}&
\includegraphics[scale=.1]{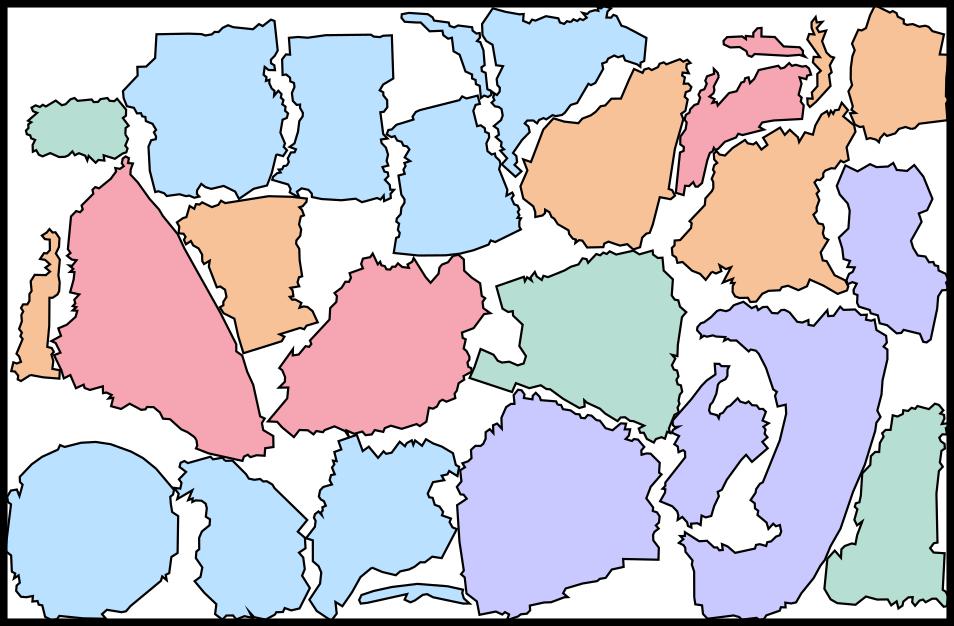}\\
 & \scalebox{0.7}{pr: 68.0\%} &\scalebox{0.7}{pr: 64.7\%} & \scalebox{0.7}{pr: 64.2\%} & \scalebox{0.7}{pr: 74.3\%}\\
\midrule
\includegraphics[scale=.1]{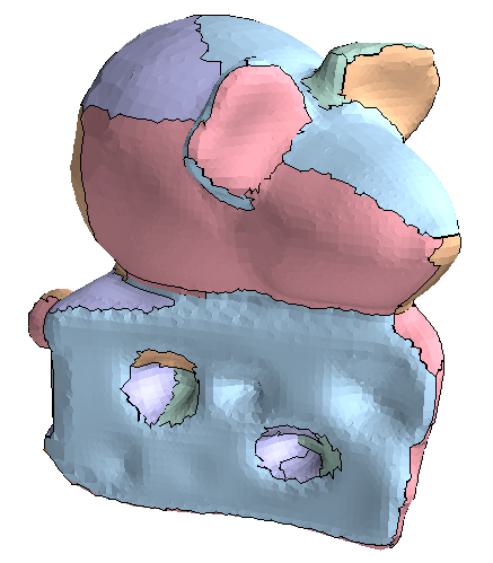}&
\includegraphics[scale=.1]{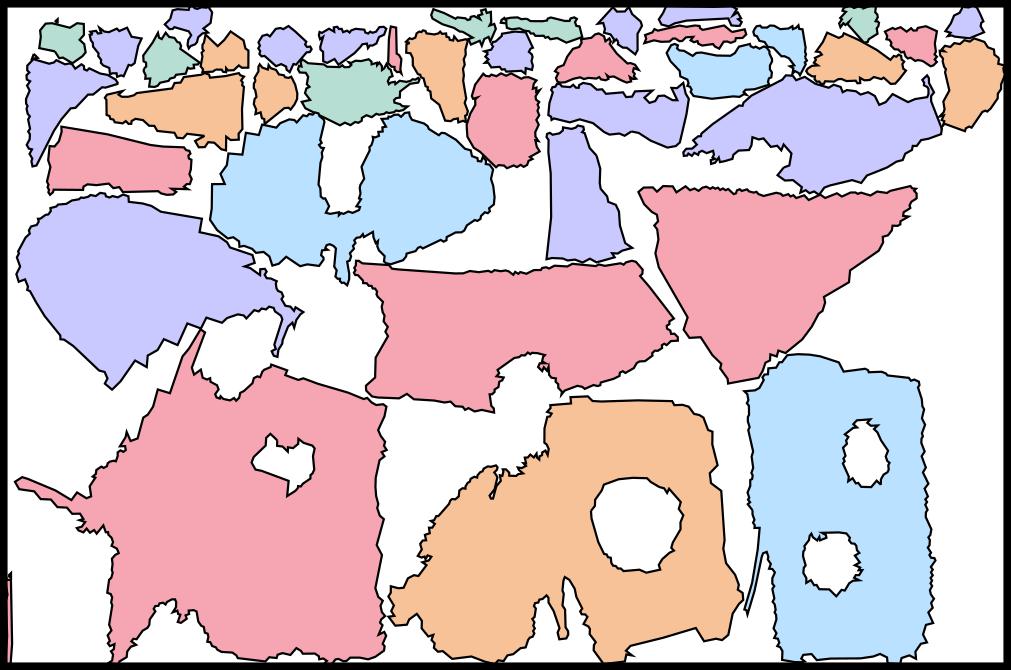}&
\includegraphics[scale=.1]{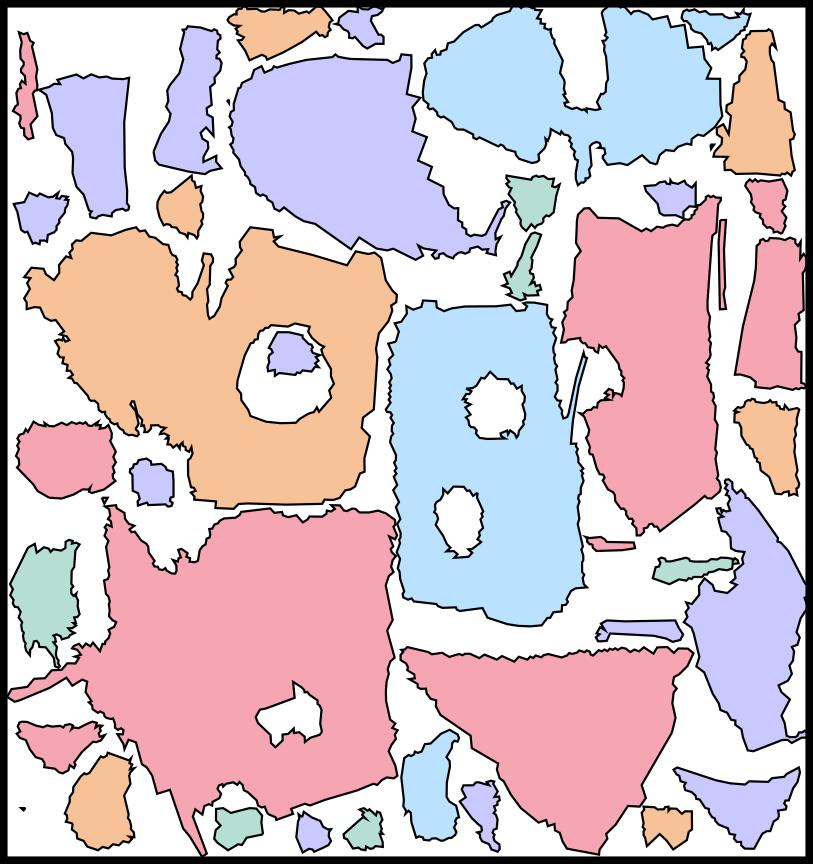}&
\includegraphics[scale=.1]{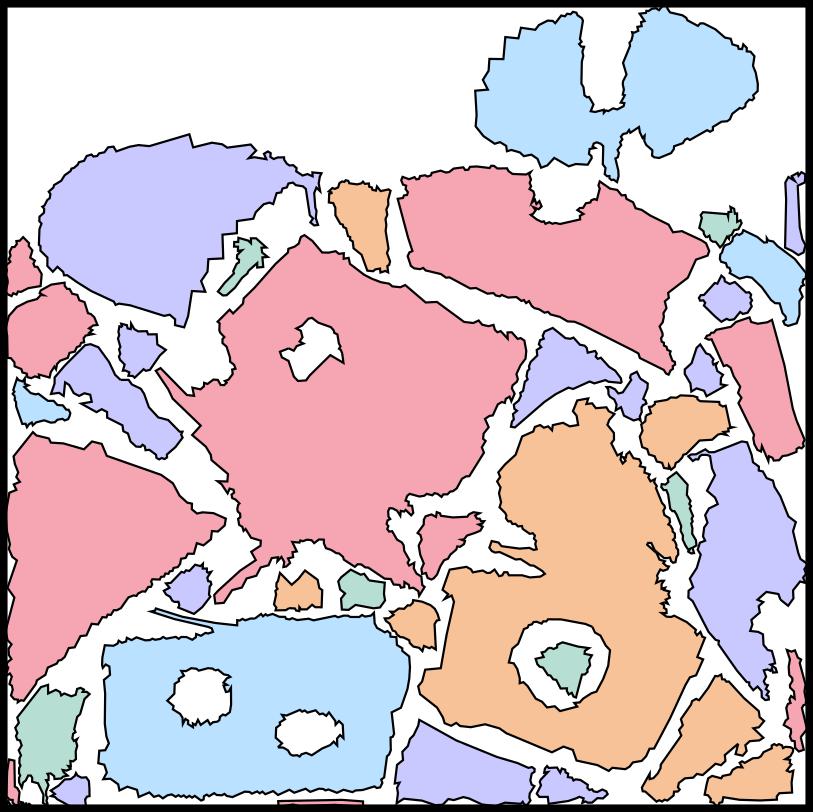}&
\includegraphics[scale=.1]{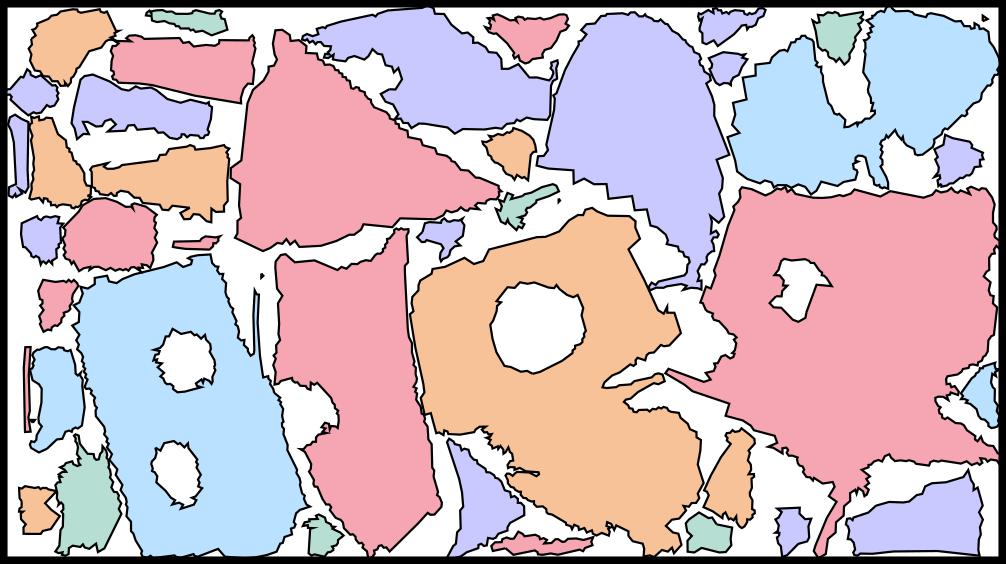}\\
 & \scalebox{0.7}{pr: 65.3\%} &\scalebox{0.7}{pr: 62.9\%} & \scalebox{0.7}{pr: 67.2\%} & \scalebox{0.7}{pr: 78.3\%}\\
\midrule
 \includegraphics[scale=.1]{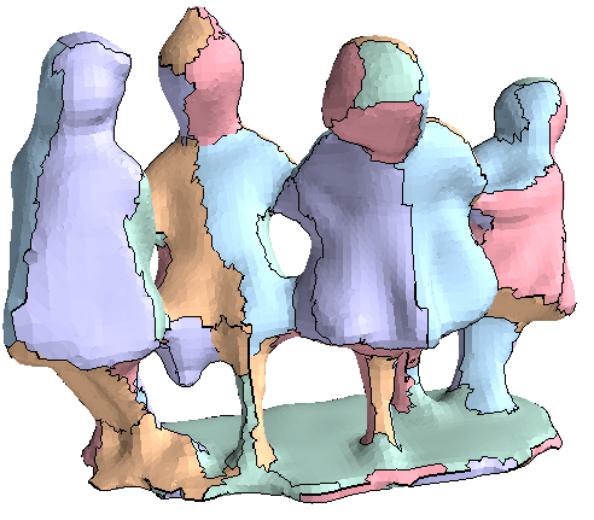}&
\includegraphics[scale=.1]{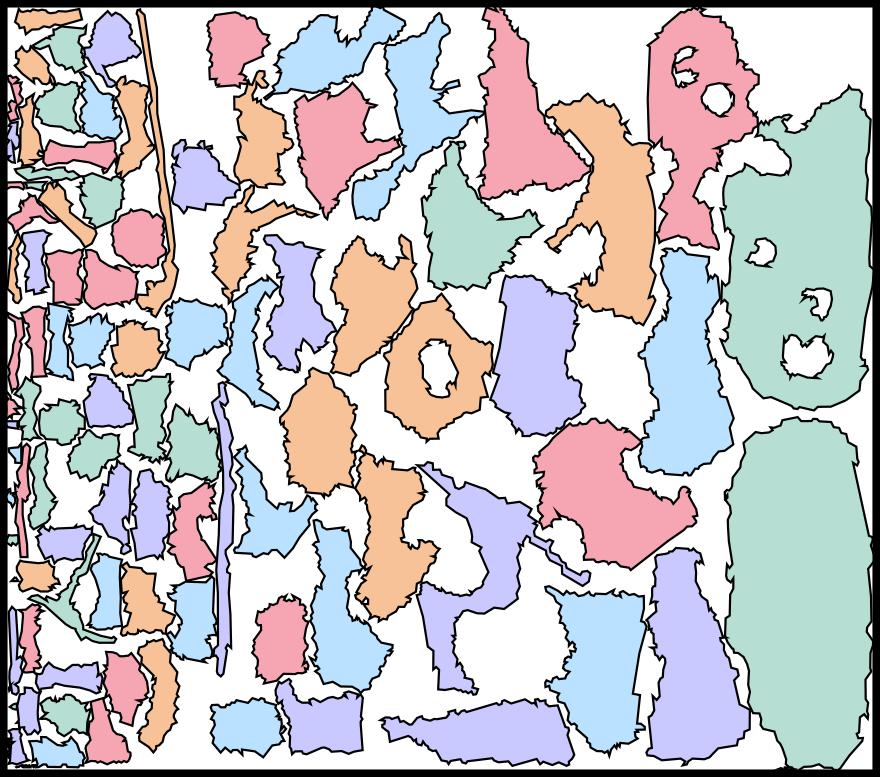}&
\includegraphics[scale=.1]{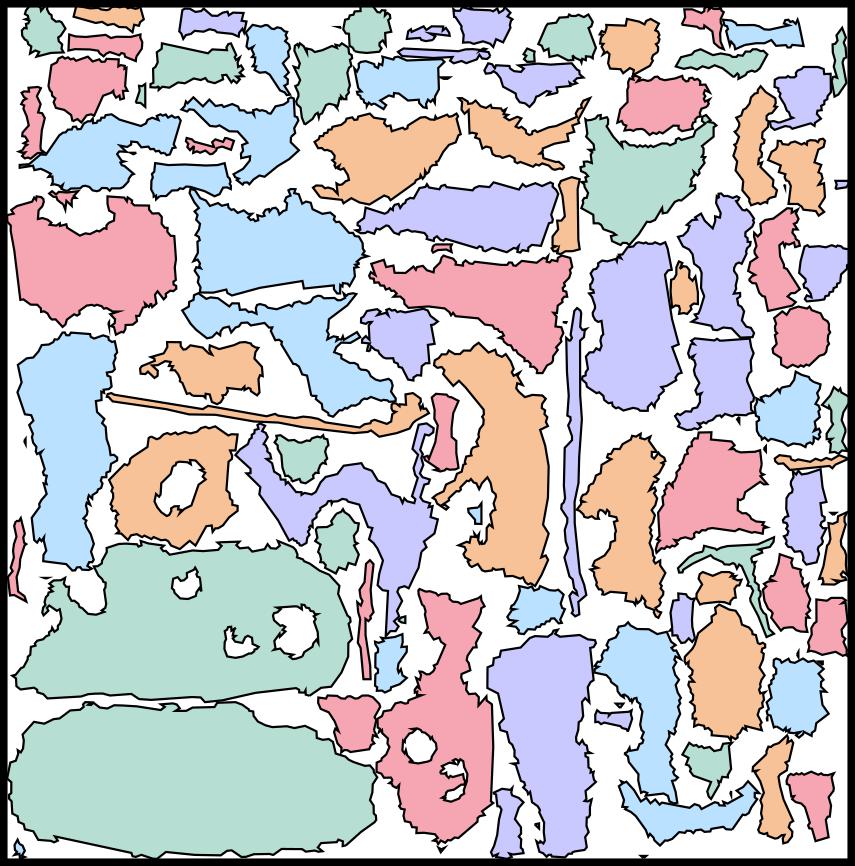}&
\includegraphics[scale=.1]{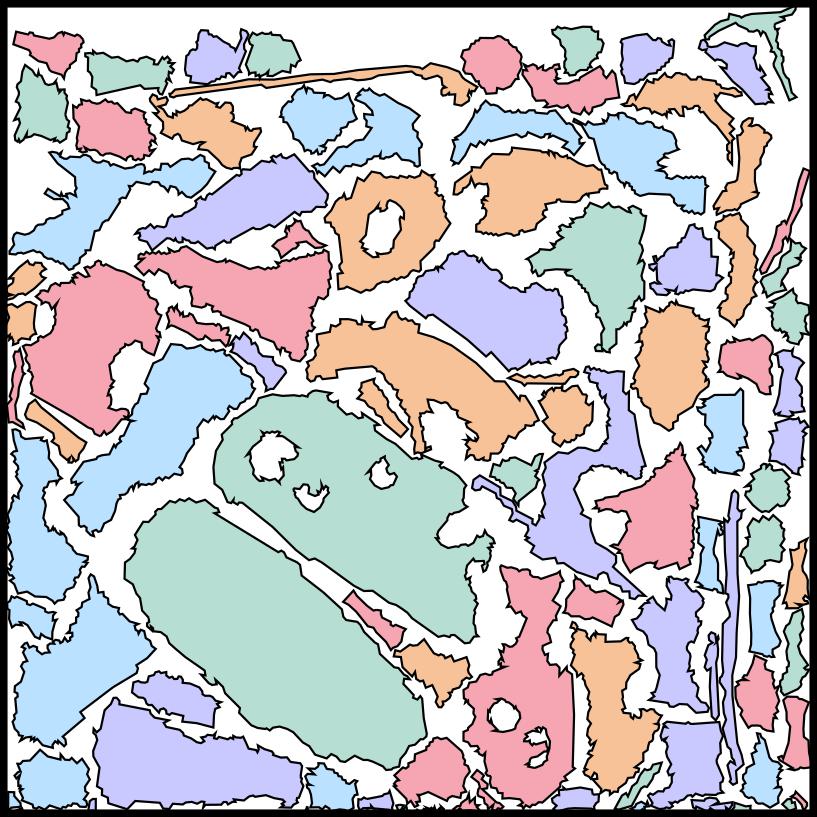}&
\includegraphics[scale=.1]{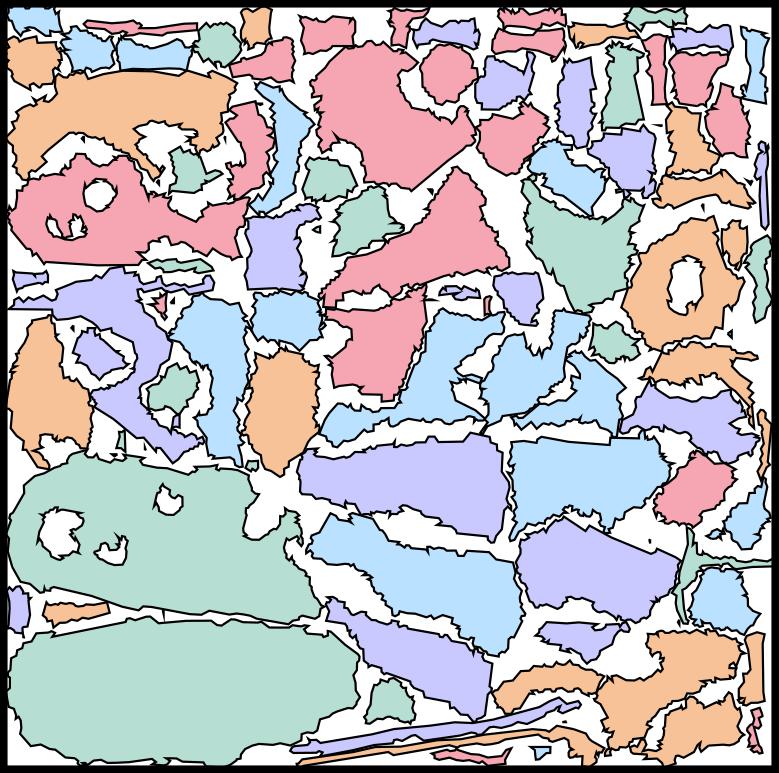}\\
 & \scalebox{0.7}{pr: 64.8\%} &\scalebox{0.7}{pr: 63.7\%} & \scalebox{0.7}{pr: 66.5\%} & \scalebox{0.7}{pr: 73.9\%}\\
\midrule
\includegraphics[scale=.1]{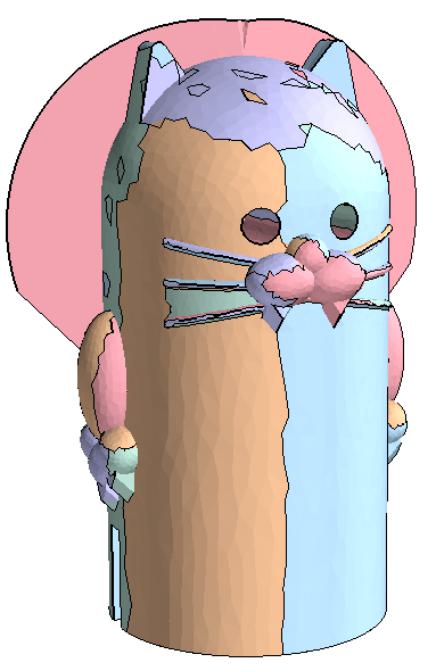}&
\includegraphics[scale=.1]{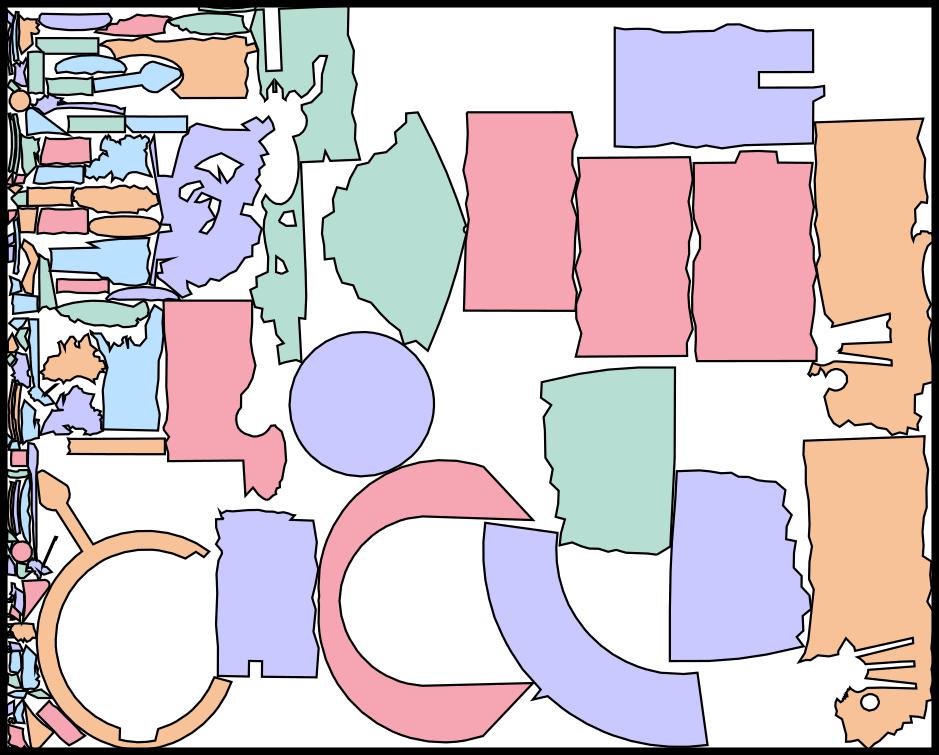}&
\includegraphics[scale=.1]{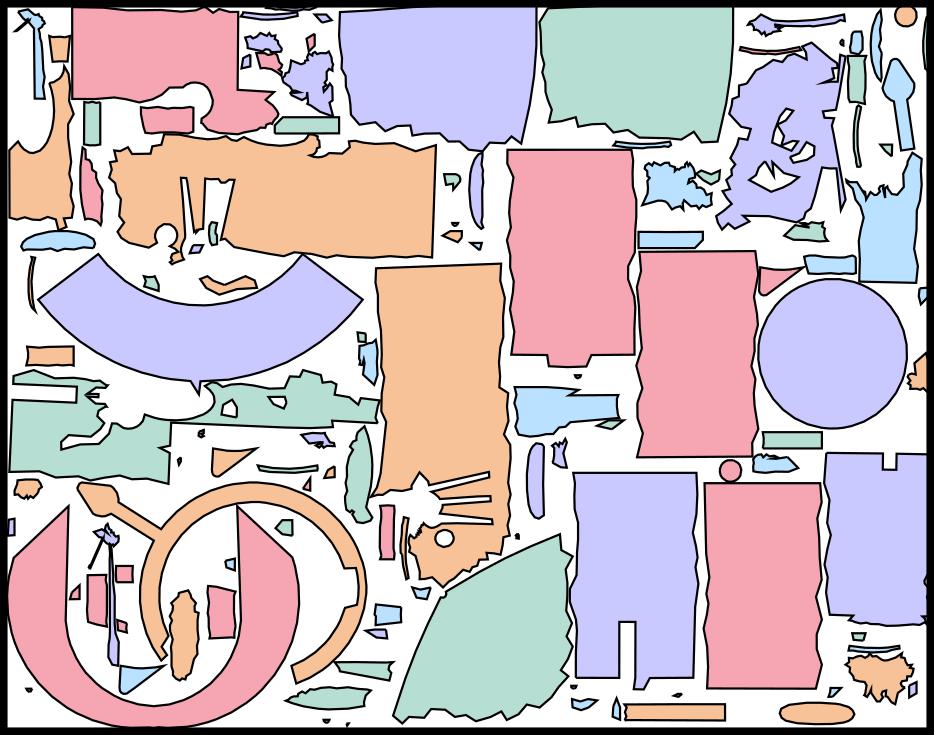}&
\includegraphics[scale=.1]{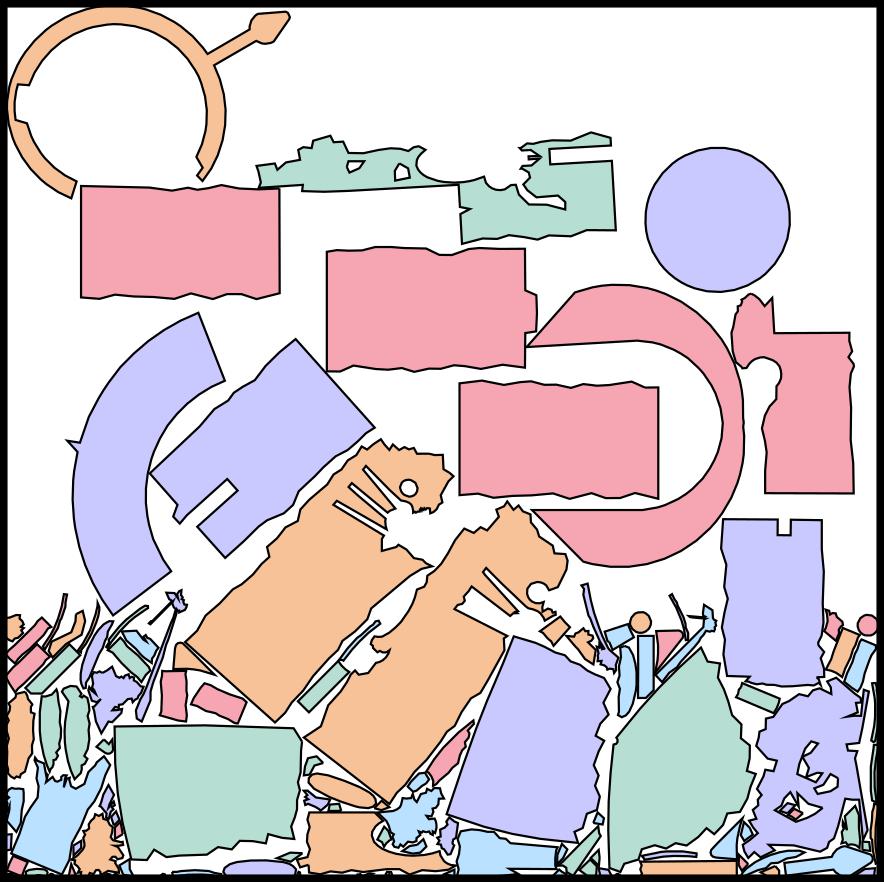}&
\includegraphics[scale=.1]{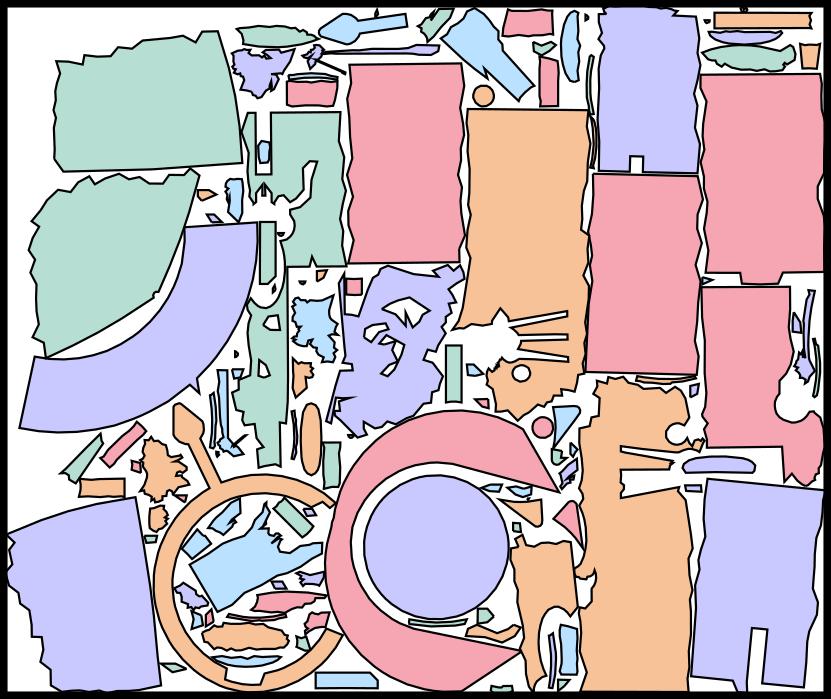}\\
 & \scalebox{0.7}{pr: 64.8\%} &\scalebox{0.7}{pr: 63.7\%} & \scalebox{0.7}{pr: 66.5\%} & \scalebox{0.7}{pr: 74.9\%}\\
\midrule
\includegraphics[scale=.1]{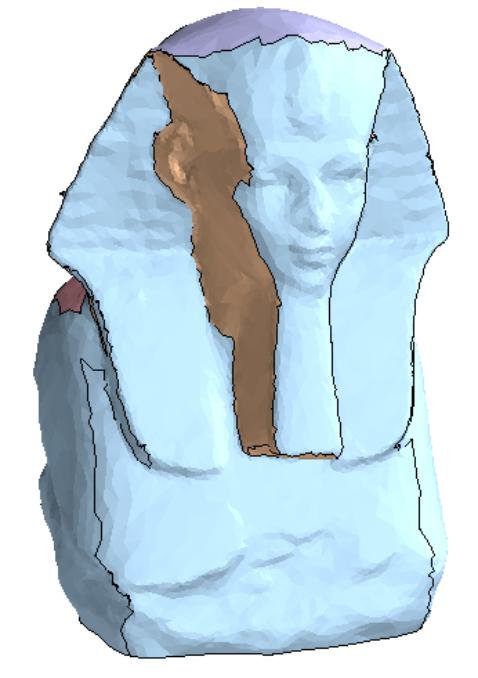}&
\includegraphics[scale=.1]{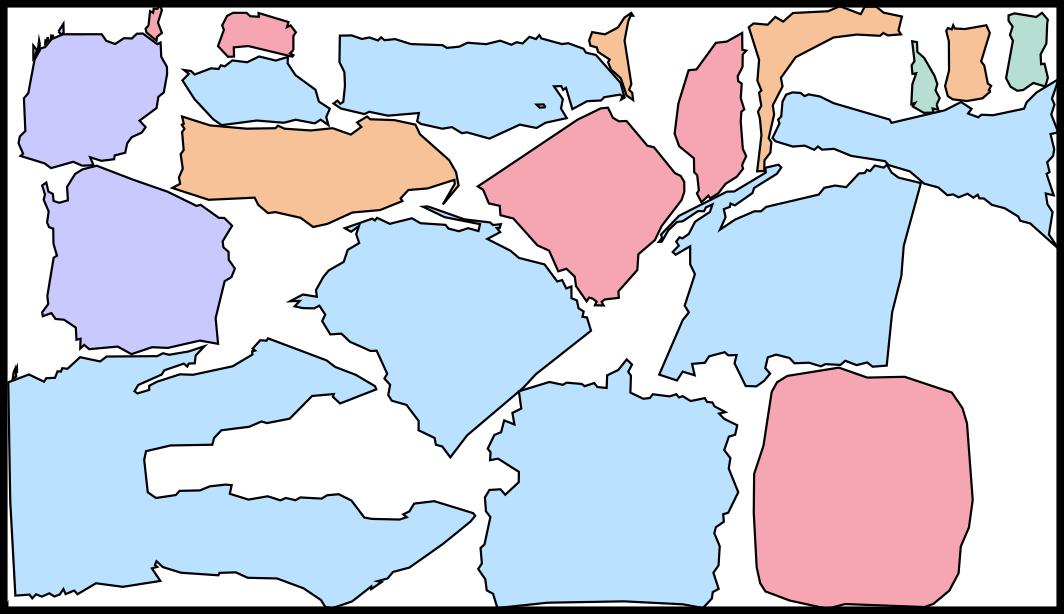}&
\includegraphics[scale=.1]{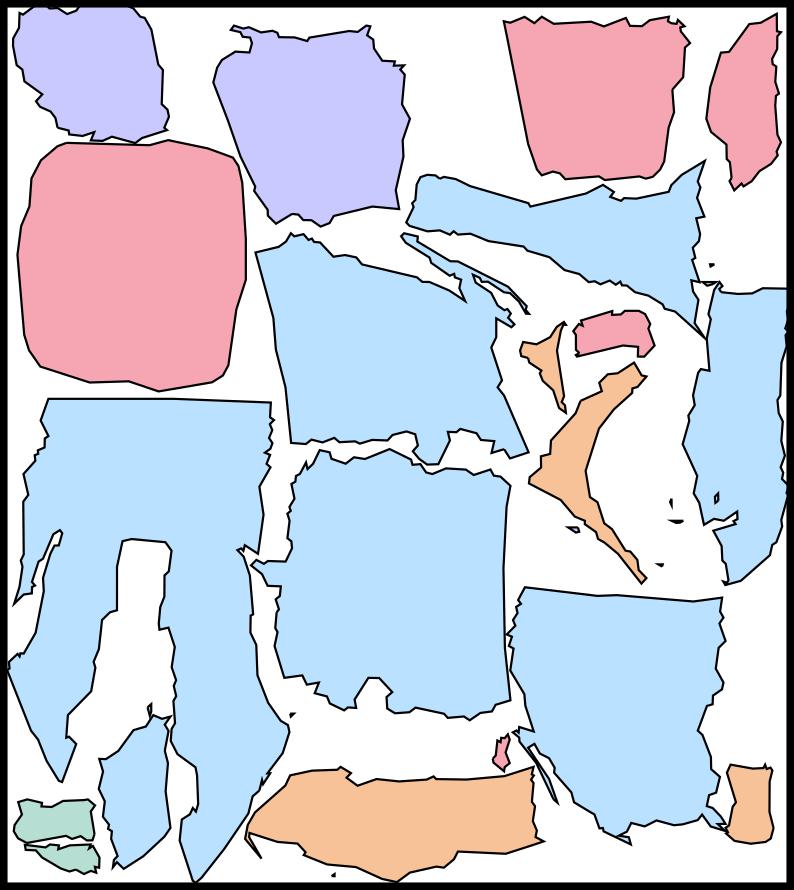}&
\includegraphics[scale=.1]{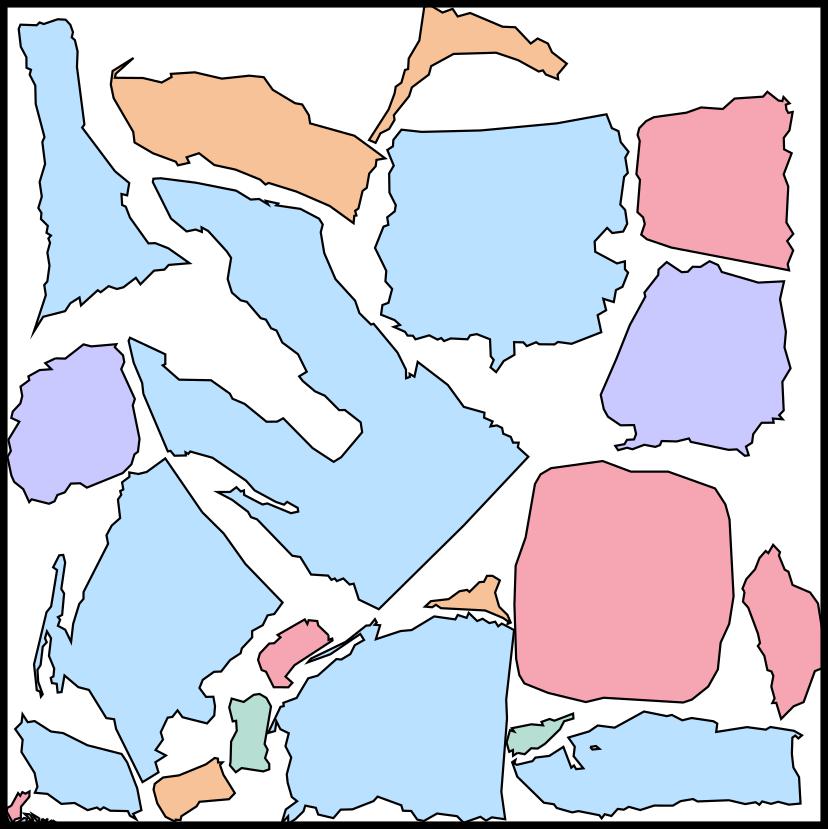}&
\includegraphics[scale=.1]{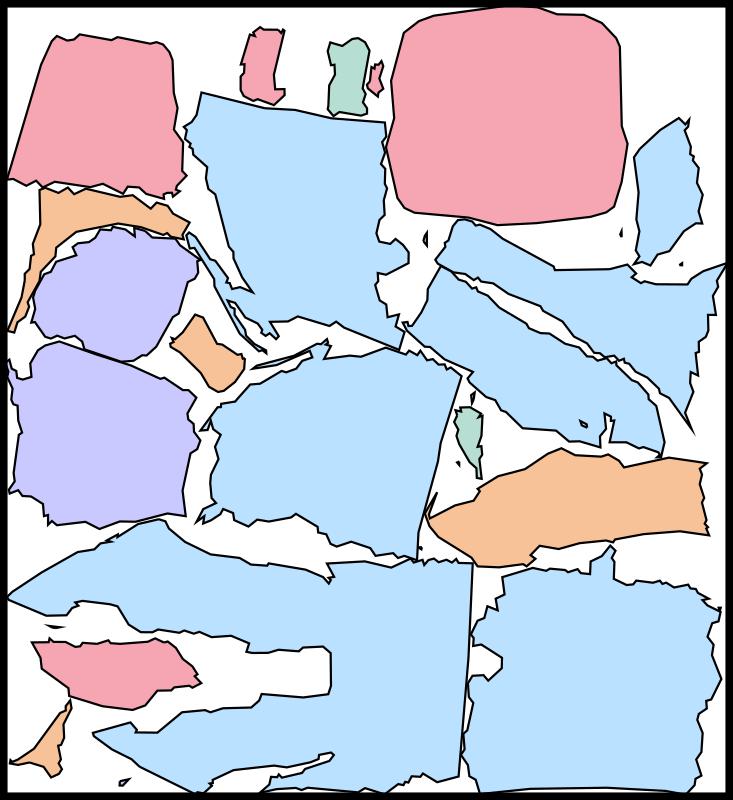}\\
 & \scalebox{0.7}{pr: 66.8\%} &\scalebox{0.7}{pr: 66.1\%} & \scalebox{0.7}{pr: 64.6\%} & \scalebox{0.7}{pr: 76.1\%}\\
 \bottomrule
\end{tabular}}
\egroup
\end{figure*}

\begin{figure*}[ht]
\centering
\bgroup
\def\arraystretch{.5}
\resizebox{\linewidth}{!}{
\begin{tabular}{ccccc}
\toprule
 & \scalebox{0.7}{\cite{sander2003multi}} & \scalebox{0.7}{XAtlas} & \scalebox{0.7}{NFP} & \scalebox{0.7}{Ours}\\
\midrule
\includegraphics[scale=.1]{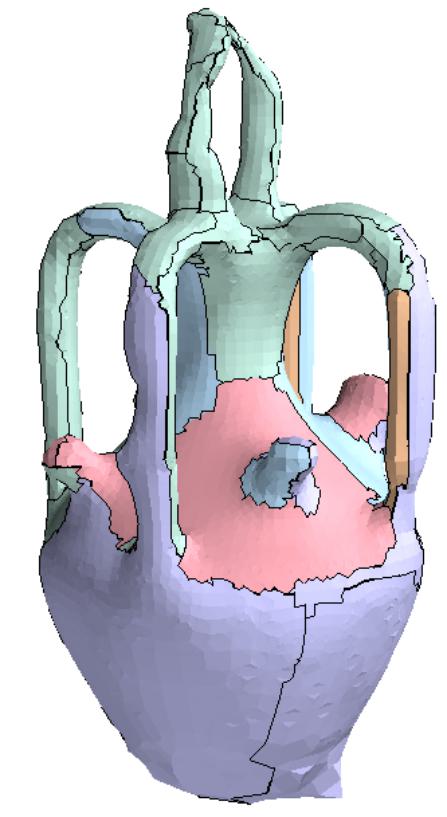}&
\includegraphics[scale=.1]{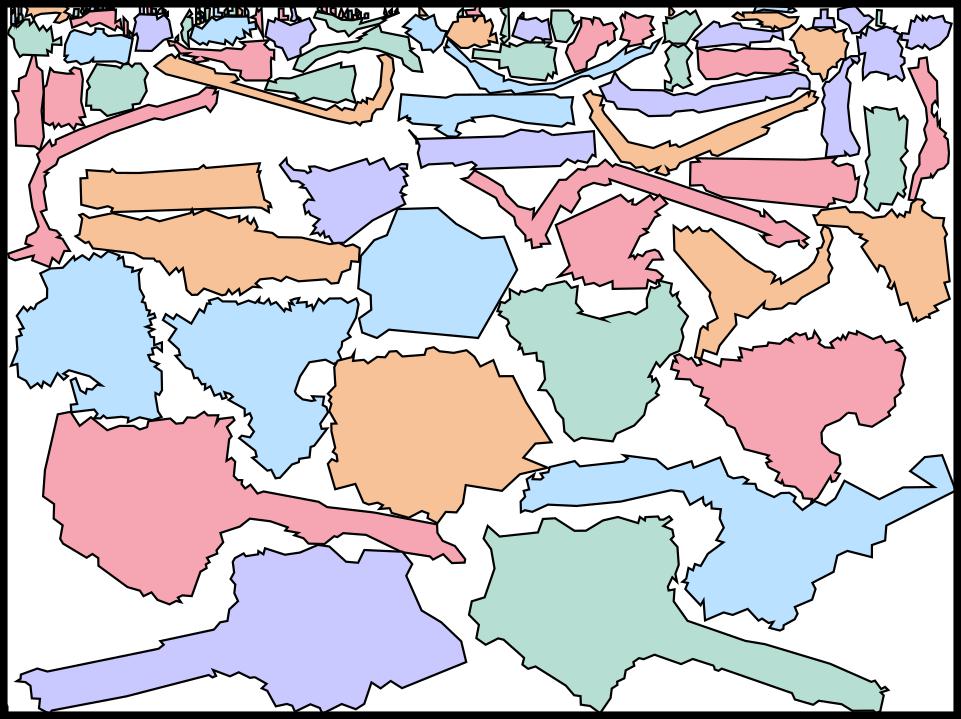}&
\includegraphics[scale=.1]{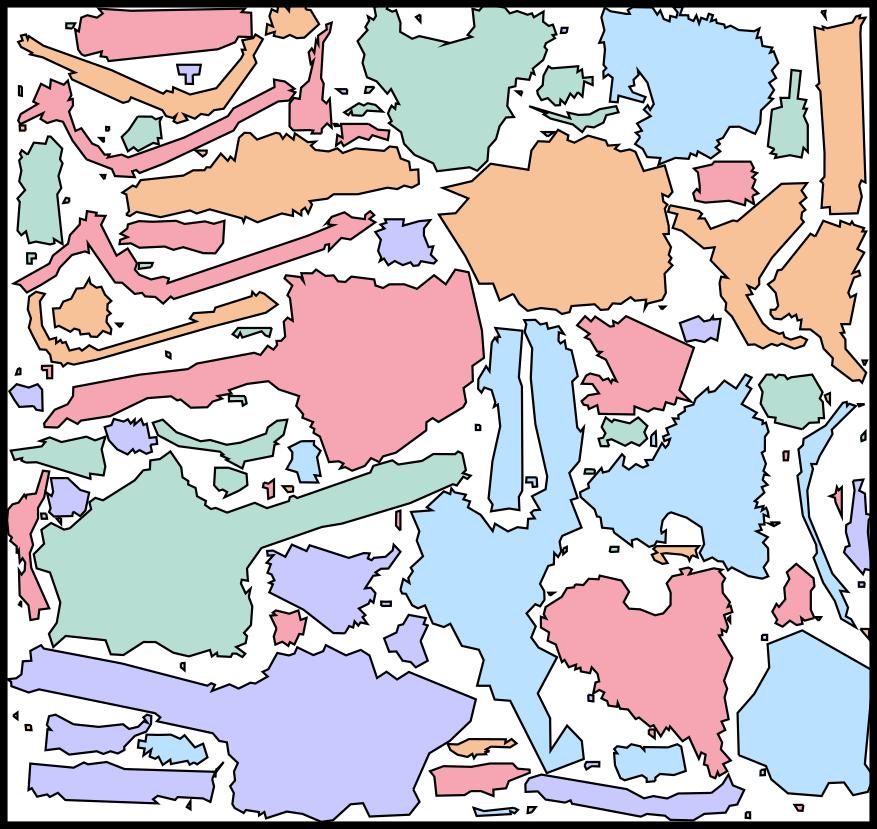}&
\includegraphics[scale=.1]{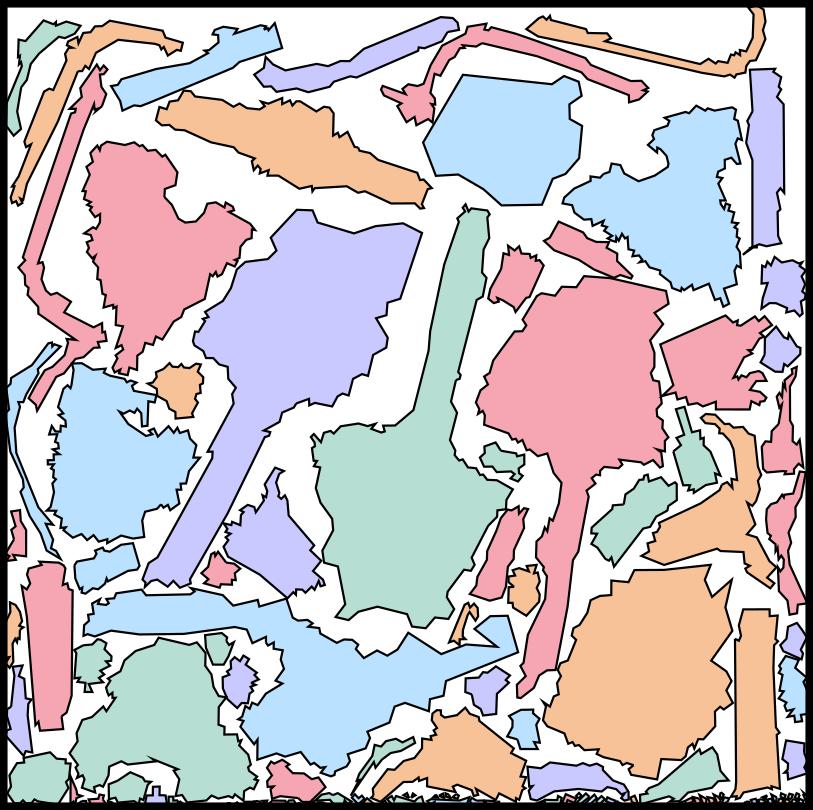}&
\includegraphics[scale=.1]{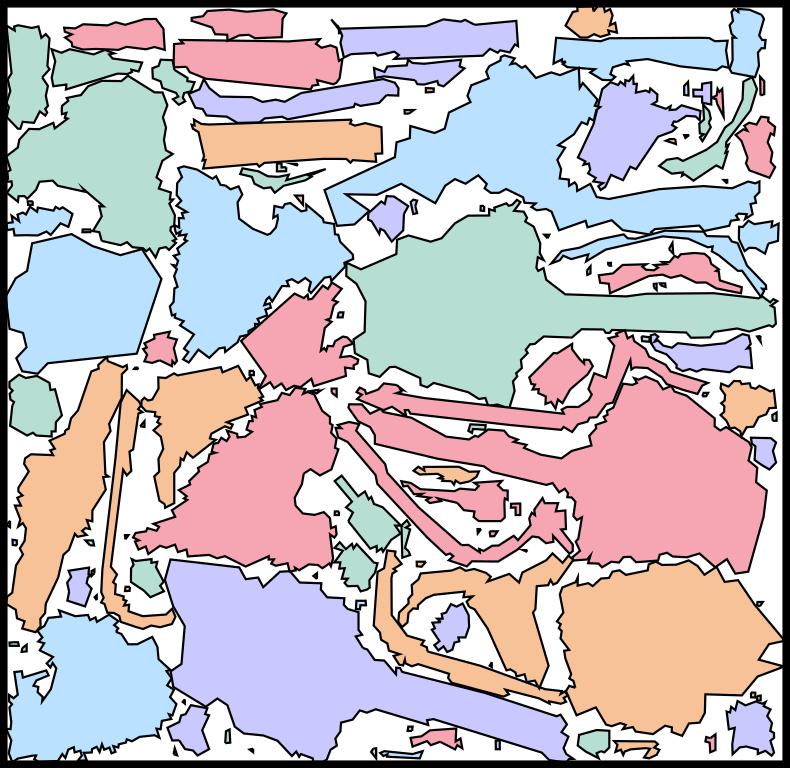}\\
 & \scalebox{0.7}{pr: 63.9\%} &\scalebox{0.7}{pr: 64.0\%} & \scalebox{0.7}{pr: 67.5\%} & \scalebox{0.7}{pr: 73.5\%}\\
 \midrule
\includegraphics[scale=.1]{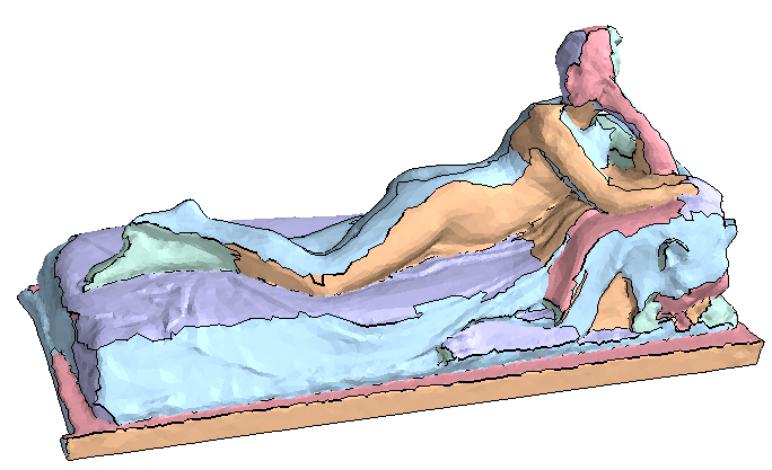}&
\includegraphics[scale=.1]{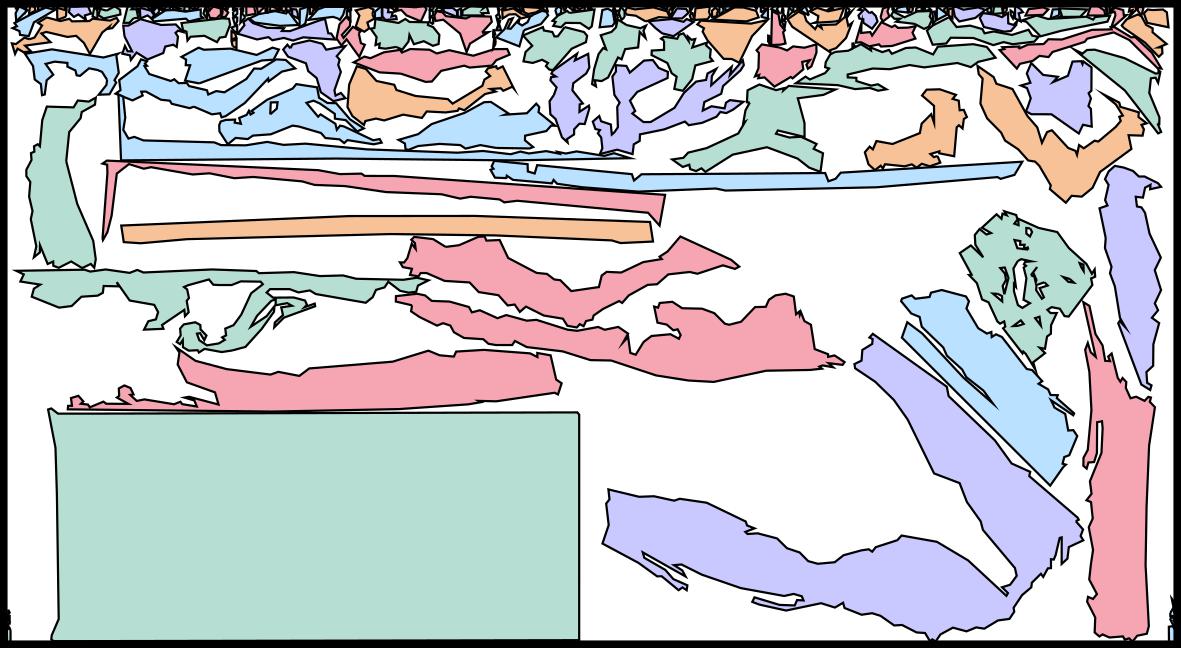}&
\includegraphics[scale=.1]{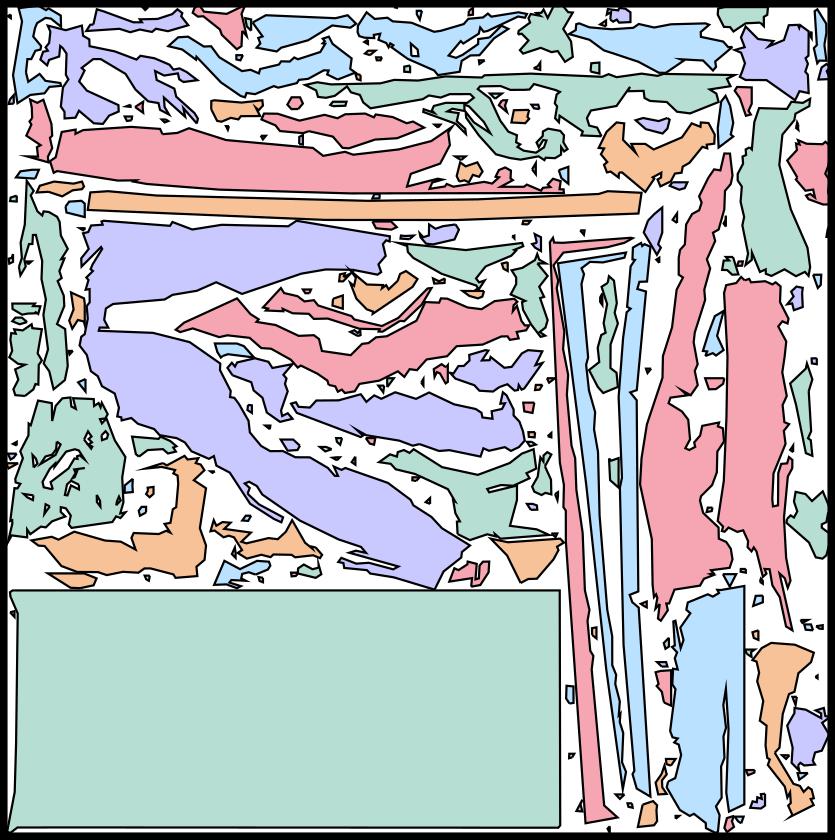}&
\includegraphics[scale=.1]{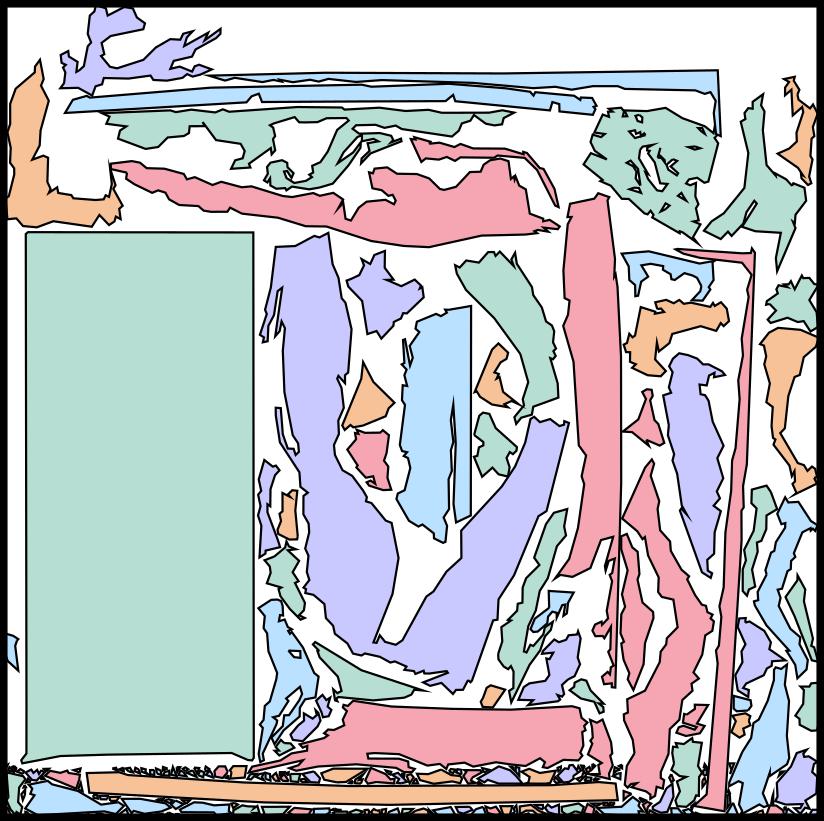}&
\includegraphics[scale=.1]{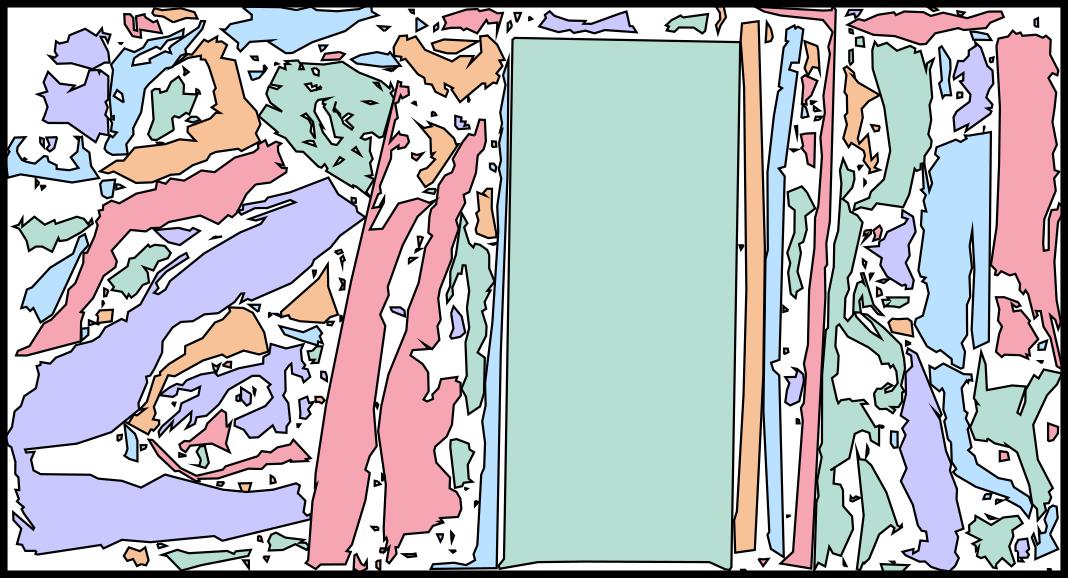}\\
 & \scalebox{0.7}{pr: 58.2\%} &\scalebox{0.7}{pr: 64.9\%} & \scalebox{0.7}{pr: 65.7\%} & \scalebox{0.7}{pr: 71.7\%}\\
 \midrule
\includegraphics[scale=.1]{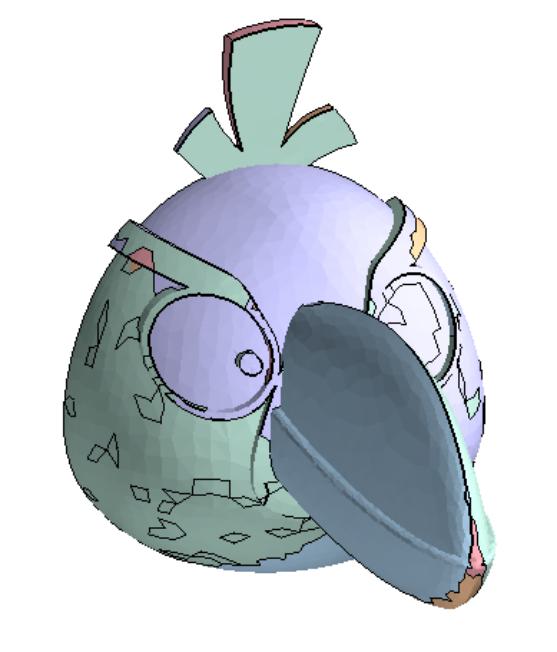}&
\includegraphics[scale=.1]{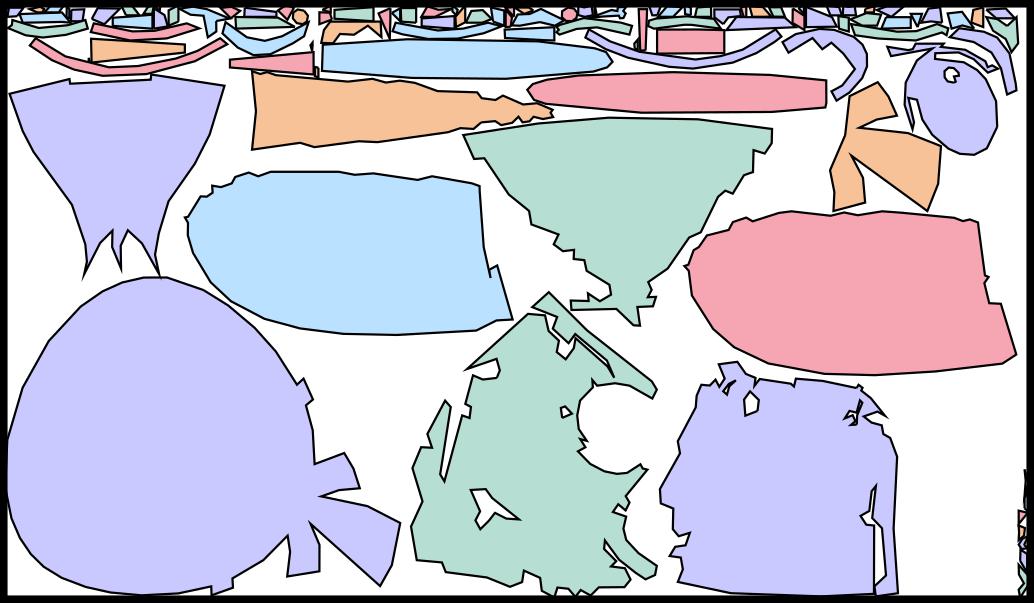}&
\includegraphics[scale=.1]{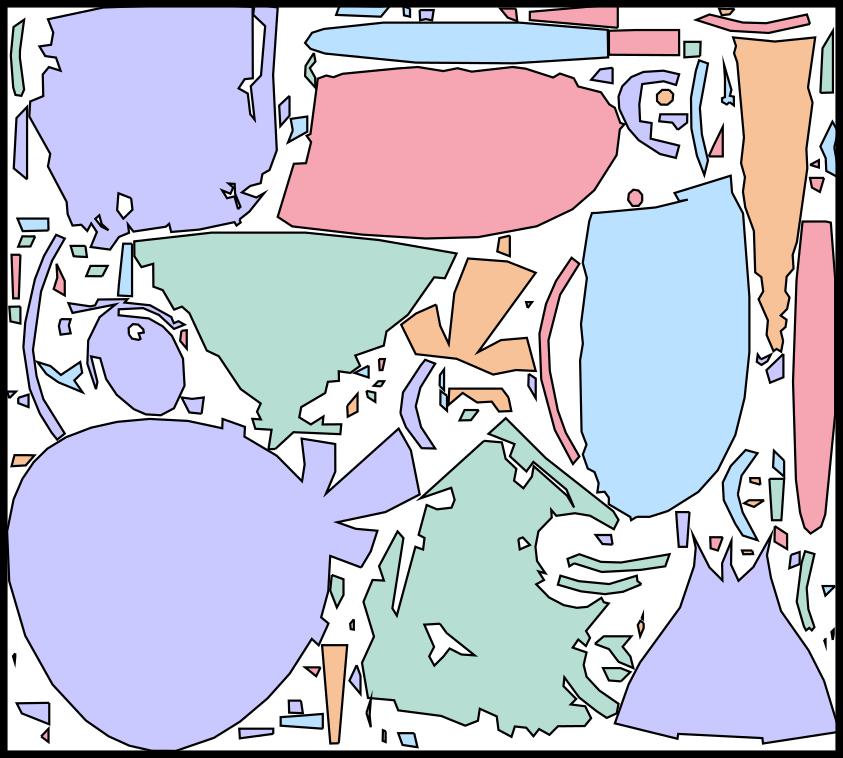}&
\includegraphics[scale=.1]{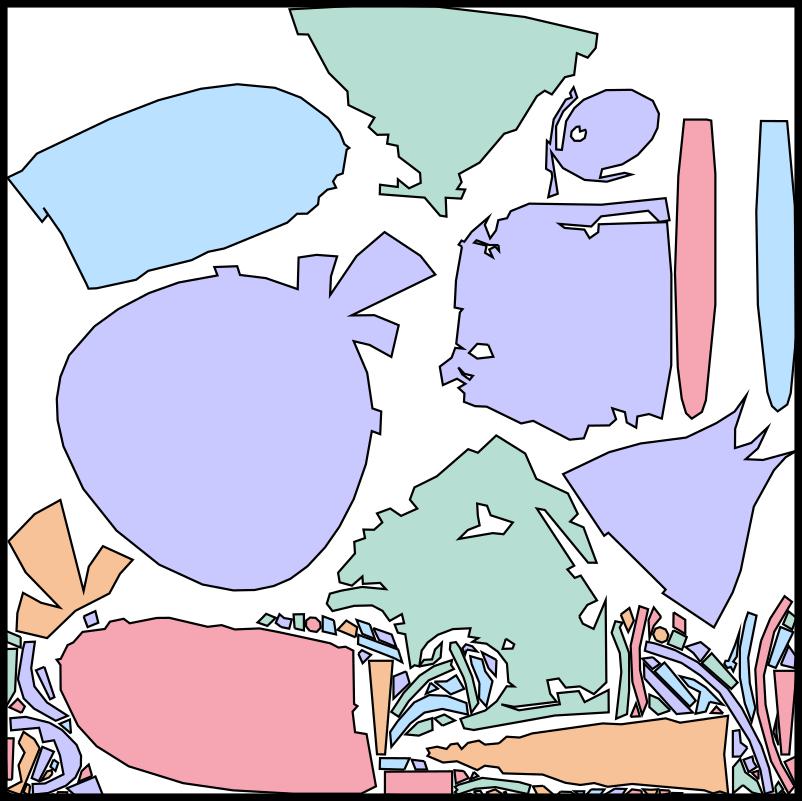}&
\includegraphics[scale=.1]{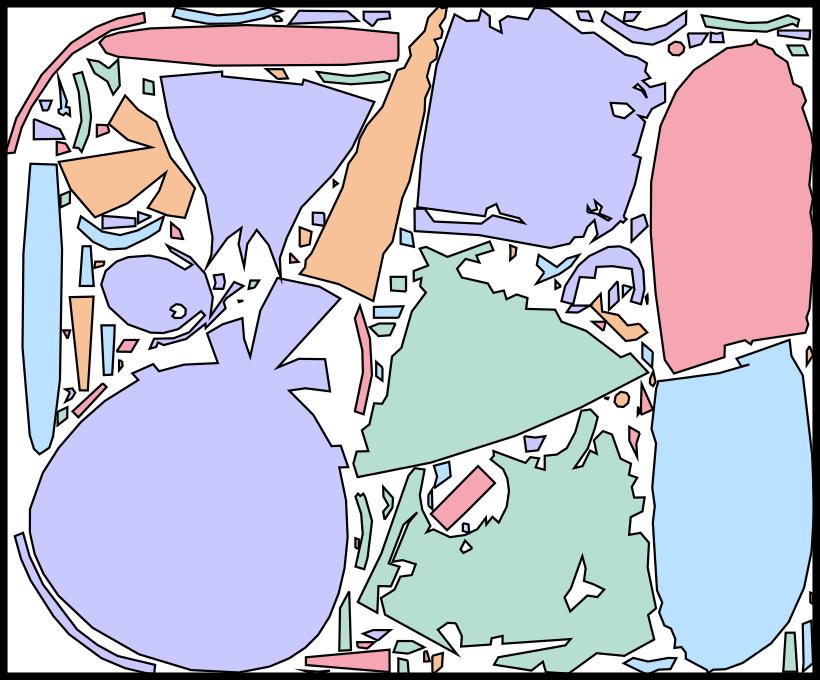}\\
 & \scalebox{0.7}{pr: 71.7\%} &\scalebox{0.7}{pr: 73.4\%} & \scalebox{0.7}{pr: 69.0\%} & \scalebox{0.7}{pr: 79.3\%}\\
 \midrule
 \includegraphics[scale=.1]{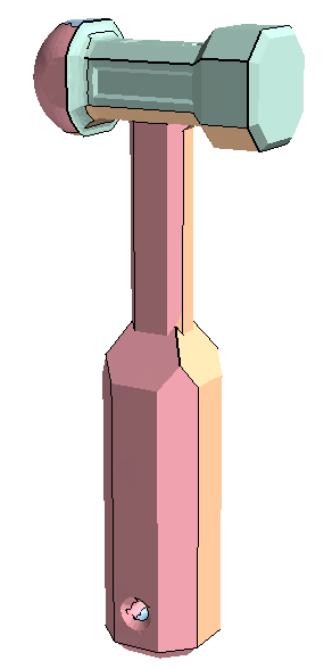}&
\includegraphics[scale=.1]{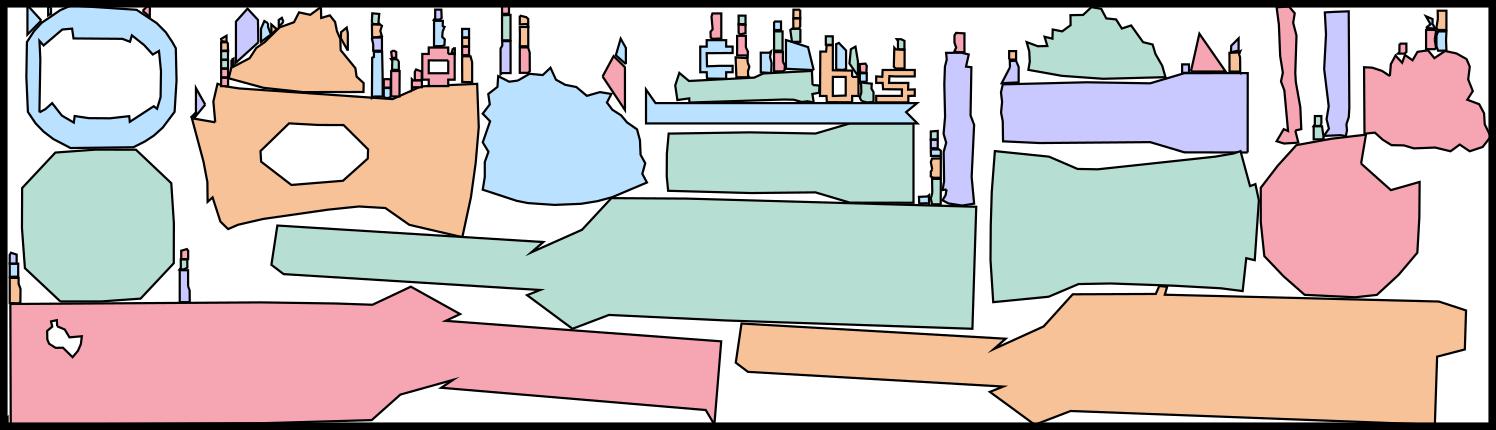}&
\includegraphics[scale=.1]{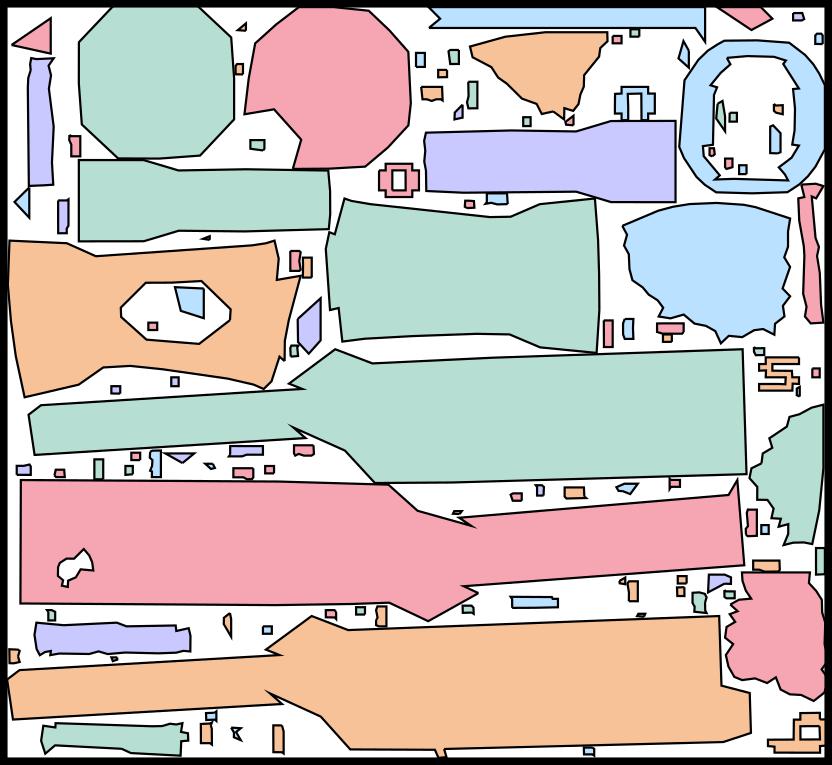}&
\includegraphics[scale=.1]{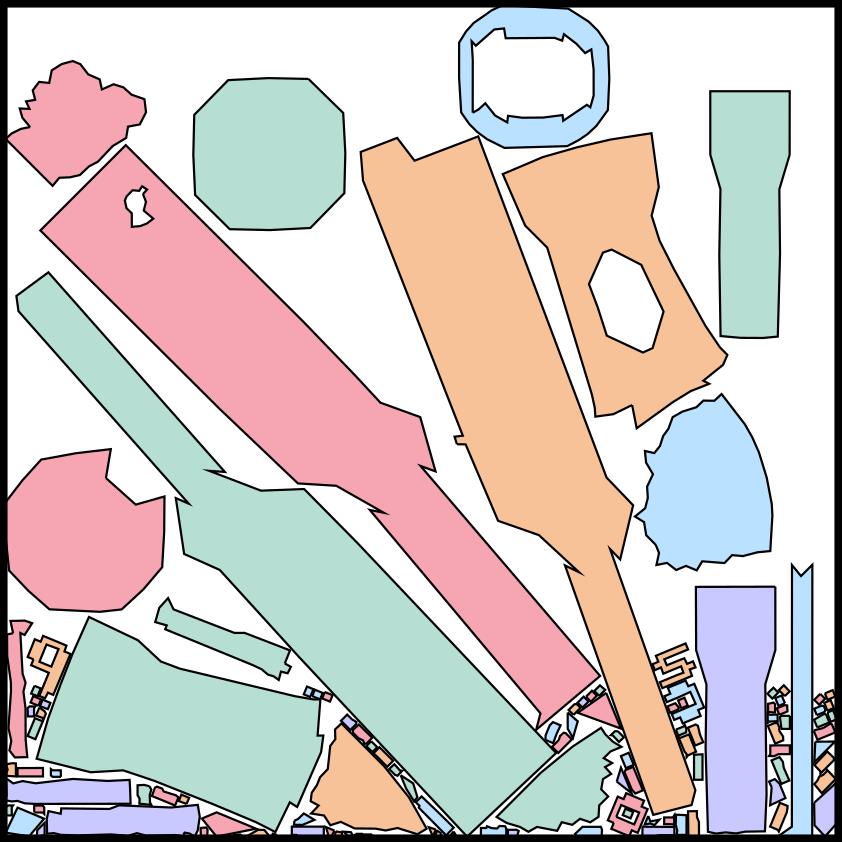}&
\includegraphics[scale=.1]{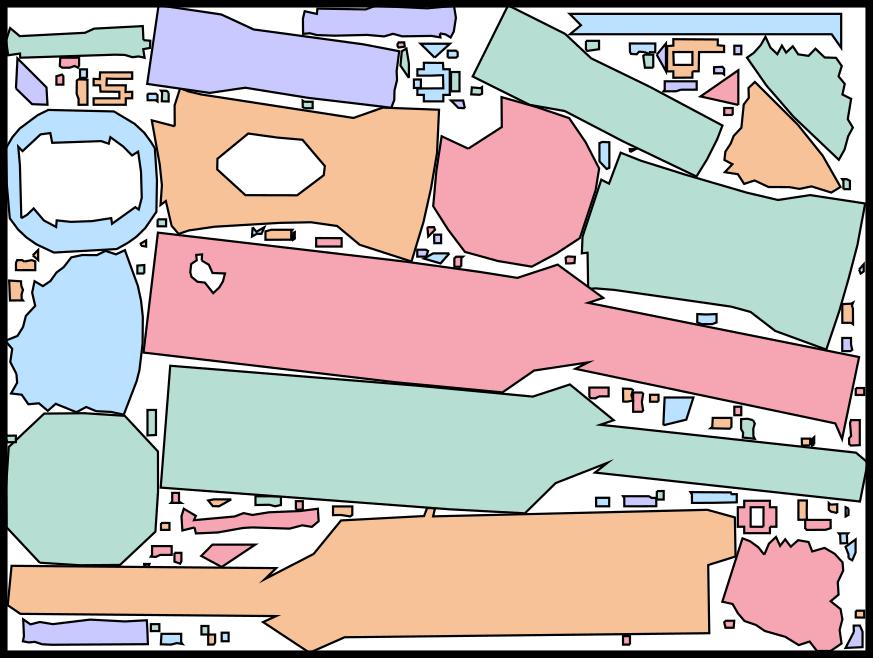}\\
 & \scalebox{0.7}{pr: 67.2\%} &\scalebox{0.7}{pr: 69.7\%} & \scalebox{0.7}{pr: 62.5\%} & \scalebox{0.7}{pr: 77.5\%}\\
 \midrule
\includegraphics[scale=.1]{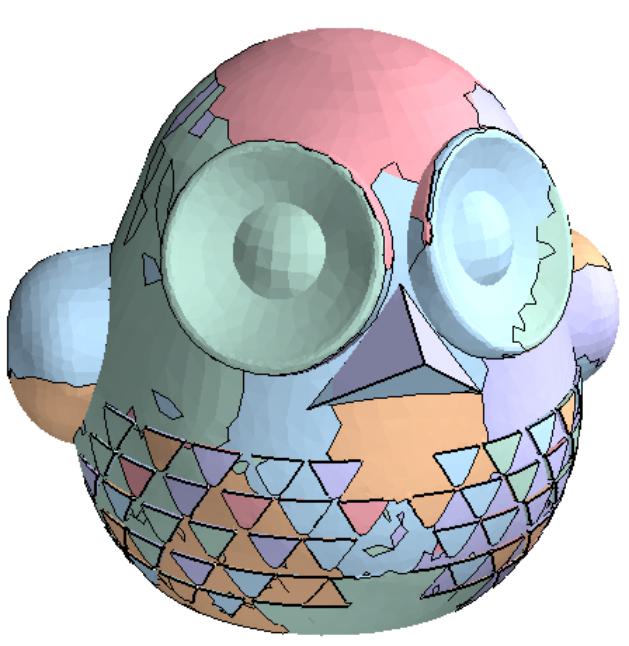}&
\includegraphics[scale=.1]{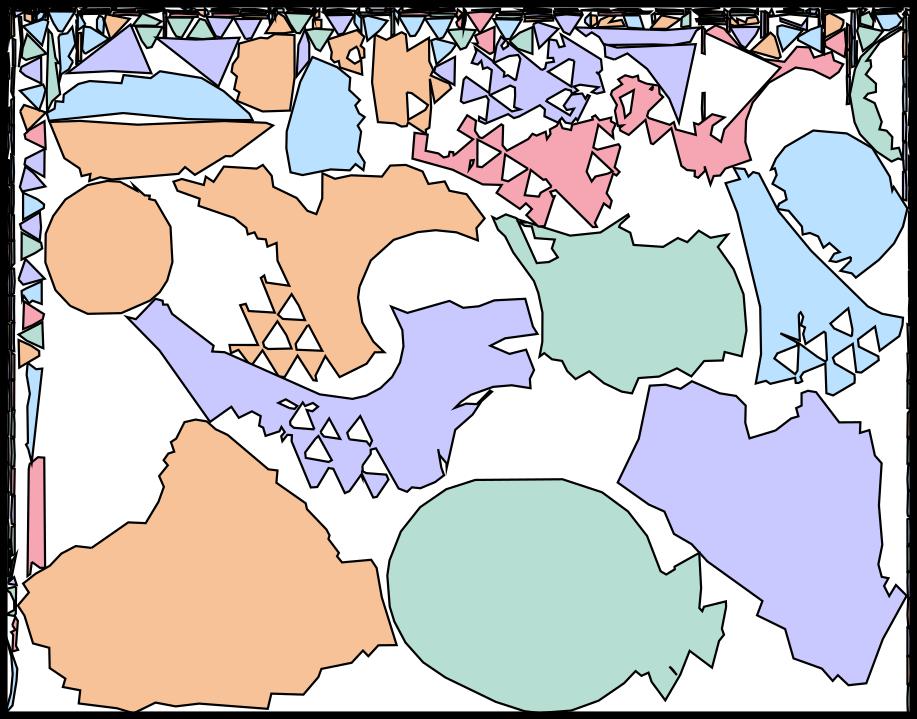}&
\includegraphics[scale=.1]{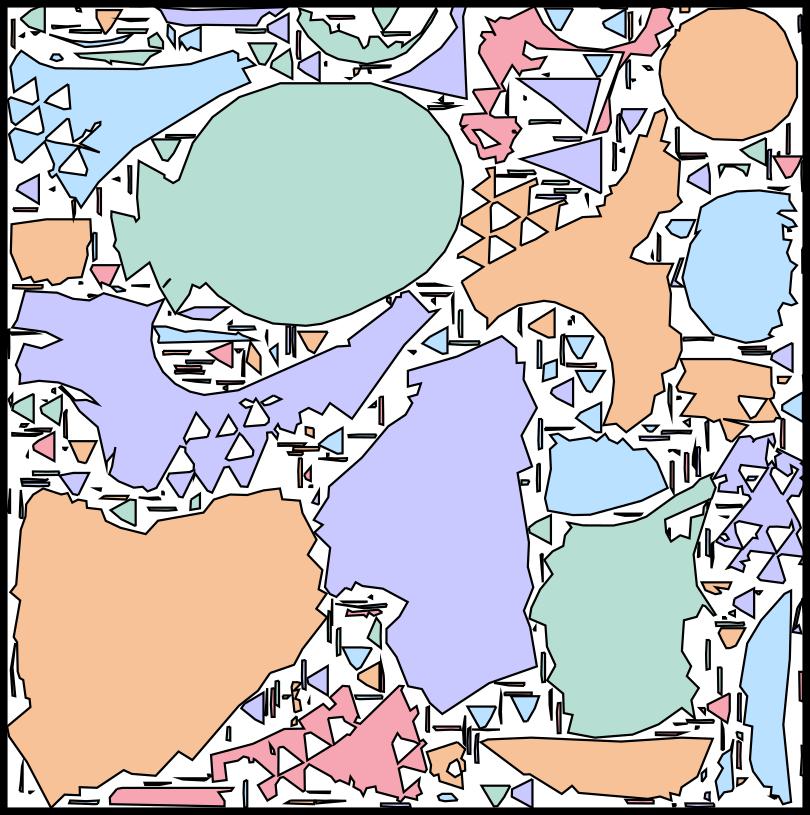}&
\includegraphics[scale=.1]{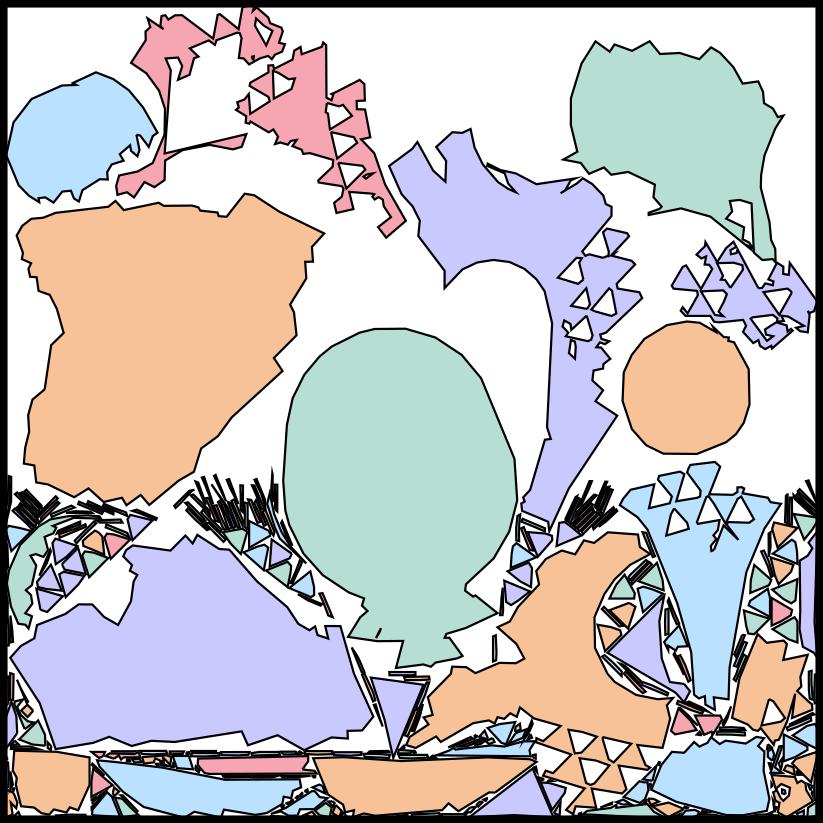}&
\includegraphics[scale=.1]{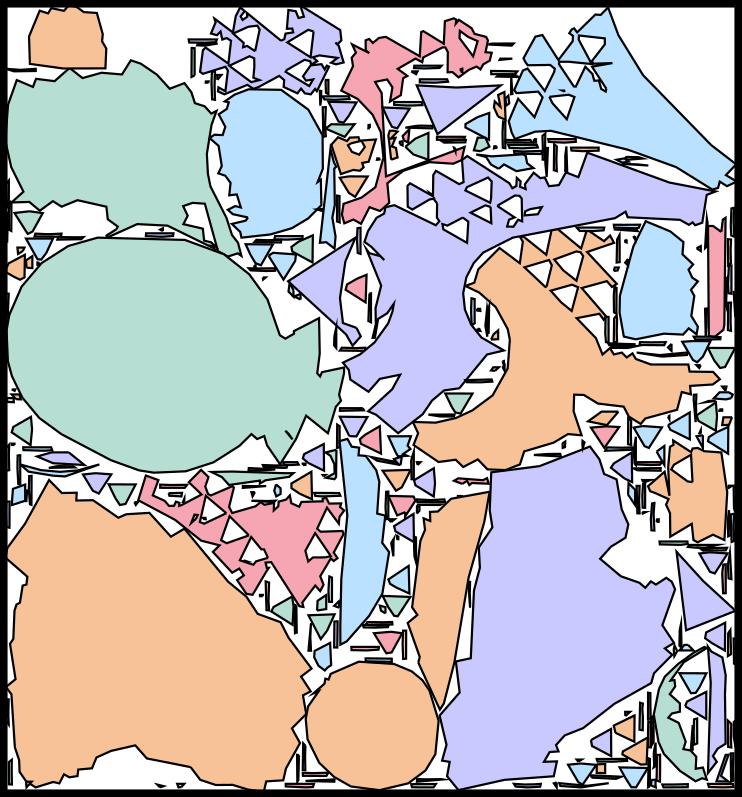}\\
 & \scalebox{0.7}{pr: 67.7\%} &\scalebox{0.7}{pr: 70.7\%} & \scalebox{0.7}{pr: 65.6\%} & \scalebox{0.7}{pr: 75.9\%}\\
 \midrule
\includegraphics[scale=.1]{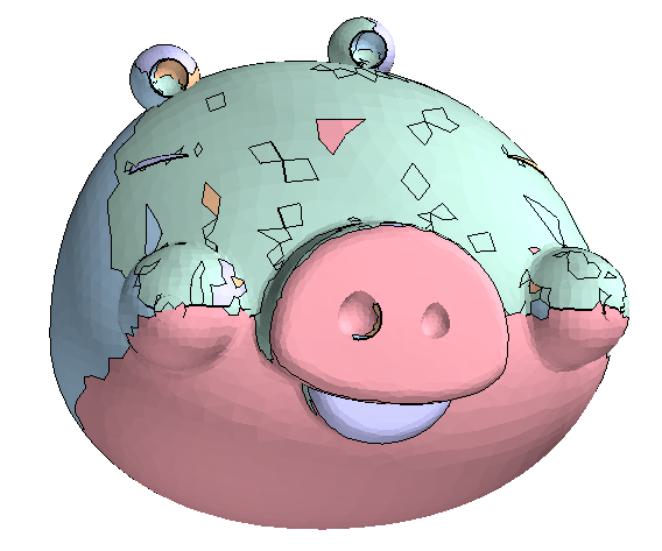}&
\includegraphics[scale=.1]{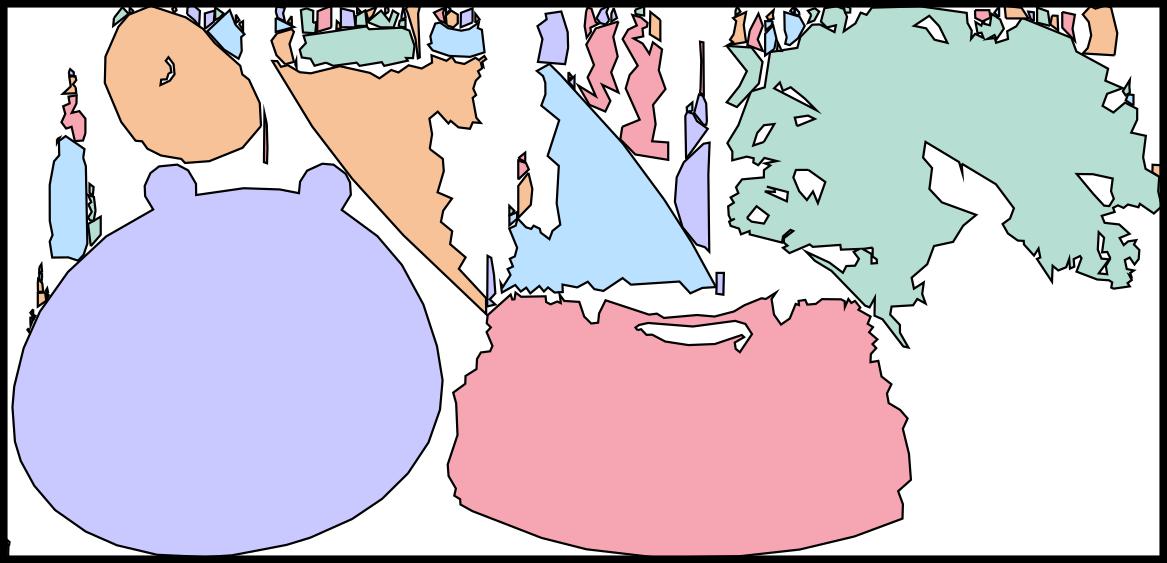}&
\includegraphics[scale=.1]{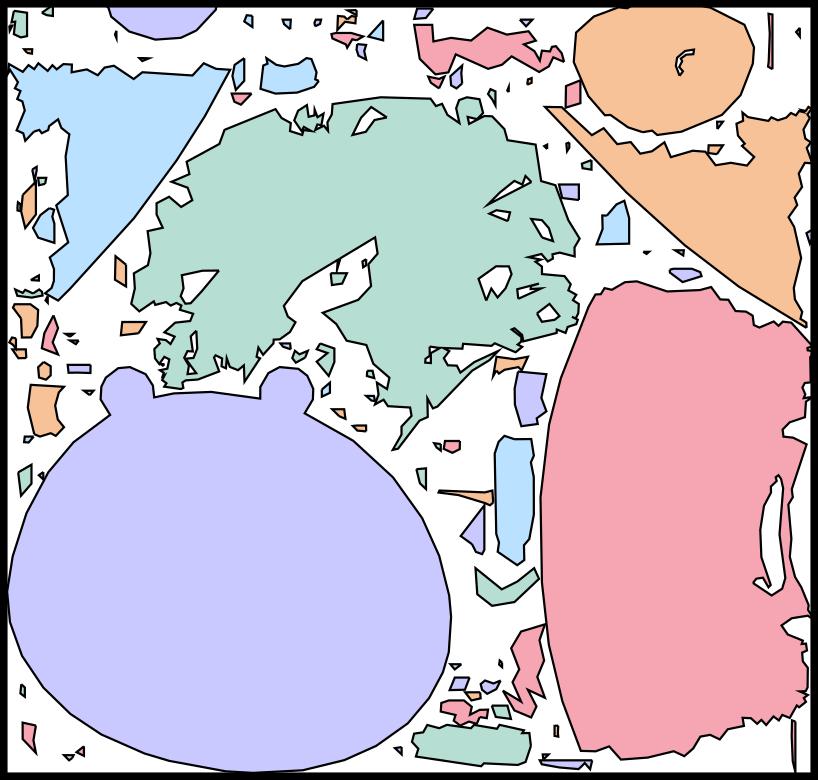}&
\includegraphics[scale=.1]{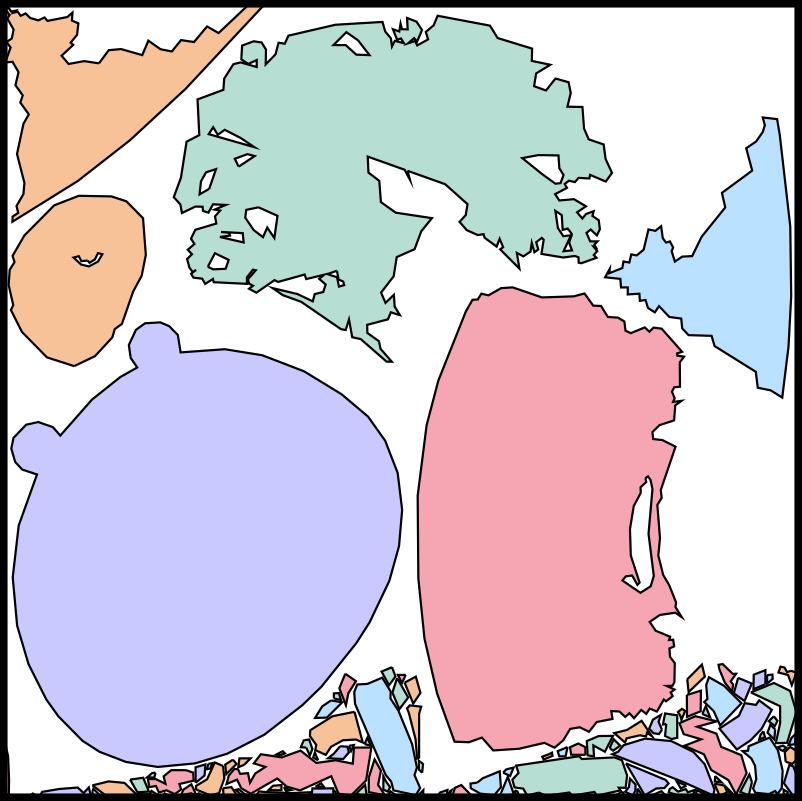}&
\includegraphics[scale=.1]{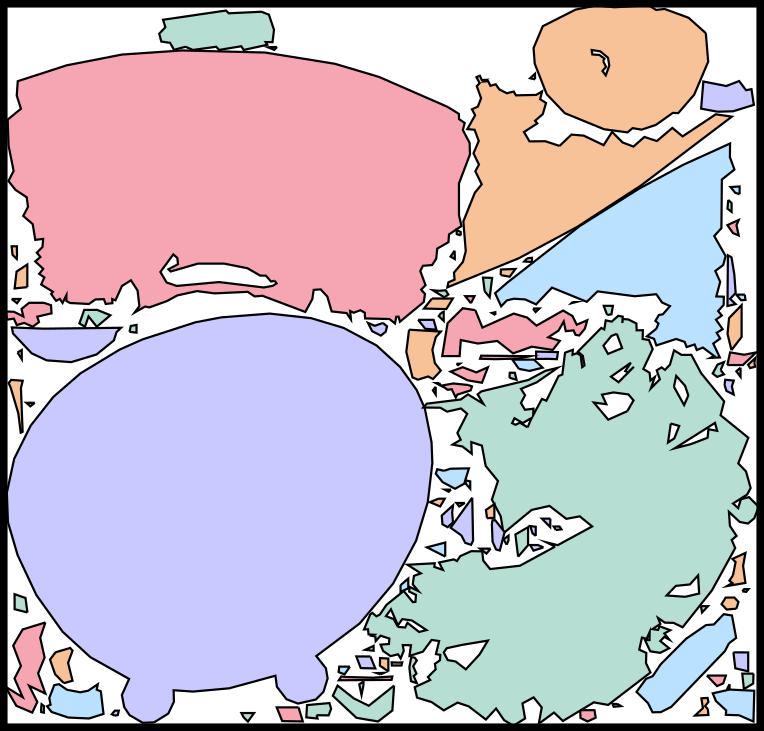}\\
 & \scalebox{0.7}{pr: 67.0\%} &\scalebox{0.7}{pr: 70.3\%} & \scalebox{0.7}{pr: 68.9\%} & \scalebox{0.7}{pr: 79.6\%}\\
 \bottomrule
\end{tabular}}
\egroup
\end{figure*}
\end{document}